\documentstyle[pre,aps,eqsecnum,amssymb,multicol,epsf,psfrag,rotate]{revtex}

\newcommand{\etal}{{\em et\/al. }}
\newcommand{\Qlaser}{{\bf K}}
\newcommand{\Qlaserscalar}{K}
\newcommand{\Iin}{I_{\text{in}}}
\newcommand{\Iout}{I_{\text{out}}}
\newcommand{\UQ}{U_{K}}
\newcommand{\HLFS}{H_{\text{LFS}}}
\newcommand{\HFS}{H_{\text{FS}}}
\newcommand{\HFSd}{H_{\text{FSd}}}
\newcommand{\Hd}{H_{\text{d}}}
\newcommand{\HSm}{H_{\text{Sm}}}
\newcommand{\HLSm}{H_{\text{LSm}}}
\newcommand{\Hel}{H_{\text{el}}}
\newcommand{\Tmdil}{T_{\text{m}}^{\text{dil}}}
\newcommand{\Tm}{T_{\text{m}}}

\newcommand{\TmSm}{T_{\text{mSm}}}
\newcommand{\TpS}{T_{\text{pS}}}
\newcommand{\TpSm}{T_{\text{pSm}}}
\newcommand{\unitx}{\hat{\bf e}_x}
\newcommand{\unity}{\hat{\bf e}_y}
\newcommand{\hatx}{\hat{\bf x}}
\newcommand{\haty}{\hat{\bf y}}
\newcommand{\kT}{k_{\text{B}}T}
\newcommand{\etaLFS}{\eta_{\bf G}}
\newcommand{\etaG}{{\overline \eta}_{\bf G}}
\newcommand{\etaK}{{\overline \eta}_{\bf K}}

\def\top#1{\vskip #1\begin{picture}(290,80)(80,500)\thinlines \put(
65,500){\line( 1, 0){255}}\put(320,500){\line( 0, 1){
5}}\end{picture}}
\def\bottom#1{\vskip #1\begin{picture}(290,80)(80,500)\thinlines \put(
330,500){\line( 1, 0){255}}\put(330,500){\line( 0, -1){
5}}\end{picture}}

\begin{document}
\tighten

\title{Novel Phases and Reentrant Melting of Two Dimensional Colloidal
Crystals}
\author{Leo Radzihovsky$^1$, Erwin Frey$^2$, and David R. Nelson$^2$} 
\address{$^1$Department of Physics,
University of Colorado, Boulder, CO 80309}
\address{$^2$Department of Physics, Harvard
University, Cambridge, MA 02138} 

\bibliographystyle{prsty}
\date{\today}
\maketitle
\begin{abstract}
  We investigate two-dimensional (2d) melting in the presence of a
  one-dimensional (1d) periodic potential as, for example, realized in
  recent experiments on 2d colloids subjected to two interfering laser
  beams. The topology of the phase diagram is found to depend
  primarily on two factors: the relative orientation of the 2d crystal
  and the periodic potential troughs, which select a set of Bragg
  planes running parallel to the troughs, and the commensurability
  ratio $p= a'/d$ of the spacing $a'$ between these Bragg planes to
  the period $d$ of the periodic potential. The complexity of the
  phase diagram increases with the magnitude of the commensurabilty
  ratio $p$.  Rich phase diagram, with ``modulated liquid'',
  ``floating'' and ``locked floating'' solid and smectic phases are
  found.  Phase transitions between these phases fall into two broad
  universality classes, roughening and melting, driven by the
  proliferation of discommensuration walls and dislocations,
  respectively.  We discuss correlation functions and the static
  structure factor in these phases and make detailed predictions of
  the universal features close to the phase boundaries. We predict
  that for charged systems with highly screened short-range
  interactions these melting transitions are generically reentrant as
  a function of the strength of the periodic potential, prediction
  that is in accord with recent 2d colloid experiments.  Implications
  of our results for future experiments are also discussed.
\end{abstract}
\vspace{-0.5cm}
\begin{multicols}{2}
\columnwidth3.4in

\narrowtext 

\section{Introduction}
\label{sec:introduction}

\subsection{Motivation and background}
\label{sec:motivation_background}

Two-dimensional (2d) melting and the mathematically related systems,
such as for example the normal-to-superfluid and planar
paramagnet-to-ferromagnet transitions in films (described by the 2d XY
model) are striking examples of the increased importance of thermal
fluctuations in low dimensional
systems.\cite{halperin_kyoto,nelson_domb-green} In contrast to their
bulk, three- (and higher-) dimensional analogs, where, typically,
fluctuations only lead to {\em quantitative} modifications of
mean-field predictions (e.g., change values of critical exponents),
here the effects are {\em qualitative} and drastic.  Located exactly
at the lower-critical dimension ($d_{\text{lc}}=2$), below which the
distinction between the high and low temperature phases is erased by
fluctuations, two-dimensional melting can proceed via a subtle,
two-stage, {\em continuous} transition, driven by unbinding of
topological defects (dislocations and disclinations). This mechanism,
made possible by strong thermal fluctuations, therefore provides an
alternative route to a direct first-order melting, argued by Landau's
mean-field analysis~\cite{landau:37} to be the {\em exclusive}
scenario.

Despite of its long history, dating back to the work of Kosterlitz and
Thouless~\cite{kosterlitz-thouless:73}, Halperin and
Nelson,\cite{nelson-halperin:79}, and Young\cite{Young:79} (KTHNY)
(which in turn built on a large body of ideas dating back to Landau
and Peierls\cite{Landau-Peierls}), interest in 2d melting and related
problems persist. On the theoretical side this, in part, is due to the
fact that the theory of 2d melting is an unusual example of a
nontrivial and quite exotic critical point that lends itself to an
asymptotically {\em exact} description. Furthermore, the KTHNY class
of transitions (2d melting and related disordering of a 2d XY model)
provides a rare example of a thermodynamically sharp phase transitions
between phases, both of which lack long range
order.\cite{Radzihovsky-Toner}

Although evidence for defect driven phase transitions has appeared
in numerous experiments on liquid crystals\cite{huang} and
Langmuir-Blodgett films\cite{knobler}, finding simple model systems
which exhibit these phenomena in experiments or simulations has proven
to be more controversial. Some system parameters appear to fall in the
range in which instead it is discontinuous melting that converts a
solid directly into a liquid. However, it appears, that two-stage
continuous melting has been recently experimentally observed by Murray
\etal\cite{murray-springer-wenk:90} and Zahn \etal\cite{zahn-etal:99}
in beautiful melting experiments on two-dimensional colloids confined
between smooth glass plates and superparamagnetic colloidal systems,
respectively. In these experiments, an orientationally
quasi-long-range ordered but translationally disordered hexatic
phase\cite{nelson-halperin:79} was observed.  This phase, intermediate
but thermodynamically distinct from the 2d solid and isotropic liquid,
is an important signature of defect driven two-stage melting. In these
two-dimensional colloids, particle positions and the associated
topological defects can be directly imaged via digital
video-microscopy, allowing precise quantitative tests of the theory.
Colloids are thus ideal experimental model systems to explore the
details of two-dimensional melting and related phenomena, many of
which are the focus of the theory presented here.\cite{ourPRL}

Soon after the initial development of the theory of two-dimensional
melting, theoretical efforts turned to the studies of the effects of
substrate, an important ingredient in many physical systems. These
studies\cite{coppersmith,pokrovsky-talapov-bak:86} uncovered a rich
phenomenology stemming from the interplay between the underlying
periodic substrate and a quasi-long-range ordered solid film
interacting with it. While many experiments have been undertaken, with
a krypton film on a graphite substrate (see e.g.
Ref.\onlinecite{pokrovsky-talapov-bak:86} for a review) being one of
the best studied, these systems are far from ideal in exploring this
rich phenomenology, because of the lack of substrate tunability; in
these systems it is difficult to change the substrate period,
dimensionality and pinning strength.

A series of pioneering experiments by Chowdhury, Ackerson and Clark
\cite{chowdhury-ackerson-clark:85} constituted an important new
development.  In these studies strongly interacting colloidal
particles, confined to two dimensions, were subjected to a
one-dimensional periodic potential, induced by the interference
fringes from two laser beams crossed in the sample. The light-induced
polarization in these micron-size dielectric particles interacts with
the laser electric field, leading to a radiation pressure
force,\cite{Ashkin:71} directed toward the regions of high laser
intensity, the antinodes maxima in the laser standing wave pattern.

One of many interesting phenomena discovered by Chowdhury \etal is the
fixed-temperature freezing transition driven by increasing the
strength of the laser potential, dubbed ``light induced freezing''
(LIF).  Qualitatively, LIF is due to the suppression of thermal
fluctuations transverse to the imposed periodic pinning laser
potential. This intuition is also supported by a more quantitative
analysis based on Landau's free energy expansion in the translational
order parameters (density Fourier modes) $\rho_{{\bf G}_i}$, with
$\{ {\bf G}_i \}$ the three smallest reciprocal lattice vectors of a
triangular lattice. In the simplest geometry, with e.g., ${\bf G}_1$
commensurate with the laser potential, $\langle \rho_{{\bf G}_1}
\rangle$ is trivially induced by the potential even in the liquid
phase. Such finite $\langle \rho_{{\bf G}_1} \rangle$ then converts
the Landau's cubic coupling $\rho_{{\bf G}_1}\rho_{{\bf
    G}_2}\rho_{{\bf G}_3}$, (which, in mean-field theory, is
responsible for melting always being first-order) into a simple upward
shift in the melting temperature for the only remaining critical mode
$\psi\equiv\rho_{{\bf G}_2}-\rho_{{\bf G}_3}$.  Not surprisingly the
resulting Landau expansion contains only even powers of this complex
order parameter $\psi$, which therefore generically orders via a {\em
  continuous} transition in the XY universality class.  Hence, within
the mean-field description discussed by Chowdhury
\etal\cite{chowdhury-ackerson-clark:85}, one expects to reach a
tricritical point upon increasing the light intensity, beyond which
the LIF transition becomes continuous.

However, because of the dominant role of thermal fluctuations in
two-dimensional systems, such ``soft-spin'' Landau expansions in order
parameter amplitudes (and the related density functional
theories\cite{chakrabarti-etal:94}) will have difficulties to capture
the subtleties of the continuous topological phase transitions
possible in these two-dimensional systems. Unfortunately, results from
Monte-Carlo simulations are inconclusive.  Although earlier
simulations~\cite{chakrabarti-etal:95} claimed to have found a
tricritical point at intermediate laser intensities, consistent with
density functional theory, recent studies from the same
laboratory~\cite{das-sood-krishnamurthy:99} refute these results.
These difficulties are perhaps unsurprising, given that even much
larger scale simulations have, so far, failed to completely resolve
the nature of 2d melting, even {\em without} a periodic external
potential.\cite{bagchi-andersen-swope:96}

An alternative (but complementary and in principle equivalent)
``hard-spin'' defect description (with order parameter amplitude
fluctuations represented by defect cores), extended to include a
one-dimensional periodic pinning potential may be necessary to
correctly capture the rich phenomenology of the early experiments by
Chowdhury \etal\cite{chowdhury-ackerson-clark:85} and the recent ones
by Wei \etal\cite{wei-bechinger-rudhardt-leiderer:98} and
others.~\cite{note:murray} Developing such a theoretical framework and
exploring its details to interpret these experiments is the goal of
the work presented here.

Our interest in this problem was stimulated by the experiments of Wei
\etal\cite{wei-bechinger-rudhardt-leiderer:98}, which extended the
light-induced melting experiments to higher laser intensities than
those studied in Ref.\onlinecite{chowdhury-ackerson-clark:85}.  One
other notable difference is that in contrast to the strong long-range
interaction of unscreened charged colloids in highly deionized
solution,\cite{chowdhury-ackerson-clark:85} in the Wei \etal
experiments colloidal particles were interacting via a short-ranged
Debye potential, with ions in the solution screening the long-ranged
Coulomb interaction. In addition to the light-induced freezing,
observed at low light intensities, the authors of
Ref.\onlinecite{wei-bechinger-rudhardt-leiderer:98} discovered a {\em
  reentrant} melting phenomena, ``light-induced melting'' (LIM),
driven by the increased strength of the laser-induced one-dimensional
periodic potential. As discussed below, this fascinating reentrance
phenomena {\em generically} emerges from our theoretical analysis in
the limit of a short Debye screening length.

The goal of this paper is to investigate the phenomena of
two-dimensional melting in the presence of a one-dimensional periodic
potential, and to answer many basic questions stimulated by these
recent experiments. What is the nature of such melting transition, if
not preempted (as it can always be) by the first-order transition?
More generally, how is the standard phase diagram for 2d melting on a
homogeneous substrate (which includes the 2d crystal, hexatic and
liquid phases) modified by the periodic laser potential?  Which of the
phases survive the light field and what new ones emerge in its
presence? The answers to these and many other questions, provided
below, lead to results consistent with experimental observations, and
have many testable consequences for the possible future experiments.

\subsection{Summary of the results}
\label{sec:results}

Even in the liquid phase at high temperatures the laser interference
fringes, which we choose to run along the $x$-axis, induce a periodic
density modulation in the colloidal liquid. As a consequence the
static structure function $S({\bf q})$ displays Bragg peaks at
$\Qlaser_n\equiv n \, (2\pi/d) \, \haty$, the integer multiples ($n
\in {\cal Z}$) of the reciprocal lattice vector $\Qlaser = (2 \pi/d)
\, \haty$ of the imposed one-dimensional periodic potential with
trough spacing $d$.\cite{Penn_group} The liquid phase density exhibits
a {\em finite} linear response to such a periodic perturbation with
amplitude $\UQ$, which is proportional to the input laser intensity
$\Iin$. This is consistent with the observations of Chowdhury
\etal\cite{chowdhury-ackerson-clark:85} who found the scattered laser
intensity, $\Iout$, at these directly induced Bragg peaks, to scale as
a cube of the input laser power $\Iin$.\cite{IinIout_liquid} These
explicitly induced features of the modulated liquid persist throughout
the phase diagram, with the additional structure emerging as a result
of numerous {\em spontaneous} symmetry breakings, which we discuss
below.

The laser-induced periodic potential also {\em explicitly} breaks
continuous 2d rotational symmetry down to ${\cal{Z}}_2$ symmetry
(rotations by $\pi$). Consequently, the one-dimensional periodic
potential induces nematic, square, hexatic and higher orientational
harmonics long-range orders, respectively characterized by a 2n-atic
bond orientational order parameter $\psi_{2n} = \langle e^{i
2n\theta({\bf r})} \rangle$, which, independent of any other details,
are nonzero throughout the phase diagram.  Therefore, in particular,
the laser potential eliminates the continuous transition from an
isotropic liquid to a hexatic liquid phase, expected in
two-dimensional liquids in the absence of an external
potential.\cite{nelson-halperin:79} This situation is analogous to a
ferromagnet in a magnetic field, where the qualitative distinction
between paramagnetic and ferromagnetic phases is erased by the
external magnetic field, with both phases displaying a finite induced
magnetization.

Since the hexatic orientational order is {\em explicitly} induced by
the laser potential, it must vanish as the laser field is tuned to
zero. Analogously to a power-law vanishing of the magnetization with
an external magnetic field in a ferromagnet at its critical point, we
predict that at low input light intensities, $\Iin$, the orientational
order parameter vanishes as a {\em universal} power of $\Iin$,
\begin{equation}
\psi_6\sim \Iin^{1/\delta_{\psi_6}}\;,\label{psi6}
\end{equation}
with $1/\delta_{\psi_6}=6$\cite{delta_psi6comment} in the liquid phase
and
\begin{equation}
1/\delta_{\psi_6}={6{\overline\eta}_{6}\over 4-{\overline\eta}_{6}}\;,
\label{delta_psi6}
\end{equation}
in the hexatic phase, where ${\overline \eta}_6$ is the exponent
describing the algebraic decay of bond orientational order in the
absense of the laser-induced periodic
potential.\cite{nelson-halperin:79} We expect $\psi_6$ to approach a
nonzero $\Iin$-independent constant in the solid phase, consistent
with the {\em spontaneous} long-range hexatic order of the 2d crystal,
even in the absence of a periodic potential.

All other details of the phase diagram and the properties of the
phases for our system strongly depend on the level of commensurability
between the two-dimensional colloidal crystal, in the absence of the
laser field, and the one-dimensional periodic potential that it
induces. This in turn is determined by two ingredients: (i) the
orientation of the triangular colloidal lattice relative to that of
the periodic potential troughs, which selects a set of Bragg planes
that run parallel to the troughs, (ii) the commensurability ratio of
the spacing $a'$ between these Bragg planes to the period $d$ of the
laser potential, defined by $p\equiv a'/d$. In this paper we will
primarily focus on the commensurate case defined by $p\in {\cal Z}$
and defer the rich phenomenology of the incommensurate case and the
commensurate-incommensurate transitions to a later
publication.\cite{fnr_future}

For these commensurate densities, independent of the order of
commensurability, $p$, at the lowest temperatures we always find that
our system freezes into a novel type of a crystal, which we call a
``locked floating solid'' (LFS). This phase derives its apparently
contradictory name from its novel highly anisotropic properties: while
the solid is pinned transversely to the troughs of the periodic
potential, executing only massive optical phonon-like excitations in
that direction, it is able to slide freely along the potential minima
with acoustic phonon excitations within the troughs. Upon integrating
out the massive $u_y$-modes and using standard renormalization group
methods \cite{nelson-halperin:79} to eliminate bound dislocation pairs
in the LFS phase, we are left with a free energy with temperature and
potential strength dependent {\em effective} elastic constants,
\begin{eqnarray}
  \HLFS = \frac{1}{2} \int d^2 r \Bigl\{
  K_{\text{eff}} \, (\partial_x u_x)^2 +
  \mu_{\text{eff}} \, (\partial_y u_x)^2 \Bigr\} \; .
\label{eq:KT_free_energy}
\end{eqnarray}
The structure function of LFS is quite unusual. Like the high
temperature modulated liquid discussed above, the LFS displays a set
of delta-function Bragg peaks (reduced by the Debye-Waller factor)
located at the multiples of the laser potential reciprocal lattice
vector $\Qlaser = (2\pi/d) \haty$, which coexist with other {\em
  spontaneously} induced Bragg and quasi-Bragg peaks.

The more detailed properties of the LFS and other phases exhibited by
our system, strongly depend on the choice of the infinite set of
colloidal crystal orientations relative to the light interference
fringes. While we will explore these numerous possibilities in their
full generality in the main body of the manuscript, in this subsection
we summarize our results only for the simplest orientation studied in
the experiments of
Refs.\cite{chowdhury-ackerson-clark:85,wei-bechinger-rudhardt-leiderer:98},
in which the periodic potential troughs run parallel to the {\em
  primary} Bragg planes.\cite{comment_define}

Experimentally, we expect our system to display a considerable amount
of irreversibility, with the choice of the relative orientation highly
dependent on the way the system is taken into the crystal state: if
the laser potential is turned on in the liquid phase (field-cooled),
the crystal will freeze into the lowest energy orientation consistent
with the imposed colloidal density (or the chemical potential) and
laser fringe spacing;\cite{levitov91} in contrast, in zero-laser-field
cooling experiments, an already formed crystal may be unable to
reorient significantly, and will therefore lock into a metastable
orientation, determined by the plane of the two interfering laser
beams.

\begin{figure}[bht]
  \narrowtext \centerline{ \epsfxsize=0.9 \columnwidth
    \epsfbox{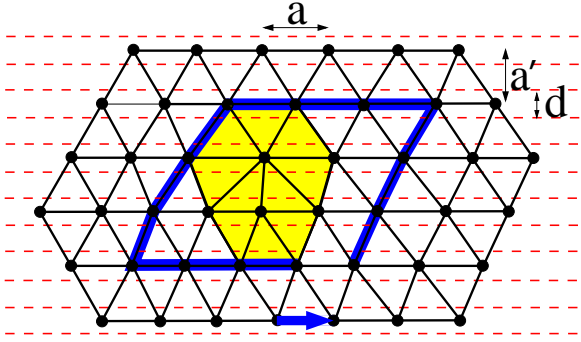} } \vspace{0.5cm}
\caption{Triangular lattice with lattice constant $a$ subject to a
periodic potential (maxima indicated by dashed lines) for $p
d=a^\prime$ with $a^\prime = \sqrt{3}a/2$ and $p=2$. Also shown is the
low energy dislocation with Burgers vector ${\bf b}$ parallel to the
corrugation of the external potential.}
\label{fig:dislocation}
\end{figure} 
Once we focus on the primary orientation, illustrated for $p=2$ in
Fig.\ref{fig:dislocation}, the phenomenology of our system is
completely determined by the integer commensurability ratio $p$. As we
will show, for commensurate densities, our system admits three phase
diagram topologies, corresponding to the three ranges of the values of
$p$: (i) $p=1$,\cite{comment_p=1} (ii) $1<p\leq p_c$, and (iii)
$p>p_c$, with the critical value of $p_c\approx 3.7$ for the primary
orientation.

\subsubsection{Commensurability ratio $p=1$:}
\label{sec:p=1}

\begin{figure}[bth]
  \narrowtext \centerline{\epsfxsize=0.9\columnwidth
    \epsfbox{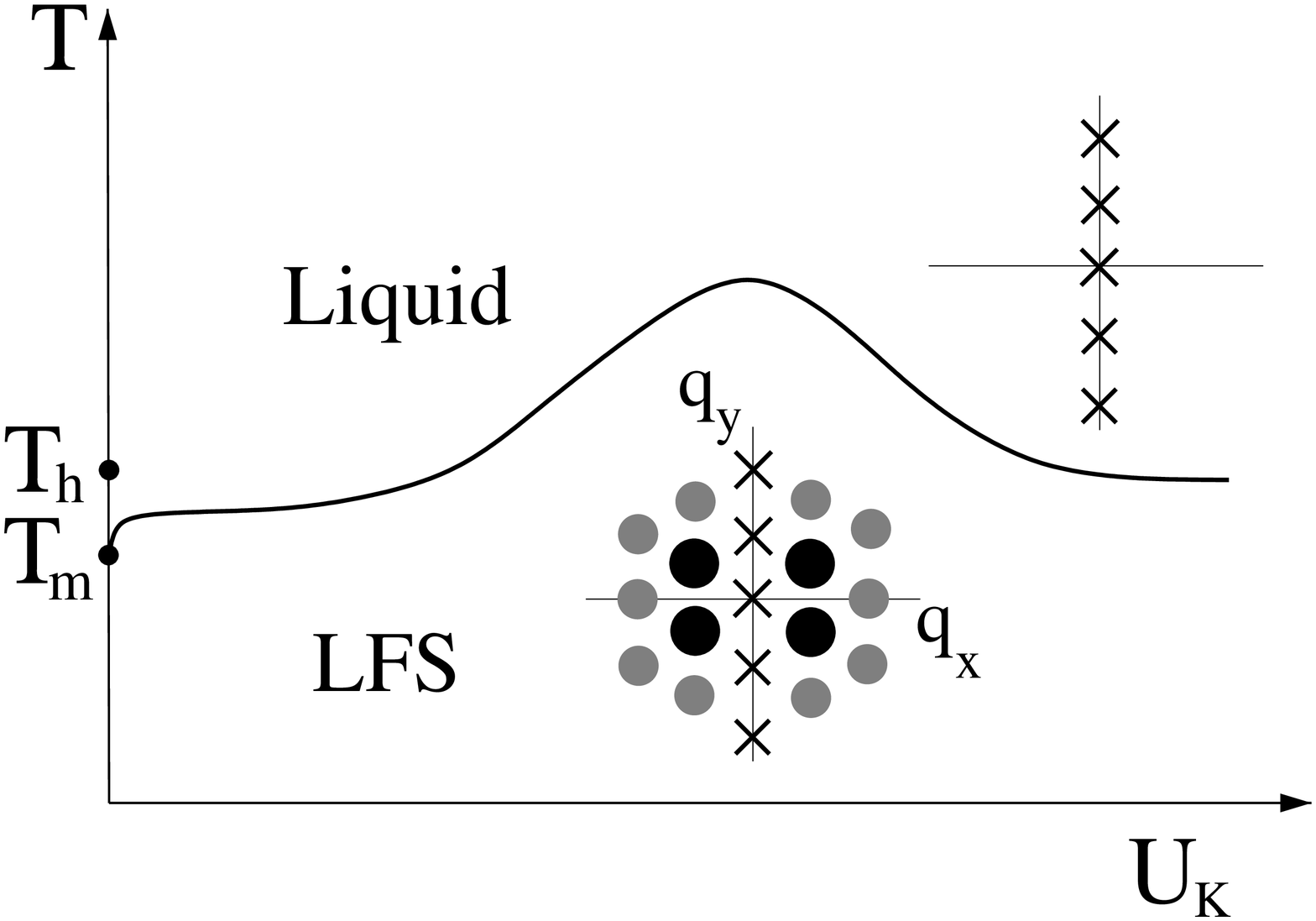}} \vspace{0.5cm}
\caption{Schematic phase diagram for a primary commensurate
orientation with commensurability ratio $p=1$. $T_h$ indicates the
transition temperature from the hexatic to the isotropic liquid phase at
$\UQ=0$. {\em Insets:} Schematic structure functions in the various
phases.  The $\times$'s indicate delta-function Bragg peaks and the
shaded circles algebraic peaks.}
\label{fig:phase_diagram_p=1}
\end{figure}

For $p=1$,\cite{comment_p=1} we find the phase behavior of the 2d
colloidal system as summarized by the phase diagram illustrated in
Fig.\ref{fig:phase_diagram_p=1}.

\begin{figure}[bth]
  \narrowtext \centerline{\epsfxsize=0.99\columnwidth
    \epsfbox{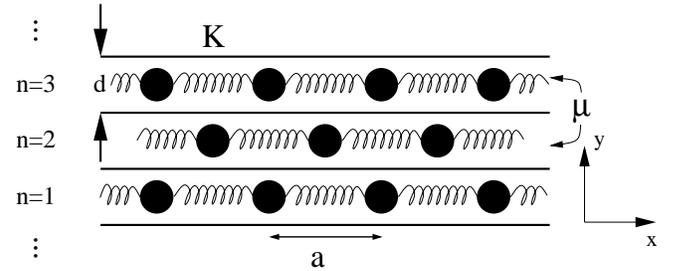}} \vspace{0.5cm}
\caption{Colloidal particles in channels, labeled by $n$, with 
  intrachannel compressional modulus $K$ and interchannel shear
  coupling $\mu$.}
\label{fig:1d_channels}
\end{figure}

Because the sharp distinction between the hexatic and isotropic liquid
phase is absent in the presence of a periodic potential, this phase
diagram contains only two thermodynamically distinct phases at finite
$\UQ$: the modulated liquid and the simplest $p=1$ ``locked floating
solid'' (LFS). We can estimate the order of magnitude of the
transition temperature between the LFS and liquid phases in terms of
microscopic elastic constants (similar to those appearing in
Eq.\ref{eq:KT_free_energy}) as follows: In the limit of strong laser
potential the particles are confined to a parallel array of equally
spaced 1d channels of spacing $d$, illustrated in
Fig.\ref{fig:1d_channels}.  If $u_n(x)$ is the particle displacement
field along the n-th channel, we can write the energy of these weakly
coupled one-dimensional rows of particles as
\begin{eqnarray}
H&=&d\sum_n\int dx\biggl\{{1\over2}K \left({d u_n\over d x}\right)^2\nonumber\\
&-&\mu\left({a\over2\pi d}\right)^2
\cos\left[{2\pi\over a}(u_{n+1}(x)-u_n(x))\right]\biggr\}\;,
\label{H-weak_coupling}
\end{eqnarray}
where $K$ is the bare compressional elastic modulus within each
channel and $\mu$ is the microscopic coupling between the channels
determining the shear modulus of the 2d system.\cite{d_comment} At
high temperatures or weak microscopic coupling $\mu$, the colloid
decomposes into an orientationally ordered two-dimensional liquid of
decoupled one-dimensional channels. At temperature $T$, the phonon
fluctuations within a channel then grow according to
\begin{equation}
\langle|u_n(x)-u_n(0)|^2\rangle = {k_B T\over d K} x\;,
\end{equation}
as can be seen from the equipartition theorem. Upon choosing $x$ such
that the root mean square phonon fluctuations equal the intrachannel
particle spacing $a$, we determine a translational correlation length
$\xi_T(T)$, which diverges at low temperatures
\begin{equation}
\xi_T(T) = {K d\over k_B T} a^2\;.
\end{equation}
The channels will couple to form a coherent two-dimensional LFS when
the effective coupling $d \xi_T(T)\mu (a / 2\pi d)^2$ between
correlated 1d regions of size $\xi_T(T)$ surpasses the thermal energy
$k_B T$ which decorrelates the 1d channels. We associate this
characteristic temperature with the melting temperature $T_m$ of LFS,
which is therefore given by
\begin{equation}
k_B T_m={\mbox{const.}} \, a^2\sqrt{\mu K}\;.
\label{T_coupling}
\end{equation}
A similar argument leads to the estimate for freezing into the
three-dimensional locked floating solid phases discussed by
Carraro~\cite{carraro} for rare gas atoms adsorbed into bundles of
carbon nanotubes. As we describe in
Appendix~\ref{variational_appendix}, in terms of the weakly coupled
model, Eq.\ref{H-weak_coupling}, freezing into LFS takes place at
strong coupling $\mu$, and therefore does not allow a rigorous
renormalization group treatment of the transition. Nevertheless an
approximate variational treatment is possible and is presented in
Appendix~\ref{variational_appendix}.

Instead, here we take an alternative route to the study of the LFS
melting and other transitions by working within a continuum elastic
model Eq.\ref{eq:KT_free_energy}, which is equivalent to the strong
coupling (between the channels) limit of the descrete model in
Eq.\ref{H-weak_coupling}. Such approach allows a more refined and
asymptotically exact renormalization group analysis (presented below),
within which we find that for $p=1$ the melting of the LFS phase is in
the universality class of the XY model, and is driven by unbinding of
dislocation pairs with Burgers vectors ${\bf b}=a\hatx$ along the
troughs of the periodic potential.  Consequently, in contrast to the
conventional 2d melting transition, at the melting temperature $\Tm$,
we predict a {\em universal} ratio of the jump in the geometric mean
of the long wavelength effective shear and bulk moduli,
$\mu_{\text{eff}}(\Tm^-)$ and $K_{\text{eff}}(\Tm^-)$ (describing the
elasticity of the LFS phase) to $\Tm$,

\begin{equation}
\frac{\sqrt{K_{\text{eff}}(\Tm^-) \,\mu_{\text{eff}}(\Tm^-)}}
{k_B \Tm}={8\pi\over |{\bf b}|^2} \; .
\label{universal_jump}
\end{equation}
This is in agreement, up to constants of order $1$, with the rough
estimate of the melting temperature, Eq.\ref{T_coupling} sketched
above, and with the variational method presented in
Appendix\ref{variational_appendix}. The most striking feature of the
$p=1$-LFS melting transition is the shape of the phase boundary
$\Tm(\UQ)$, whose universal features guarantee a generically reentrant
melting, under conditions such as the experiments of Wei
\etal\cite{wei-bechinger-rudhardt-leiderer:98}. At low light
intensities, i.e., small $\UQ$, we find that the melting curve has a
universal, cusp shape:
\begin{equation}
\Tm (\UQ)\sim \Tm(0) +  [\ln (k_{\rm B} \Tm / \UQ)]^{-1/{\overline \nu}}\;,
\label{Tm_smallU}
\end{equation}
with ${\overline \nu} \approx 0.36963$.  On the other hand for large
$\UQ$, i.e., for $k_B \Tm(\UQ)/\UQ\ll1$, we find that for short-range
particle interactions ($\kappa a \gtrsim 5.8$), $\Tm(\UQ)$ generically
{\em increases} with decreasing amplitude $\UQ$ of the periodic
modulation, according to
\begin{equation}
  \Tm (\UQ) = \Tm^\infty 
  \left\{ 1 + \frac{5[(\kappa a)^2\!-\!31]}{64 \pi^2}  
                           \left(1\!+\! \frac{13}{3 \kappa a} \right) 
                           \frac{k_B \Tm^\infty}{p^2 \UQ} 
  \right\}
\label{TmU_K}
\end{equation}
thus implying reentrant melting for a band of temperatures as a
function of potential strength (see Fig.~\ref{fig:phase_diagram_p=1}).
In Eq.\ref{TmU_K} above, $\kappa$ is the inverse of the Debye
screening length, tunable by adjusting the solution salt concentration,
and $\Tm^\infty\equiv \Tm(\UQ\rightarrow\infty)$, which, for the
system studied in Ref.\onlinecite{wei-bechinger-rudhardt-leiderer:98},
we estimate to be approximately $1.3 \, \Tm(\UQ=0)$.

The structure function for the $p=1$-LFS, illustrated in
Fig.\ref{fig:phase_diagram_p=1} is also quite unusual. In addition to
the set of Bragg peaks, directly induced by the laser field, $S({\bf
q})$ also displays an independent set of {\em quasi}-Bragg peaks at
the off-$q_y$-axis reciprocal lattice vectors ${\bf
G}$,\cite{nelson-halperin:79}
\begin{equation}
S({\bf q}) \sim {1\over\mid {\bf q} - {\bf G} \mid^{2-\etaLFS}}\;,
\end{equation}
which distinguishes the LFS from the modulated liquid state.  The
corresponding density-density correlation function $C_{\bf G} ({\bf
  r})= \langle \rho^{\phantom{*}}_{\bf G} ({\bf r}) \rho^*_{\bf G} (0)
\rangle$ for reciprocal lattice vectors with $G_x \neq 0$ shows a
power-law decay
\begin{equation}
 C_{\bf G} ({\bf r}) \sim 
 \left| \left(\frac{\mu_{\rm eff}}{K_{\rm eff}}\right)^{1/2} x^2 
 + \left(\frac{K_{\rm eff}}{\mu_{\rm eff}}\right)^{1/2} y^2 
 \right|^{-\etaLFS/2}\;,
\label{rhorho_power_law}
\end{equation}
where $\mu_{\rm eff}$ and $K_{\rm eff}$ are the effective shear and
bulk elastic moduli in Eq.\ref{eq:KT_free_energy} for the deformations
along the troughs ($x$-axis) of the periodic potential. The exponent
$\etaLFS$ depends on the relative orientation of the colloidal crystal
and the troughs.  Unlike conventional 2d
melting\cite{nelson-halperin:79}, it is {\em universal} at the melting
transition, and is given by
\begin{equation}
\etaLFS^* \equiv \etaLFS(\Tm^-) =({\bf G}\cdot{\bf b}/4 \pi)^2\;,
\label{eq:LFS_exponent}
\end{equation}
where ${\bf b}$ is the smallest allowed Burgers vector in the trough
direction.  For the primary orientation, illustrated in
Fig.\ref{fig:dislocation}, with $b=a$, the exponent characterizing the
algebraic order in the off-axis peaks (see
Fig.~\ref{fig:phase_diagram_p=1}) closest to the $q_y$-axis is
$\etaLFS^* = 1/4$; for the next row of peaks with $G_x = 4 \pi / a$ we
find $\etaLFS^* = 1$, consistent with the algebraic decay observed in
Ref.~\cite{wei-bechinger-rudhardt-leiderer:98} (for a more detailed
discussion see section~\ref{sec:discussion}).

Our analysis also makes {\em exact} predictions for the structure
function peak amplitudes in the limit of low laser intensity.  Similar
to the hexatic orientational order parameter $\psi_6$, Eq.\ref{psi6},
the translational order parameter, defined by
$M_{\Qlaser_n}\equiv\langle\rho_{\Qlaser_n}\rangle$, is induced by the
periodic potential throughout the phase diagram.  However, in contrast
to the liquid phase, where it vanishes linearly with $\UQ$, in the
crystal phase for $T<\Tm(0)$, we find
\begin{equation}
M_{\Qlaser_n}
\sim\mid \UQ\mid^{1/\delta_M}\;,\label{M_Q}
\end{equation}
with $\delta_M$ defined in analogy with the critical exponent at a
ferromagnetic critical point,
\begin{equation}
1/\delta_M={{\overline\eta}_{\Qlaser_n}\over
4-{\overline \eta}_{\Qlaser_n}}\;,\label{delta_M}
\end{equation}
and where 
\begin{equation}
{\overline \eta}_{\Qlaser_n}= \frac{\kT}{4 \pi} \frac{3 \mu +
\lambda}{\mu (2 \mu + \lambda)} \Qlaser_n^2
\end{equation}
is the exponent with which the real-space density-density correlation
function decays in a 2d crystal {\em without} a substrate
potential.\cite{nelson-halperin:79} We therefore predict that for
$T<\Tm(0)$ the intensity of the on-$q_y$-axis Bragg peaks vanishes as
an {\em exact} power of the input laser intensity $\Iin$, according to
\begin{mathletters}
\begin{eqnarray}
\Iout({{\bf K}_n})&\sim&|\langle\rho_{{\bf K}_n}\rangle|^2\Iin\;,
\label{I_onaxis_a}\\
&\sim&\Iin^{1+2/\delta_M}\;.\label{I_onaxis_b}
\end{eqnarray}
\label{I_onaxis}
\end{mathletters}

In contrast, we predict the intensity $\Iout$ of the off-axis
quasi-Bragg peaks, labeled by a reciprocal wavevector ${\bf G}$, to
vanish as
\begin{equation}
\Iout({\bf G})
\sim \Iin^{1+2\hat{\eta}_{\bf G}/(4-\hat{\eta}_{\bf G})} 
L^{2-({\overline\eta}_{\bf G} - \hat{\eta}_{\bf G})}\;,\label{I_offaxis}
\end{equation}
where 
\begin{equation}
 \hat{\eta}_{\bf G} \equiv {\overline\eta}_{\bf G} (1-G_x^2/G^2)\;,
\label{hat_eta}
\end{equation}
and $L$ is the system size.

We can also define the translational correlation length by the widths
of the off-$q_y$-axis Lorentzian peaks in the structure function. As
the melting temperature $\Tm$ is approached from above, given the XY
nature of the $p=1$-LFS melting transition, we expect the correlation
lengths parallel and perpendicular to the troughs to diverge according
to
\begin{equation}
\xi_{x,y} \sim e^{c/\mid T-\Tm\mid^{1/2}}\;,
\end{equation}
where $c$ is a constant of order unity.

\subsubsection{Intermediate commensurability ratios: $1<p<p_c$}
\label{sec:1<p<p_c}

\begin{figure}[bht]
  \narrowtext \centerline{ \epsfxsize=0.9 \columnwidth
    \epsfbox{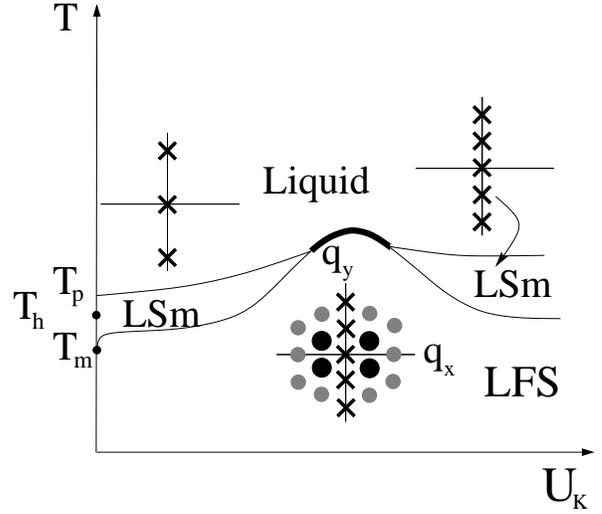} } \vspace{0.5cm}
\caption{Schematic phase diagram for a primary commensurate
  orientation with commensurability parameter in the range $1<p<p_c$
  (the case $p=2$ is shown here). Thin lines indicate continuous phase
  transitions. The thick line between the LFS and modulated liquid
  phase is most likely a first order phase boundary. {\em Insets:}
  Schematic structure functions. As in
  Fig.\ref{fig:phase_diagram_p=1}, the $\times$'s indicate
  delta-function Bragg peaks and the shaded circles algebraic peaks.}
\label{fig:phase_diagram_1_p_pc}
\end{figure}

For $1<p<p_c$, the phase diagram, illustrated in
Fig.\ref{fig:phase_diagram_1_p_pc}, generically includes an additional
symmetry-allowed ``locked smectic'' (LSm) phase.

The LSm distinguishes itself from the modulated liquid by
spontaneously breaking the liquid's discrete translational symmetry by
$d$ down to translations by $p d$.\cite{comment_otherSm} In contrast
to the LFS, however, the LSm exhibits only short-range correlations
between colloidal positions lying in different troughs, and therefore
does not resist shear deformations for displacements along the
potential minima.  Correspondingly, as illustrated in
Fig.\ref{fig:phase_diagram_1_p_pc}, the structure function of the LSm
phase displays spontaneously induced Bragg peaks at $\Qlaser_n/p$, in
addition to the Bragg peaks at $\Qlaser_n$, directly induced by the
laser interference fringes.  For $1<p<p_c$, the LFS also displays
these spontaneous Bragg peaks on the $q_y$ axis at ${\bf
  q}=\Qlaser_n/p$.

Symmetry dictates that the freezing of the modulated liquid into the
LSm is in the {\em p-state clock} model universality class. Also,
similar to the melting of the $p=1$-LFS, we find that the
$1<p<p_c$-LFS melts into the LSm through a transition in the XY
universality class and will therefore also exhibit the usual
Kosterlitz-Thouless phenomenology.\cite{kosterlitz-thouless:73} We
have also added in the phase diagram the possibility of a direct
transition from the LFS to the modulated liquid at intermediate
potential strength. We expect this transition to be different than the
LFS-liquid transition for $p=1$. Whereas the $p=1$ transition is in
the XY universality class, for $1<p<p_c$ the LFS-liquid transition is
associated with simultaneous loss of XY and descrete (Ising for $p=2$)
order. Because at this latter transition two unrelated symmetries are
simultaneously restored, we expect it to be first order. At the
multicritical point, where the liquid, LFS and LSm phases meet, the
critical behavior is presumably described by a two-dimensional
compressible Ising model (for $p=2$)\cite{Ising_compressible} of the
form
\begin{eqnarray}
H_{\rm I-XY}[u,S] = 
&&\int d^2 r \left[\frac12(\nabla S)^2 + \frac12 r S^2 +
 v S^4\right] + \HLFS[u]\nonumber \\  
  &&+ \int d^2 r \left( \gamma_x \partial_x u_x + 
\gamma_y \partial_y u_x \right)S^2\;.
\end{eqnarray}
$S$ is a continuous Ising order parameter, that distinguishes the LSm
from the liquid, $\gamma_{x,y}$ are ``magnetoelastic'' parameters,
which couple elastic strain to ``magnetization'' $S$, and where the
parameters of the model are tuned to the tricritical point at which
both order parameters vanish simultaneously. It would be interesting
to study the properties of such tricritical point, which to our
knowledge has not been previously explored.

\subsubsection{Large commensurability ratios and floating phases: $p>p_c$}
\label{sec:p>pc}

\begin{figure}[bht]
  \narrowtext \centerline{ \epsfxsize=0.9 \columnwidth
    \epsfbox{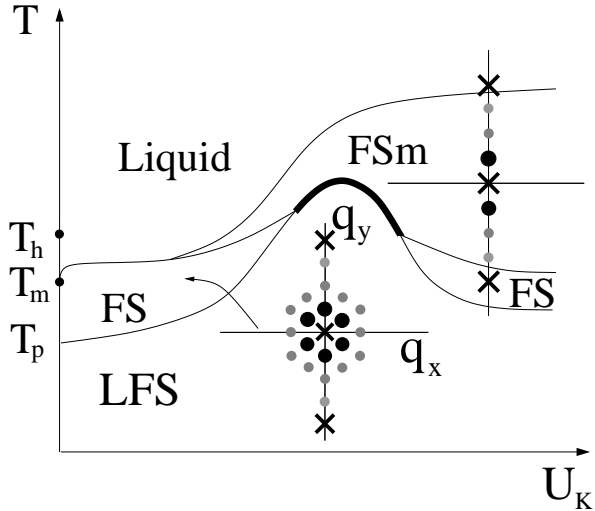}} \vspace{0.5cm}
\caption{Schematic phase diagram for a primary commensurate
  orientation with commensurability parameter $p>p_c$ (The case $p=4$
  is shown here). As in Fig.~\ref{fig:phase_diagram_p=1} the thick
  line indicates a first order transition.  {\em Insets:} Schematic
  structure factors. As in Fig.\ref{fig:phase_diagram_p=1}, the
  $\times$'s indicate delta-function Bragg peaks and the shaded
  circles algebraic peaks.}
\label{fig:phase_diagram_pc_p}
\end{figure}

For these higher values of $p$, the complexity of the 2d colloidal
phase diagram (displayed in Fig.\ref{fig:phase_diagram_pc_p}) further
increases, allowing two new phases, the ``floating solid'' (FS) and
the ``floating smectic'' (FSm).

The new phases are distinguished from their ``locked'' counter parts,
the LFS and LSm, by their ability to slide (float) {\em across} the
troughs of the periodic potential; technically, the periodic potential
is irrelevant (in the renormalization group sense) and therefore can
be treated perturbatively inside the FS and FSm phases.  As
illustrated in Fig.\ref{fig:phase_diagram_pc_p}, all the {\em
  spontaneously} induced structure function peaks of these floating
phases are {\em quasi}-Bragg peaks, and therefore the corresponding
density correlation functions display real-space {\em power-law}
decay, similar to Eq.\ref{rhorho_power_law}. Although, in principle,
the critical values $p_c^{\rm S}$, $p_c^{\rm Sm}$ for the appearance
of each of these floating phases are most likely distinct, for
simplicity of the presentation we have taken $p_c^{\rm S}=p_c^{\rm
  Sm}\equiv p_c$.  If in reality these critical values are
sufficiently distinct, and $p_c^{\rm S}<p_c^{\rm Sm}$, then we expect
an intermediate range of $p$ values, $p_c^{\rm S}<p<p_c^{\rm Sm}$, for
which no FSm appears.

We find that phase transitions between the corresponding locked and
floating phases (LFS-FS and LSm-FSm) are in the same universality
class as the well-known thermal roughening
transition,\cite{roughenning} the dual of the Kosterlitz-Thouless
transition, with an identical phenomenology. Similar to the XY-melting
LFS-LSm transition discussed above, the melting of the FS into the FSm
proceeds via unbinding of dislocation pairs with $x$-directed Burgers
vectors. However, because of the presence of massless spectator phonon
modes in $y$-direction (transverse to the troughs of the periodic
potential), the melting of the FS into the FSm is in a new
universality class.

The direct transition from the LFS to the FSm phase is most likely
first order.  Here the order of the $u_x$-modes changes from
quasi-long-range to short-range (via unbinding of type I dislocations,
i.e., those with Burgers vector parallel to the troughs of the
periodic potential.)  and for the $u_y$-modes from long-range to
quasi-long-range (via depinning from the laser potential, i.e., a
roughening transition). If both order parameters become critical at
the same point in the phase diagram, which will be the case at the
multicritical points where the FS, LFS, and FSm phases meet, we have a
phase transition corresponds a simultaneous transition of the KT type
and its dual analog.

The remainder of this paper is organized as follows: in Section
\ref{sec:model}, we introduce and motivate our model for 2d solids
subjected to a 1d periodic potential and discuss the details specific
to the experiments on 2d colloids in the laser
potential.\cite{wei-bechinger-rudhardt-leiderer:98} In
Sec.\ref{sec:phases} we give a detailed analysis of all the phases
which are allowed by symmetry. In particular, the static structure
factors and correlation functions are discussed. The mechanisms,
dislocation unbinding and soliton proliferation, driving the phase
transitions are investigated in Sec.\ref{sec:phase_transitions}. In
Sec.\ref{sec:reentrance} we derive the universal features of the
melting phase boundary, demonstrating that for sufficiently
short-range interactions it generically exhibits a reentrant melting
observed in the experiments of Wei \etal
.\cite{wei-bechinger-rudhardt-leiderer:98} Some aspects of the
response of the translational and bond-orientational order parameter
to a small external 1d periodic potential are analyzed in
Sec.\ref{sec:response} using a renormlization group crossover
analysis. In Sec.\ref{sec:discussion} we elaborate on some
implications of our results to experiments and for computer
simulations.

\section{Basic Ingredients}
\label{sec:model}
\subsection{``Microscopic'' model}
\label{sec:microscopic_model}

\begin{figure}[bht]
  \narrowtext \centerline{ \epsfxsize=0.9 \columnwidth
    \epsfbox{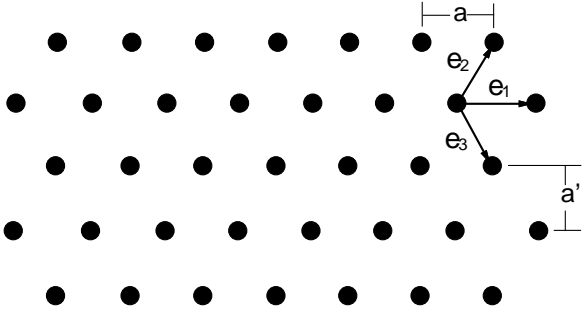}} \vspace{0.5cm}
\caption{Perturbation-free ideal hexagonal colloidal crystal, 
  characterized by fundamental lattice vectors ${\bf e}_i$.}
\label{fig:lattice_free}
\end{figure}

In the absence of external perturbations, we expect that, at
sufficiently low temperatures the 2d colloidal system freezes into a
hexagonal 2d crystal illustrated in Fig.\ref{fig:lattice_free}.
\noindent Its lattice sites ${\bf r}_n=n_1{\bf e}_1+n_2{\bf e}_2$, 
with $n_{1,2}\in \cal{Z}$ are spanned by a set of lattice vectors
\begin{mathletters}
\begin{eqnarray}
{\bf e}_1&=&a\unitx\;,\label{e1}\\
{\bf e}_2&=&{a\over2}(\unitx+\sqrt{3}\unity)\;,\label{e2}\\
{\bf e}_3&=&{a\over2}(\unitx-\sqrt{3}\unity)\;.\label{e3}
\label{lattice_vectors}
\end{eqnarray}
\end{mathletters}
or equivalently, in Fourier space, the lattice is characterized by a
set of three fundamental reciprocal lattice vectors
\begin{mathletters}
\begin{eqnarray}
{\bf G}_1&=&(2\pi/a')\unity\;,\label{G1}\\
{\bf G}_2&=&(\pi/a')(\sqrt{3}\unitx - \unity)\;,\label{G2}\\
{\bf G}_3&=&(\pi/a')(\sqrt{3}\unitx + \unity)\;,\label{G3}
\end{eqnarray}
\label{reciprocal_lattice_vectors}
\end{mathletters}
with $a'=a\sqrt{3}/2$ and $a$ the mean colloidal spacing related to
the particle density $\rho$ by $\rho = 2/\sqrt{3} a^2$.
\begin{figure}[bht]
  \narrowtext \centerline{ \epsfxsize=0.4 \columnwidth
    \epsfbox{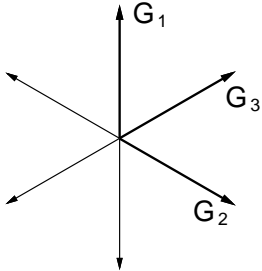}} \vspace{0.5cm}
\caption{A set of three fundamental reciprocal lattice vectors 
  ${\bf G}_i$, which completely characterize a perfect hexagonal
  lattice.}
\label{fig:G123}
\end{figure}
At sufficiently long scales\cite{comment_longscales} and to quadratic
order in the elastic strain
\begin{equation}
u_{ij}={1\over2}(\partial_i u_j + \partial_j u_i)\;,
\end{equation}
associated with the colloidal displacement field ${\bf u}(x,y)$
relative to the equilibrium position in the unconstrained solid, the
elastic energy of a 2d hexagonal crystal is well described by the
continuum isotropic elastic Hamiltonian
\begin{equation}
  H_0 = 
  \frac 12 \, \int d^2 r \left( 2 \mu u_{ij}^2 + \lambda u_{kk}^2 \right)\;.
\label{H_0}
\end{equation}
The Lam\'e coefficients $\mu$ and $\lambda$, with $\mu$ the usual
shear modulus are the only two elastic constants necessary to
completely characterized the elastic energy associated with small
deformations of an unperturbed 2d hexagonal solid.

An applied 1d periodic potential, which in
experiments\cite{chowdhury-ackerson-clark:85,wei-bechinger-rudhardt-leiderer:98}
with dielectric colloidal spheres is conveniently created by two
interfering laser beams, is easily incorporated as an additional
energetic contribution $H_\Qlaser$
\begin{equation}
H_\Qlaser=-\UQ{\sqrt{3}\over2}\sum_n
\cos\left[\Qlaser\cdot\bigl({\bf r}_n+{\bf u}({\bf r}_n)\bigr)\right]\;,
\label{H_Q}
\end{equation}
where we have focussed on the energetically most important {\em
lowest} harmonic $\Qlaser$ of such a laser-induced potential. The
coupling $\UQ$ measures the strength of the $\Qlaser$th harmonic of
the laser potential, proportional to the input laser intensity $\Iin$.

For the purpose of the discussion in this section we have chosen an
``internal'' reference frame $(\unitx, \unity)$ where the orientation
of the hexagonal lattice is kept fixed. Later, beginning with
Sec.\ref{sec:phases} we will switch to a ``laboratory frame''
$(\hatx,\haty)$ where the orientation of the laser potential is fixed
with $\Qlaser \parallel \haty$, i.e. the troughs are running parallel
to the $\hatx$-axis.

\subsection{Commensurability and reciprocal lattice}
\label{sec:commensurability}

For a general wave vector $\Qlaser$, the periodic (laser) potential
--- characterized by a plane wave, $e^{i \Qlaser\cdot{\bf r}}$ ---
will not be commensurate with the hexagonal lattice.  Only for a
particular orientation and magnitude of $\Qlaser$ will the spacing
between the potential minima match with the periodicity of the
hexagonal lattice. It is this special set of {\em commensurate}
periodic potentials that is the focus of our work here. The
characteristic set of commensurate wave vectors is easy to find since
the reciprocal lattice is {\em defined} to be the set of all wave
vectors ${\bf G}$ that yield plane waves with the periodicity of a
given Bravais lattice. Hence, commensurability is equivalent to the
condition that $\Qlaser$ coincides with one of the reciprocal lattice
vectors ${\bf G}$.

In other words, the planes defined by the minima of the external
potential ($\cos (\Qlaser\cdot {\bf r})$) are a superset of the family
of lattice planes (Bragg planes) defined by the shortest reciprocal
lattice vector, ${\bf G}_{\vec m} = m_1 {\bf G}_1 + m_2 {\bf G}_2$
with Miller indices $m_1$ and $m_2$, parallel to the wave vector of
the external potential $\Qlaser$,
\begin{equation}
  \Qlaser = p {\bf G}_{\vec m} = p_1 {\bf G}_1 + p_2 {\bf G}_2 \;.
\label{commensurateQ}
\end{equation}
Note that here we focus on situations where the colloidal particles
are allowed to sit in the minima of the external potential only. More
generally, one could also consider situations where the
commensurability parameter is less than $1$ with $p$ a rational
fraction.\cite{comment_p=1}

With $d= 2 \pi / \mid \Qlaser \mid$ being the periodicity of the
potential and $a_{\vec m}' = 2 \pi / \mid {\bf G}_{\vec m} \mid$
defining the distance between the lattice planes, the {\em
  commensurability ratio} $p$ is given by
\begin{mathletters}
\begin{eqnarray}
p &\equiv& \frac{a_{\vec m}'}{d} 
   = \frac{\mid \Qlaser \mid}{\mid {\bf G}_{\vec m} \mid} \\
  &=& \frac{\sqrt{3} a / 2}{d} \, (m_1^2+m_2^2-m_1m_2)^{-1/2} \;.
\end{eqnarray}
\label{commensurability_parameter}
\end{mathletters}
This allows us to characterize the laser potential by a set of Miller
indices ${\vec m} = (m_1,m_2)$ and the commensurability ratio $p$,
i.e., in summary by a commensurability vector ${\vec p} = p {\vec m} =
(p_1,p_2)$.  Equivalently, the orientation of the Bragg planes can
also be characterized by the shortest direct lattice vector pointing
parallel to the troughs of the external potential,
\begin{equation}
  {\bf R}_{\vec n} \equiv n_1 {\bf e}_1 + n_2 {\bf e}_2 
\label{eq:n1n2}
\end{equation}
with the condition $ {\bf R}_{\vec n} \cdot{\bf G}_{\vec m} = 0$,
i.e., $(n_1,n_2)=(m_1,-m_2)$ a set of integers (direct lattice Miller
indices), with no common factor complementary to the Miller indices.

\begin{figure}[bht]
  \narrowtext \centerline{ \epsfxsize=0.9 \columnwidth
    \epsfbox{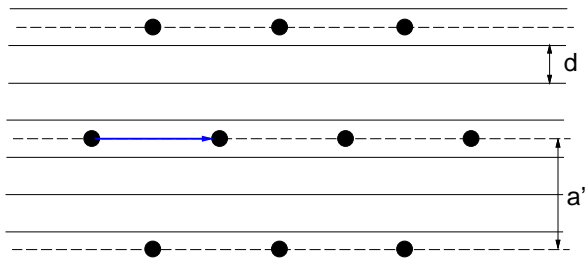}} \vspace{0.5cm}
\caption{2d hexagonal colloidal crystal in the presence of a
commensurate 1d periodic potential with period $d$, commensurability
vector $\vec{p}=3(1,0)$, and potential maxima indicated by solid
lines. Dashed lines denote the Bragg rows picked out by the laser
potential minima.}
\label{fig:lattice_pinned3_0}
\end{figure}

In Fig.\ref{fig:lattice_pinned3_0} we displayed an example for the
simplest set of relative orientations between the periodic potential
and the colloidal crystal. In our notation it corresponds to an
orientation $(m_1,m_2)=(1, 0)$ (or equivalently $(n_1,n_2)=(1,0)$) and
a commensurability ratio $p=3$, i.e., $\Qlaser = 3 {\bf
  G}_{(1,0)}=3{\bf G}_1$ and a Bragg row spacing $a_{\vec
  m}'=a'=a\sqrt{3}/2$.  Because in such $\vec{m}=(1,0)$ orientations,
it is {\em primary} Bragg rows\cite{comment_define} that run parallel
to the periodic potential troughs, we call these relative orientations
``primary''.  Aside from the simplicity of these configurations, our
interest in them is driven by experiments in
Refs.\cite{chowdhury-ackerson-clark:85,wei-bechinger-rudhardt-leiderer:98},
where a primary $p=1$ orientation was studied.

\begin{figure}[bht]
  \narrowtext \centerline{ \epsfxsize=0.9 \columnwidth
    \epsfbox{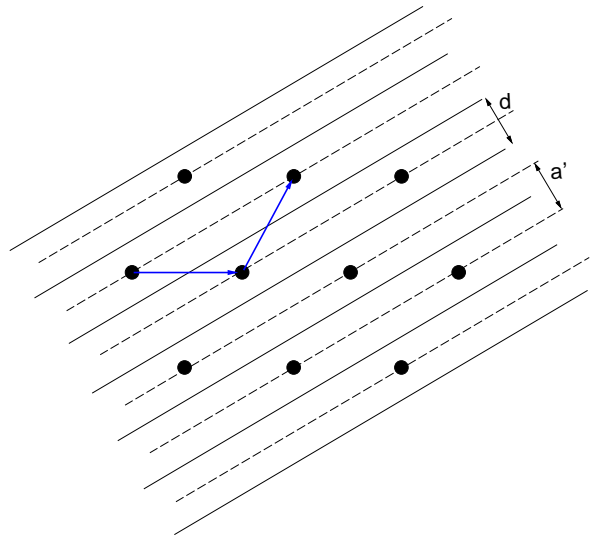}} \vspace{0.5cm}
\caption{2d hexagonal colloidal crystal in the presence of a commensurate 1d 
  periodic potential with period $d$, commensurability vector
  $\vec{p}=(1,-1)$, and potential maxima indicated by solid
  lines. Dashed lines denote the Bragg rows picked out by the laser
  potential minima.}
\label{fig:lattice_pinned1_-1}
\end{figure}
In addition to these primary $\vec{p}=p(1,0)$ configurations, we will
also make detailed predictions for the next simplest $\vec{p}=p(1,-1)$
set of relative lattice--laser potential configurations, illustrated
for $p=1$ in Fig.\ref{fig:lattice_pinned1_-1}. We call these
orientations ``dual-primary'', because they correspond to Bragg rows
running perpendicular to a {\em fundamental real space} lattice vector with
$\Qlaser=p({\bf G}_1-{\bf G}_2)=-{\bf e}_3 4\pi/a^2$, rather than to
one of the three {\em fundamental reciprocal} lattice vectors.  In
terms of the direct lattice these dual-primary orientations correspond
to $(n_1,n_2) = (1,1)$ and Bragg row spacing $a_{\vec m}'=a/2$.
\begin{figure}[bht]
  \narrowtext \centerline{ \epsfxsize=0.9 \columnwidth
    \epsfbox{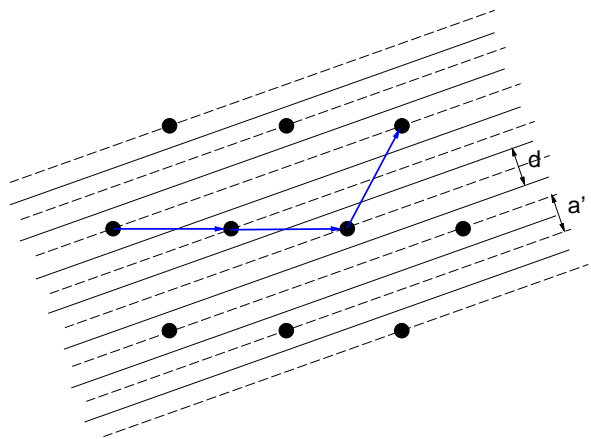}} \vspace{0.5cm}
\caption{2d hexagonal colloidal crystal in the presence of a commensurate 1d 
  periodic potential with period $d$, commensurability vector
  $\vec{p}=(2,-1)$, and potential maxima indicated by solid
  lines. Dashed lines denote the Bragg rows picked out by the laser
  potential minima.}
\label{fig:lattice_pinned2_-1}
\end{figure}
Using the definition of commensurate configurations,
Eq.\ref{commensurateQ}, in Eq.\ref{H_Q}, we find that $H_\Qlaser$
reduces to
\begin{mathletters}
\begin{eqnarray}
H_\Qlaser&=&-\UQ{\sqrt{3}\over2}\sum_{{\bf r}_n}
\cos\left[\Qlaser\cdot{\bf u}({\bf r}_n)\right]\;,\label{H_Qsimple1}\\
&=&-\UQ a^{-2}\int d^2 r
\cos\left[\Qlaser\cdot{\bf u}({\bf r})\right]\;,\label{H_Qsimple2}
\end{eqnarray}
\label{H_Qsimple}
\end{mathletters}
where in going from Eq.\ref{H_Qsimple1} to Eq.\ref{H_Qsimple2} we went
over to a continuum description, an innocuous approximation for the
smooth, $|{\bf u}({\bf r}_{n+1})-{\bf u}({\bf r}_{n})| \ll a$
distortions, that we study here.

An equivalent ``soft-spin'' continuum description of the above
interaction is in terms of the elementary translational order
parameters $\rho_{{\bf G}_i}({\bf r})=\rho_{{\bf G}_i}^0 e^{i{\bf
    G}_i\cdot({\bf r}+{\bf u}({\bf r}))}$ with $i=1,2$.  The
laser-induced periodic potential $h_\Qlaser({\bf r})=
Re\left[h_\Qlaser^0 e^{i \Qlaser\cdot{\bf r}}\right]$, acts like an
ordering field on the $\vec{p}=\pm(p_1,p_2)$th harmonics of the
fundamental order parameters $\rho_{{\bf G}_i}({\bf r})$, allowing a
linear coupling to $\rho_{\pm\bf G}({\bf r})=\rho_{\pm\bf G}^0 e^{\pm i{\bf
    G}\cdot({\bf r}+{\bf u}({\bf r}))}$
\begin{mathletters}
\begin{eqnarray}
H_\Qlaser&=&-\alpha\int d^2r\left[h_\Qlaser^*({\bf r})
\rho_{\bf G}({\bf r})
+c.c.\right]\\ 
&=&-\alpha\rho_{\bf G}^0 h_\Qlaser^0\int d^2r
\left[e^{i({\bf G}-\Qlaser)\cdot{\bf r}}
e^{i {\bf G}\cdot{\bf u}}+c.c.\right]
\label{H_Qsoft_spin}
\end{eqnarray}
\end{mathletters}
which is finite at long scales only if the condition
Eq.\ref{commensurateQ} is satisfied, in which case it reduces to the
expression given in Eq.\ref{H_Qsimple2}, with the identification
$\UQ/a^2=2\alpha\rho_{\bf G}^0 h_\Qlaser^0$.  Hence the periodic
potential explicitly breaks translational symmetry and therefore
induces a finite translational order parameters $\rho_{\pm
\Qlaser}(r)$ throughout the phase diagram.

\subsection{Broken rotational symmetry and anisotropic elasticity}
\label{sec:rotational-symmetry}

The imposed 1d periodic potential also {\em explicitly} breaks,
throughout our phase diagram, the 2d rotational symmetry down to
${\cal Z}_2$ (Ising) symmetry, corresponding to rotations by $\pi$.
We can see this more explicitly by noting that the laser potential
$h_\Qlaser({\bf r})= Re\left[h_\Qlaser^0 e^{i \Qlaser\cdot{\bf
      r}}\right]$ generates a set of even-rank tensor fields,
\begin{equation}
h_{i_1...i_{2n}}^{(2n)}
=\overline{\partial_{i_1} h_\Qlaser({\bf r})
\partial_{i_2} h_\Qlaser({\bf r})\ldots\partial_{i_{2n}} h_\Qlaser({\bf r})}\;,
\label{h_2n}
\end{equation}
where in above the overline denotes a spatial average.  The lowest
order, rank $2$, tensor field is given by
\begin{mathletters}
\begin{eqnarray}
h_{ij}^{(2)}
&=&\overline{\partial_i h_\Qlaser({\bf r})\partial_j h_\Qlaser({\bf r})}\;,
\label{h_ija}\\
&=&\frac12 |h_\Qlaser^0|^2 \Qlaserscalar_i \Qlaserscalar_j\;,
\label{h_ijb}
\end{eqnarray}
\end{mathletters}
It is clear from their definition, that these laser-generated
$2n$-rank tensor fields have strengths proportional to
$(\UQ)^{2n}\propto (\Iin)^{2n}$. They act as external ordering
fields, which explicitly break rotational invariance (modulo
rotations by $\pi$) of our system and therefore induce throughout our
phase diagram finite $2n$-adic orientational order parameters. These
can be characterized by rank $2n$ symmetric traceless tensors, which
are real irreducible representations of the rotation group and are the
``angular harmonics'' of the lowest order, rank 2, nematic order
parameter
\begin{equation}
Q_{ij}^{(2)}= S(\hat{n}_i \hat{n}_j-{1\over2}\delta_{ij})\;.
\label{Q_ij}
\end{equation}
The unit vector $\hat{\bf n}$, defines the principle axis of the
nematic order, and, given Eq.\ref{h_ijb}, points parallel or perpendicular
(depending on the sign of the coupling between $h_{ij}^{(2)}$ and
$Q_{ij}^{(2)}$) to the periodic potential wavevector $\Qlaser$. In two
dimensions these $2n$-rank tensor representations are well-known to be
equivalent to the one-dimensional complex irreducible representations
\begin{equation}
\psi_{2n}=e^{i 2n\theta}\;.
\end{equation}
Since in the presence of these laser-induced ordering fields all
$\psi_{2n}$ orientational order parameters are finite throughout our
phase diagram, no sharp continuous {\em orientational} ordering phase
transitions are possible in our system. This is in contrast to the
melting of the unperturbed lattice, where a thermodynamically sharp
orientational phase transition is allowed between the isotropic and
the anisotropic (e.g, hexatic, in a hexagonal lattice)
liquids.\cite{nelson-halperin:79} Therefore, throughout this
paper we confine our attention only to phases and phase transitions
that {\em spontaneously} break the {\em translational} symmetry of the
explicitly orientationally ordered, modulated colloidal liquid phase.

The existence of these orientational ordering fields $h_{2n}$ has
important consequences to the form of the colloidal crystal elastic
energy.  To deduce the form of the appropriate elastic Hamiltonian it
is instructive to first consider a 2d hexagonal lattice in the absence
such {\em explicit} symmetry breaking fields. Such state is
characterized by a finite value of the hexagonal orientational order
parameter $\psi_6$,\cite{nelson-halperin:79} with the full 2d
rotational symmetry broken down to the symmetry of discrete rotations
by $2\pi/6$. Nevertheless to a quadratic order in the strain tensor
$u_{ij}$, the energy is invariant under a full 2d rotation group.

In the absence of a periodic potential the hexagonal orientational
order can be further {\em spontaneously} broken down to a lower
symmetry. A physically important example is a uniaxially distorted
hexagonal 2d crystal of anisotropic, orientationally ordered
molecules, as, for instance found in a nematic liquid crystal. Such a
system exhibits a {\em spontaneous} nematic order parameter
$Q_{ij}^{(2)}$, which modifies the isotropic elasticity $H_0$,
Eq.\ref{H_0}. To a quadratic order in the strain $u_{ij}$ three
additional energetic contributions
\begin{equation}
\delta H_{0}=\int d^2r\left[\alpha_1 u_{ij}Q_{ij}^{(2)} +
\alpha_2 \bigl(u_{ij}Q_{ij}^{(2)}\bigr)^2 + 
\alpha_3 u_{ij}u_{jk}Q_{ki}^{(2)}\right]\;
\label{deltaH_0}
\end{equation}
are allowed. Because the nematic order is induced spontaneously,
simultaneous rotations of the lattice degrees of freedom and of the
nematic axis (encoded in $Q_{ij}^{(2)}$), relative to an arbitrary
frame fixed in the lab, is clearly a symmetry of such uniaxially
distorted lattice. It is not difficult to show that this rotational
freedom allows us to eliminate $\alpha_1$ coupling linear in $u_{ij}$
\begin{eqnarray}
H_{\alpha_1}&=&\alpha_1 S \int d^2r 
\Bigg[u_{xx}\left(\sin^2\theta-{1\over2}\right)+
u_{yy}\left(\cos^2\theta-{1\over2}\right)\nonumber\\
&+&u_{xy}2\sin\theta\cos\theta\Bigg]\;,
\label{H_alpha1}
\end{eqnarray}
by a judicious choice of the rotation angle $\theta$ and a uniaxial
area-preserving distortion
\begin{equation}
u_i\rightarrow u_i+\phi_i\;.  
\label{distortion}
\end{equation}
It is important to note that this is only possible because the three
independent degrees of freedom, $\theta$, $\phi_x$ and $\phi_y$, at
our disposal, are sufficient to cancel the three independent linear
terms $u_{xx}$, $u_{yy}$, and $u_{xy}$ in $H_{\alpha_1}$,
Eq.\ref{H_alpha1}.

Adding $H_{\alpha_2}$ and $H_{\alpha_3}$ contributions,
Eq.\ref{deltaH_0}, to the Hamiltonian of an undistorted hexagonal
lattice, Eq.\ref{H_0}, we find a general elastic Hamiltonian for a
{\em spontaneously} uniaxially distorted hexagonal lattice (in the
absence of an external potential)\cite{ostlund-halperin:81} is given
by
\begin{eqnarray}
H_0^{a}&=&\int d^2r\Big[2\mu u_{xy}^2 + {1\over2}\lambda_{xx}
u_{xx}^2+{1\over2}\lambda_{yy}u_{yy}^2\nonumber\\
&+&\lambda_{xy}u_{xx}u_{yy}\Big]\;,
\label{H_0a}
\end{eqnarray}
where we have chosen a coordinate system in which the $x$ and $y$ axes
coincide with the $\hat{\bf n}$ and $\hat{\bf z}\times\hat{\bf n}$
principle axes of the orientational nematic order parameter
$Q_{ij}^{(2)}$. The two {\em additional} elastic constants, a total of
four in $H_0^a$, are consistent with two new couplings $\alpha_2$ and
$\alpha_3$ allowed by the finite orientational nematic order parameter
$Q_{ij}^{(2)}$.  The four independent elastic constants also coincide
with the expectation, that with a symmetry between $x$ and $y$ broken,
the elastic energies associated with the strain tensor components
\begin{mathletters}
\begin{eqnarray}
u_{xx}&=&\partial_x u_x\;,\\
u_{yy}&=&\partial_y u_y\;,\\
u_{xy}&=&{1\over2}(\partial_x u_y+\partial_y u_x)
\end{eqnarray}
\label{u_xy}
\end{mathletters}
are clearly independent. Although rotations relative to the
orientational uniaxial order is no longer a symmetry of $H_0^a$,
because only the {\em symmetric} strain tensor $u_{ij}$ enters the
elastic energy, $H_0^a$ is still invariant under ``atomic''
displacements
\begin{equation}
{\bf u}=\theta\hat{\bf z}\times{\bf r}\;,
\label{u_rotate}
\end{equation}
which correspond to {\em global} rigid rotations of the 2d solid, by
an infinitesimal angle $\theta$ about the $z$-axis.  This latter
symmetry is clearly present in an anisotropic lattice {\em without} an
external pinning potential.

In contrast, however, in our system, the 1d periodic potential has a
{\em fixed} orientation in the laboratory frame.  Hence, in addition
to the uniaxial lattice anisotropy, such a potential also explicitly
breaks symmetry of rotations relative to the lab (and therefore to the
periodic potential) frame. It therefore picks out a special coordinate
system relative to which the angle $\theta$ is measured. 

As discussed above such external potential acts as an external
$2n$-rank tensor fields and explicitly breaks the corresponding
orientational symmetry. The appropriate elastic energy can be deduced
by focusing on the lowest order nematic ordering field
$h_{ij}^{(2)}$. It allows the following additional energetic
contributions
\begin{equation}
H_{h_2}=-\int d^2r\left[u_{ij}h_{ij}^{(2)} +
Q_{ij}^{(2)}h_{ij}^{(2)}\right]\;.
\label{H_h2}
\end{equation}
that explicitly break symmetry of rotations relative to the frame
picked out by the periodic potential.

For the purposes of classification of the relative orientations
discussed in previous subsection, Sec.\ref{sec:commensurability}, it was
more convenient to keep the lattice fixed and to rotate the periodic
potential into a particular orientation, uniquely labeled by an
integer 2d Miller index vector $\vec{p}=(p_1,p_2)$. However, once an
orientation has been selected and classified by $\vec{p}$, to analyze
the continuum elastic model and its thermodynamics that follows it is
more convenient to work in a coordinate system in which, instead, the
troughs of the 1d periodic potential run along the new $x$-axis. For
such a choice of a lab coordinate system,
\begin{equation}
h_{ij}^{(2)}={1\over2}|h_\Qlaser^0|^2 \Qlaserscalar^2 {\hat y}_i {\hat y}_j\;,
\end{equation}
Using this expressions for $h_{ij}^{(2)}$ together with $Q_{ij}$,
Eq.\ref{Q_ij} inside Eq.\ref{H_h2}, and combining it with
$H_{\alpha_1}$, Eq.\ref{H_alpha1}, we find the following symmetry
breaking energetic contribution, which, in the presence of a 1d
periodic potential must be added to $H_0^a$, Eq.\ref{H_0a}
\begin{eqnarray}
H_{\alpha_1+h_2}&=&\int d^2r\Big[\alpha_1 S u_{xx}(\sin^2\theta-{1\over2})+
u_{yy}(\alpha_1 S\cos^2\theta\nonumber\\
&-&{\alpha_1 S\over2}-h)+
\alpha_1 S u_{xy}\sin{2\theta}-{1\over2}h S\cos{2\theta}\Big]\;,
\label{H_alpha1+h2}
\end{eqnarray}
where $h\equiv{1\over2}\Qlaserscalar^2|h_\Qlaser^0|^2$ and angle
$\theta$ measures the deviation of the nematic axis $\hat{\bf n}$ away
from $\Qlaser$ set by the orientation of the periodic potential.  While
it is still possible to eliminate the terms linear in $u_{xx}$ and
$u_{yy}$ by a lattice distortion Eq.\ref{distortion}, in the presence
of the external potential it is no longer possible to shift away the
$u_{xy}$ term. Selecting $\phi_i$ so as to cancel $u_{xx}$, $u_{yy}$
and combining the resulting $H_{\alpha_1+h_2}$ with $H_0^a$ we find
\begin{eqnarray}
H^{a}&=&\int d^2r\Big[2\mu u_{xy}^2 + {1\over2}\lambda_{xx}
u_{xx}^2+{1\over2}\lambda_{yy}u_{yy}^2\nonumber\\
&+&\lambda_{xy}u_{xx}u_{yy}+
\alpha u_{xy}\sin{2\theta}-{\gamma\over2}\cos{2\theta}\Big]\;,
\label{H_a}
\end{eqnarray}
where we defined rotational symmetry breaking couplings
$\alpha\equiv\alpha_1 S$ and $\gamma\equiv h S$. It is clear from the
above $\alpha$ and $\gamma$ terms in $H^a$ that, in the absence of
strain, $u_{xy}=0$, the energy is minimized by $\theta=0$,
corresponding to the nematic axis alignment with $\Qlaser$, imposed by
the periodic potential. In the presence of fluctuations $\theta$ will
be small but finite. Expanding $H^a$, above, in these small
fluctuations, we obtain a final form of the elastic Hamiltonian
characterizing our system
\begin{eqnarray}
\Hel&=&\int d^2r\Bigg[2\mu u_{xy}^2 + {1\over2}\lambda_{xx}
u_{xx}^2+{1\over2}\lambda_{yy}u_{yy}^2\nonumber\\
&+&\lambda_{xy}u_{xx}u_{yy}+
2\alpha u_{xy}\theta + 2\gamma \theta^2\Bigg]\;.
\label{H_el_anisotropic}
\end{eqnarray}

\begin{figure}
\centering
\setlength{\unitlength}{1mm}
\begin{picture}(150,75)(0,0)
\put(-20,-20){\begin{picture}(150,0)(0,0)
\includegraphics{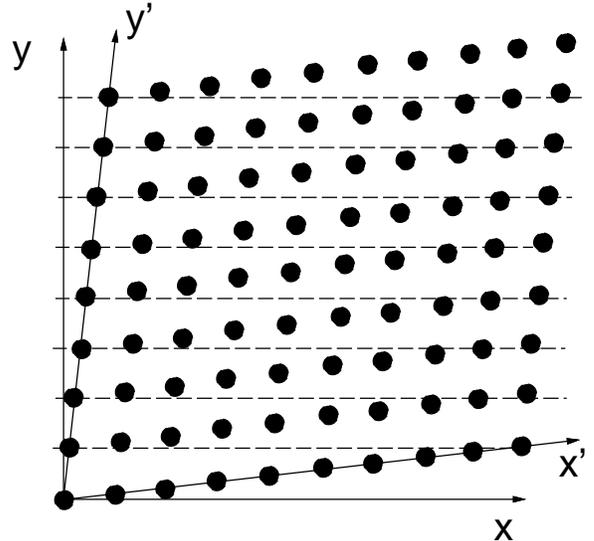}
\end{picture}}
\end{picture}
\caption{A $u_{xy}\neq 0$ shear deformation (shown for simplicity for
  a square lattice) with principle axes along the $(1,1)$ and $(1,-1)$
  directions in the xy-plane. In the presence of a trough potential
  (dashed lines) parallel to the x-direction, the particle array, with
  axes $x'$ and $y'$, can lower its interaction energy with the
  periodic potential by rotating in a clockwise direction to bring the
  particles into better alignment with the minima in the trough
  potential.}
\label{shear_rotate}
\end{figure}

Now to complete our derivation we must relate the angle $\theta$,
characterizing the orientation of the nematic order to the elastic
$u_i$ degrees of freedom. We expect that the orientations of the
nematic and hexatic order parameters, present in our uniaxially
distorted hexagonal lattice, are locked together. Since in the
crystalline phase, fluctuations in this bond orientational order are
in turn locked to the local rotation angle induced by the phonon
displacements, in the Hamiltonian Eq.\ref{H_el_anisotropic} we can
make the well-known identification
\begin{equation}
\theta = \frac{1}{2}(\partial_x u_y - \partial_y u_x)\;,
\end{equation}
thereby completing our derivation. We find that the resulting elastic
Hamiltonian, which characterizes a hexagonal lattice in the presence
of a 1d periodic potential, involves $6$ elastic constants.  While a
similar form was suggested, based on symmetry, by Ostlund and
Halperin~\cite{ostlund-halperin:81} in their analysis of melting of
distorted hexagonal crystal films, the $\alpha$ term appearing in our
Hamiltonian, Eq.\ref{H_el_anisotropic} was missed in their
expression. As illustrated in Fig.\ref{shear_rotate}, physically, the
$\alpha$ term is present because, with troughs running along the
x-direction, an xy-strain will bring particles in Bragg planes lying
in the troughs out of alignment with the minima of the periodic
potential. This generates a torque which attempts to rotate the
lattice and improve the alignment.

The elastic Hamiltonian $\Hel$, Eq.\ref{H_el_anisotropic}, together
with the commensurate pinning potential $H_\Qlaser$,
Eq.\ref{H_Qsimple2}, defines our model 2d colloidal system in the
presence of a commensurate periodic laser potential. Our aim, in the
remainder of the paper is to analyze the symmetry allowed phases and
the nature of the transitions between them embodied in this model.
%
%
%
%

\section{Symmetry-Allowed Phases}
\label{sec:phases}

The starting point of our analysis is the model Hamiltonian
$H=\Hel+H_\Qlaser$, obtained from combining Eqs.\ref{H_Qsimple2} and
\ref{H_el_anisotropic}. Here we have chosen (without loss of
generality) $\Qlaser$ to lie along the $y$-axis, i.e., the periodic
potential troughs running parallel to the $x$-axis, a convention that
we will stick to throughout the remainder of the paper. This
Hamiltonian admits a rich variety of thermodynamically distinct
phases. As discussed in the Introduction, the phase diagrams depend on
the commensurability ratio $p$, or more specifically, in which of the
three regimes $p=1$, $1<p\leq p_c$, or $p>p_c$, $p$ actually falls.
The complexity of the phase diagram is highest for $p>p_c$, and so in
order to discuss all the phases possible in our system, we focus on
these high $p$ commensurability ratios.

It is convenient to enumerate the five allowed phases starting with
the most ordered, which naturally occurs at the lowest temperatures,
and proceeding toward the higher temperature disordered phases, by
invoking two types of disordering mechanisms, {\em dislocation
unbinding} and {\em soliton proliferation}.  A detailed investigation
of these mechanisms is deferred to the subsequent section, where we
discuss the nature of these transitions and their hierarchy as a
function of temperature and periodic potential strength (laser
intensity). Here, we focus on the phases themselves, rather than on
their location in the resulting phase diagram. As discussed in detail
in the previous section and in the Introduction, the imposed periodic
potential {\em explicitly} breaks rotational symmetry, and therefore
all five phases exhibit true long-range orientational order. This
external potential also explicitly breaks continuous translational
symmetry along $y$ (with potential troughs taken to run along $x$)
down to a discrete symmetry of translations by the period $d$ of the
potential.  Hence all phases will trivially exhibit long-range order
in the corresponding translational order parameter, leading to true
delta-function Bragg peaks at the multiples of the reciprocal lattice
vector $(2\pi/d) \haty$ in their structure functions.

\subsection{Solid Phases}
\label{sec:solid_phases}

As in the absence of a periodic potential, the most ordered phase of
isotropic particles confined to 2d is a solid. The striking effect of
turning on an external 1d periodic potential is that it can lead to
two thermodynamically distinct uniaxially distorted hexagonal crystal
phases, which we term a ``locked floating solid'' (LFS) and a
``floating solid'' (FS). Being crystalline, both of these phases
exhibit 2d translational (quasi-long-range) order, and are
characterized by a finite shear modulus.  These emerge as a result of
breaking the translational symmetry $T^y_d\otimes T^x$ of the
``modulated liquid'' (ML) (see below), corresponding to independent
discrete translations by $d\haty$ and continuous translations along
$\hatx$, down to 2d discrete translations generated by lattice vectors
${\bf e}_1$ and ${\bf e}_2$, Eq.\ref{e1},\ref{e2}. Although in the
presence of thermal fluctuations dislocations will be thermally
nucleated, in the solid phases they are confined to finite size
dipoles with a zero Burgers ``charge''. These, therefore, can be
safely integrated out of the partition function with only weak {\em
  finite} renormalization of the elastic constants.  Consequently, a
purely elastic description in terms of phonon modes $u_x$ and $u_y$ is
appropriate in both phases.

The LFS and FS phases differ in the importance of the periodic pinning
potential. In the FS, expected to occur at temperatures higher than
the LFS, thermal fluctuations in the positions of the colloidal
particles are sufficiently large such that at long length scales they
average away most\cite{comment_irrelevant} of the long scale effects
of the periodic potential. In contrast, in the LFS the periodic
potential strongly pins the colloidal crystal transversely to its
troughs.

\subsubsection{Floating Solid (FS)}
\label{sec:FS}

The floating solid can be rigorously differentiated from its locked
counterpart as a 2d colloidal crystal phase in which the periodic
potential is {\em irrelevant} in the renormalization group sense.
This implies that at long scales, many, but not all (see below) of the
thermodynamic properties of the FS are well described by the elastic
Hamiltonian $\Hel$, Eq.\ref{H_el_anisotropic}, with two coupled {\em
  ``massless''} $u_x$ and $u_y$ degrees of freedom, and ignoring
$H_\Qlaser$.  Therefore, in many ways the FS is qualitatively quite
similar to a 2d solid without the periodic pinning potential. In
particular, this similarity extends to the lattice displacement
correlation functions which are logarithmic in $x$ and $y$.  However,
these similarities do not extend to all correlation functions, and the
periodic pinning potential has important qualitative consequences for
the FS phase that distinguish it from an ordinary 2d solid. The
``irrelevance'' of the periodic potential means only that a
perturbative expansion in $\UQ$, for a sufficiently small value is
{\em convergent}. Consequently, average quantities, that are finite at
$\UQ=0$ can be well approximated by their $\UQ=0$-values, i.e.,
working with $H\approx \Hel$, as is usually done. However, quantities
that {\em vanish} (or diverge) to this zeroth order, must be evaluated
to the next lowest order in $\UQ$ to obtain a nontrivial (finite)
result.

To illustrate this point, recall that the periodic potential {\em
explicitly} breaks rotational and translational symmetry, despite its
irrelevance in the FS phase. While the former leads to uniaxial
anisotropy in the hexagonal lattice, the latter is responsible for the
true long-range order in the translational order parameter $\rho_{\bf
G}({\bf r})$, with ${\bf G}$ integer multiples of the wave vector
$\Qlaser$ characterizing the periodicity of the external potential.
In the presence of the periodic potential, even the most disordered
modulated liquid phase (see below) displays true long-range
translational and orientational order.  Clearly then, a more ordered
FS will also break these symmetries.

As a concrete example of how the periodic potential affects the FS
phase, consider the real-space 2-point correlation function of the
translational order parameter
\begin{equation}
\rho_{\bf G}({\bf r})\equiv e^{i{\bf G}\cdot {\bf u}({\bf r})}\;,
\end{equation}
defined by
\begin{mathletters}
\begin{eqnarray}
C_{\bf G}({\bf r})&\equiv&\langle \rho_{\bf G}({\bf r})\rho_{\bf G}^*(0)\rangle
\;,\label{rhorho_a}\\
&=&\langle e^{i{\bf G}\cdot ( {\bf u}({\bf r})-{\bf u}(0) )}
\rangle\;,\label{rhorho_b}\\
&\equiv&C_{\bf G}^{(c)}({\bf r})+
\langle\rho_{\bf G}\rangle\langle\rho_{\bf G}^*\rangle
\;,\label{rhorho_c}
\end{eqnarray}
\end{mathletters}
where in Eq.\ref{rhorho_c}, $C_{\bf G}^{(c)}({\bf r})$ is the
connected part of $C_{\bf G}({\bf r})$. The distinguishing feature of
the FS phase is the irrelevance of the periodic potential $H_\Qlaser$.
Hence in the limit of a weak laser potential, i.e., small $\UQ$, we
can compute $C_{\bf G}({\bf r})$ in a controlled, convergent
perturbative expansion in $\UQ$.  The connected part $C_{\bf
  G}^{(c)}({\bf r})$ is nontrivial even to zeroth order in $\UQ$, and
a standard calculation gives,\cite{nelson-halperin:79}
\begin{equation}
C_{\bf G}^{(c)}({\bf r})\sim {1\over|\bf r|^{\etaG}}\;,\label{Cc}
\end{equation}
where
\begin{equation}
\etaG={|{\bf G}|^2\over 4\pi}{\kT\over\mu}
{3\mu+\lambda\over2\mu+\lambda}\;.\label{eta_G}
\end{equation}
For simplicity, we have used the isotropic elastic Hamiltonian,
Eq.\ref{H_0}, in place of the correct six elastic constant anisotropic
Hamiltonian $\Hel$, Eq.\ref{H_el_anisotropic}, which leads to a
qualitatively similar, but anisotropic power-law decay of spatial
correlations.  We use long wavelength elastic constants finitely
renormalized by thermally excited bound dislocation dipoles.
\top{-2.5cm}
\columnwidth3.4in
\narrowtext
\noindent
We can compute the persistent part of $C_{\bf G}({\bf r})$, by
calculating
\begin{equation}
\langle\rho_{\bf G}\rangle=\langle e^{i{\bf G}\cdot {\bf u}({\bf 0})}
\rangle\;,
\end{equation}
in a perturbation theory in $\UQ$, which, because of the irrelevance
of the periodic potential is convergent in the FS phase. For $\UQ=0$
the translational order parameter vanishes like $\langle\rho_{\bf
  G}\rangle= (L/a)^{-\etaG /2}$ with system size $L\rightarrow
\infty$. Upon expanding the Boltzmann weight
$e^{-(H_0+H_\Qlaser)/\kT}$ in a power series in $\UQ$, we find to
leading order in $\UQ$ and for $L\rightarrow \infty$
\end{multicols}
\widetext
\begin{equation}
\langle\rho_{\bf G}\rangle
=\sum_{n=1}^\infty{1\over n!}\left({\UQ\over 2 a^2 \kT}\right)^n
\left( \delta_{{\bf G},n\Qlaser}+ \delta_{{\bf G},-n\Qlaser} \right)
\prod_{j=1}^n\int d^2r_j\prod_{j=1}^n (|{\bf r}_j|/a)^{-n{\overline \eta}_\Qlaser}
\prod_{i < j}^n (|{\bf r}_{i}-{\bf r}_{j}|/a)^{-{\overline \eta}_\Qlaser}  \, .
\label{rhoG1} 
\end{equation}
Here we have used the fact that for $L\rightarrow \infty$
\begin{equation}
\langle 
\exp \biggl[
i \sum_\alpha {\bf q}_\alpha \cdot {\bf u} ({\bf r}_\alpha) 
\biggr]
\rangle = 
\exp \biggl[ 
\sum_{\alpha < \beta} {\bf q}_\alpha \cdot {\bf q}_\beta
\, G^{(c)} ({\bf r}_\alpha-{\bf r}_\beta) 
\biggr] 
\end{equation}
\bottom{-2.7cm}
\begin{multicols}{2}
\columnwidth3.4in
\narrowtext
\noindent
for $\sum_\alpha {\bf q}_\alpha = {\bf 0}$ and zero otherwise. We have
also introduced a phonon connected correlation function $G^{(c)} ({\bf
  r})$
\begin{equation}
G^{(c)}({\bf r}) \equiv 
\frac{1}{4}
\langle |{\bf u}(0) - {\bf u}({\bf r})|^2 \rangle_0 \;.
\label{Gc_define}
\end{equation}
Averages with elastic Hamiltonian are designated by
$\langle\ldots\rangle_0$. Upon again approximating $\Hel$ by its
isotropic form $H_0$, Eq.\ref{H_0}, a straightforward calculation in
the limit $L/a\gg1$, $r/a\gg1$, gives
\begin{equation}
G^{(c)}({\bf r}) \approx 
\frac{\etaG}{G^2} \log{(r/a)}\;.\label{ave0b}
\end{equation}
Since $\UQ$ is irrelevant in the floating solid phase the integrals in
Eq.\ref{rhoG1} are IR convergent (i.e. for $L \rightarrow \infty$).
The power-laws appearing in the integrand are implicitly understood to
be cutoff below the lattice constant $a$ scale by the obvious behavior
(see Eq.\ref{Gc_define}) of the phonon correlation function
$\lim_{r\rightarrow a} G^{(c)}({\bf r})=0$.  Upon performing the
spatial integrals, which are dominated by the behavior of the
connected phonon correlation function at small distances (UV, lattice
cutoff $a$), we obtain up to non-universal factors of order $1$
\begin{equation}
\langle\rho_{\bf G}\rangle
\approx \sum_{n=1}^\infty 
{1\over n!}\left({\UQ\over 2 a^2 \kT}\right)^n 
(\delta_{{\bf G},n\Qlaser}+\delta_{{\bf G},-n\Qlaser})
\;.
\label{rhoG2}
\end{equation}

Hence, as argued above on physical grounds, despite of the irrelevance
of the periodic potential, the FS displays true long-range order in
the translational order parameter $\rho_{\bf G}$, with ${\bf G}$
satisfying ${\bf G}=\pm n\Qlaser$, with $C_{\bf G}({\bf r})$
approaching its asymptotic value as a power-law in $r$. Other
translational order parameters, with ${\bf G}$ {\em not} satisfying
the above condition have pure power-law correlation functions, decaying
to zero at long separations. In particular, these include the
fundamental translational order parameters $\rho_{{\bf G}_i}$, which
display quasi-long-range order in the FS phase.

Having calculated the translational correlation function $C_{\bf
  G}({\bf r})$, the structure function
\begin{equation}
S({\bf q})=\sum_{{\bf r}_m} e^{-i{\bf q}\cdot{\bf r}_m}
C_{\bf q}({\bf r}_m)
\label{Sq}
\end{equation}
can now be easily obtained. Using Eqs.\ref{rhorho_a}, \ref{Cc}, and
\ref{rhoG2} and taking advantage of the Poisson summation formula to
perform the sum over the lattice sites ${\bf r}_m$, we find
\begin{equation}
S({\bf q})\approx\sum_{\bf G}{1\over|{\bf q}-{\bf G}|^{2-\etaG}}
+ \sum_{n=-\infty}^{\infty\;\prime} A_n\delta^{(2)}({\bf q}-n\Qlaser)\;,
\label{Sq2}
\end{equation}
with
\begin{equation}
A_n={1\over(n!)^2}\left({\UQ\over 2 a^2 \kT}\right)^{2n}\;,
\label{A_n}
\end{equation}
and prime on the summation in Eq.\ref{Sq2} indicating that the $n=0$
term is excluded.

\begin{figure}[bth]
  \narrowtext \centerline{\epsfxsize=0.33\columnwidth
    \epsfbox{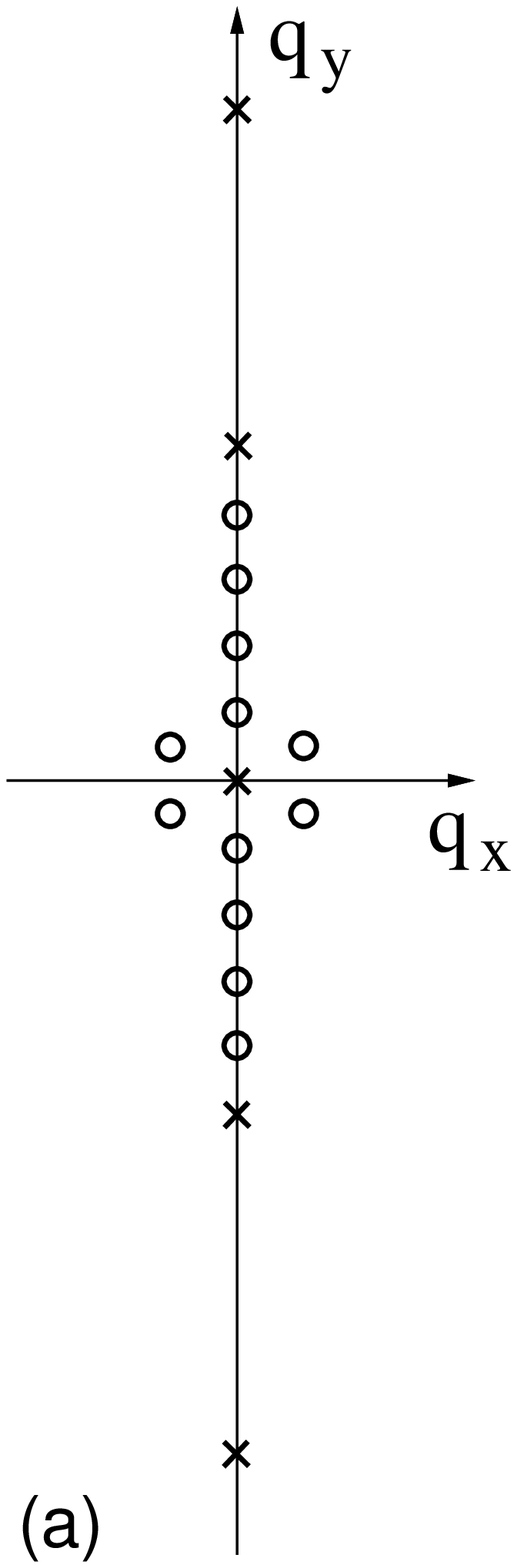} \hspace{0.04\columnwidth}
    \epsfxsize=0.33\columnwidth \epsfbox{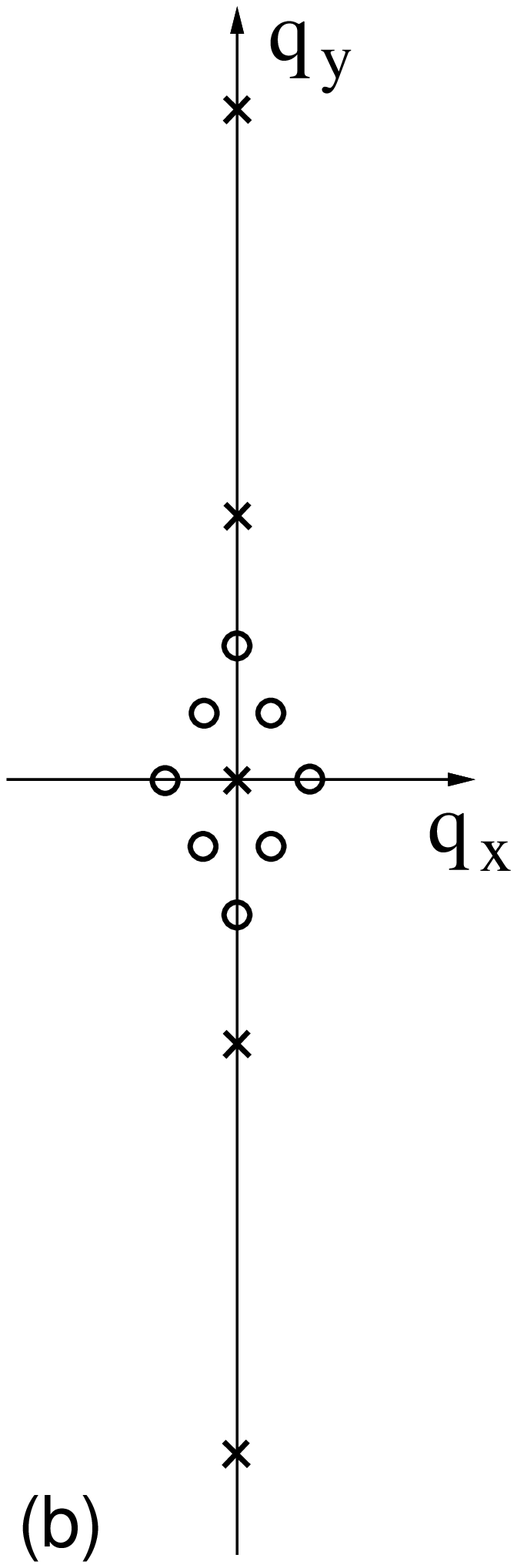}} \vspace{0.5cm}
\caption{Schematic structure function for the FS phase with the 
  commensurability vector (a) $\vec{p}=(5,0)$ and (b) $\vec{p}=(2,-2)$,
  illustrating a combination of the quasi- and true Bragg peaks, given
  by Eq.\ref{Sq}. Crosses indicate true Bragg peaks and open circles
  quasi-Bragg peaks.}
\label{fig:SqFSfig}
\end{figure}

Equation \ref{Sq2} predicts true Bragg peaks (with power-law
corrections) at multiples of the periodic potential wave vector
$\Qlaser$ and pure power-law (quasi-) Bragg peaks at {\em all other}
reciprocal lattice vectors $\bf G$, even for those with ${\bf
G}\parallel\Qlaser$.  Note that in a real physical system, the
periodic potential will {\em not} in general be a single harmonic as
assumed in our model, Eq.\ref{H_Q}.  Hence, we expect that the Bragg
peak amplitude $A_n$ observed in experiments will be a {\em sum} of
terms like those in Eq.\ref{A_n} and the square of the amplitude of
the $n$th Fourier harmonic, $U_{n\Qlaser}$ of the applied periodic
potential.  This of course will only modify the prefactors in the
different contributions to $S({\bf q})$, predicted for the FS in
Eq.\ref{Sq2}. We schematically illustrate $S({\bf q})$ for a floating
solid in Fig.\ref{fig:SqFSfig} for the commensurability vectors
$\vec{p}=(5,0)$ and $\vec{p}=(2,-2)$, respectively, with the $y$-axis
chosen to point along $\Qlaser$.
The set of on-$\hat{\bf q}_y$-axis quasi-Bragg peaks (indicated by
open circles) interleaving the true Bragg peaks (indicated by ``x'''s)
is the notable feature that distinguishes the FS from its locked
counterpart LFS, in which all on-$\hat{\bf q}_y$-axis peaks are true
Bragg peaks.

\subsubsection{Locked Floating Solid (LFS)}
\label{sec:LFS}

At sufficiently low temperatures, the periodic potential will always
be a relevant perturbation, pinning the 2d solid in the direction
perpendicular to its troughs. Because of the 1d nature of the pinning
potential, the 2d crystal will remain unpinned along the direction of
the potential minima and will be able to adjust freely in that
direction. To reflect this dual character, we therefore call this
phase the ``{\em locked floating solid}''.

At high laser intensity, such that the bare value of the pinning
energy $\UQ$ is much larger than the elastic energy $\mu a^2$ for the
shortest (and therefore all) wavelength phonon mode, our system is in
the {\em strong pinning regime}. For a commensurate periodic potential, in
this regime, fluctuations in the lattice positions perpendicular to
the troughs are small and the periodic potential $H_\Qlaser$,
Eq.\ref{H_Qsimple}, can be safely expanded in powers of the
corresponding phonon degree of freedom, $\Qlaser\cdot{\bf u}$, leading
to
\begin{mathletters}
\begin{eqnarray}
H_\Qlaser&\approx&{\mbox const.}+{1\over2}\UQ a^{-2}\int d^2 r
\Big(\Qlaser\cdot{\bf u}({\bf r})\Big)^2\;,\label{H_Qexpand1}\\
&\approx&{\mbox const.}+{1\over2}\UQ a^{-2}\Qlaserscalar^2\int d^2 r \, 
u_y^2({\bf r})\;.\label{H_Qexpand2}
\end{eqnarray}
\end{mathletters}

In contrast, a {\em weak pinning regime}, $\UQ\ll\mu a^2$, consists of
two sets of elastic modes, those with $k<k_c$ and those with $k>k_c$,
where $k_c\equiv K/b_*$ is a crossover wavevector for which the elastic
energy density $\mu (k_c a)^2$ just balances the pinning energy
density $\UQ(b_*) K^2$ at the same length scale.  Since the pinning
energy is subdominate to the elastic energy for modes with $k>k_c$, we
can simply integrate out these weakly pinned modes perturbatively in
$\UQ$. This results in an effective strength of the pinning potential
given by
\begin{equation}
\UQ(b_*)=\UQ b_*^{-\etaK/2}\;.\label{U_Qb}
\end{equation}
After equating this to the corresponding elastic energy $\mu
(a/b_*)^2$ we find
\begin{equation}
b_*=\left({\mu a^2\over \UQ}\right)^{2/(4-\etaK)}\;,\label{b_star}
\end{equation}
which, when inserted inside Eq.\ref{U_Qb} leads to 
\begin{mathletters}
\begin{eqnarray}
\UQ(b_*)&=&\UQ \left({\UQ\over \mu a^2}\right)^{\etaK/(4-\etaK)}\;,\\
&=&\UQ^{4/(4-\etaK)}/(\mu a^2)^{\etaK/(4-\etaK)}\;.
\end{eqnarray}
\label{U_Qb2}
\end{mathletters}

Since the $u_y$ fluctuations in the remaining strongly pinned elastic
modes are small, the effective pinning potential $H_\Qlaser$ can once
again be safely expanded in powers of $u_y$. Doing this we obtain a
result identical to Eq.\ref{H_Qexpand2}, but with $\UQ$ replaced by
$\UQ(b_*)$ given in Eq.\ref{U_Qb2}.

Hence, in both the strongly and weakly pinned regimes, unlike the FS
phase, the LFS is characterized at long wavelengths by one acoustic
($u_x$) and one optical ($u_y$) phonon mode, with an effective
Hamiltonian
\begin{equation}
H=\Hel+{\mu\over2\xi^2} \int d^2 r \, u_y^2({\bf r})\;.
\label{H_strong-pinning}
\end{equation}
Here, we have introduced a correlation length $\xi$ which, given
Eqs.\ref{H_Qexpand2}, \ref{U_Qb2}, reads
\begin{equation}
\xi^{-2}(\UQ)=
\cases{
\frac{\UQ}{\mu a^2} \Qlaserscalar^2, & \text{for $\frac{\UQ}{\mu a^2} \gg 1$,} \cr
\Bigl(\frac{\UQ}{\mu a^2}\Bigr)^{4/(4-\etaK)}\Qlaserscalar^2, 
& \text{for $\frac{\UQ}{\mu a^2} \ll 1$}.
}
\label{xi}
\end{equation}

At length scales longer than the crossover scale set by $\xi$,
Eq.\ref{xi}, we can safely ignore the spatial derivative of $u_y$
terms, and the LFS is well described by an effective Hamiltonian
\begin{equation}
\HLFS={1\over2}\int d^2 r\left[B_{yx}(\partial_y u_x)^2+
B_{xx}(\partial_x u_x)^2+{\mu\over\xi^{2}} u_y^2\right]\;,
\label{H_LFS}
\end{equation}
where
\begin{eqnarray}
B_{yx}&=&(\mu+\gamma-\alpha)\;,\\
B_{xx}&=&\lambda_{xx}\;.
\end{eqnarray}

We can now compute the translational order parameter correlation
function and the structure function that characterize the LFS phase.
Repeating first the calculation for the persistent part determined by
$\langle\rho_{\bf G}\rangle$, we immediately find, that, as in all the
phases in the presence of the periodic potential, $\langle\rho_{\bf
G}\rangle\neq0$ for ${\bf G}= \pm n\Qlaser$. However, the distinguishing
feature of the LFS is that this average is finite for {\em all} $\bf
G$ {\em parallel} to $\Qlaser$, by virtue of the finite pinning length
$\xi$, Eq.\ref{xi}. This result can be immediately seen by noting that
for ${\bf G}||\Qlaser$, the logarithmically divergent (with $L$)
$\langle u_x^2\rangle_0$ correlation function does {\em not} appear in
$\langle\rho_{\bf G}\rangle_0$, where the subscript ``0'' again
represents an average with the elastic Hamiltonian $H_{el}$ only.
Instead we have
\begin{mathletters}
\begin{eqnarray}
\langle\rho_{\bf G}\rangle
&=&e^{-{1\over2} G^2 \langle u_y^2\rangle}\\
&=&\left(a\over\xi\right)^{\etaG/2}\label{rhoG_final1}
\end{eqnarray}
\end{mathletters}
which only involves the ``massive'' $u_y$ degree of freedom, whose
logarithmic correlations are cutoff at $L_c=\xi$ and therefore is
finite even in the thermodynamic limit.

We can also obtain the above result via a straightforward matching
calculation.  The difficulty of computing translational correlation
functions in the weakly pinned regime of the LFS phase is that for
long length scales ($>\xi$), despite of the weakness of the pinning
potential, a direct perturbative expansion in $\UQ$ is divergent
because of its relevance (in the renormalization group sense) inside
the LFS phase. The power of the renormalization group is that it
allows us to relate this difficult weakly pinned, small $\UQ$ regime
to the strongly pinned regime, where $\UQ$ has grown to the magnitude
of the elastic energy $\mu a^2$, and can therefore be treated as a
``mass'', as in Eq.\ref{H_Qexpand2}. We can apply this matching
procedure to the computation of $\langle\rho_{\bf G}(\UQ)\rangle$, by
using a relation between the weakly and strongly pinned regimes,
namely
\begin{equation}
\langle\rho_{\bf G}(\UQ)\rangle=
b^{-\etaG/2}\langle\rho_{\bf G}(\UQ b^{2-\etaK/2})\rangle
\label{matching1}
\end{equation}
obtained using the scaling dimension of the operator $\rho_{\bf G}$
and the RG eigenvalue of $\UQ$, both easily extracted from
Eq.\ref{Cc}. Choosing the arbitrary rescaling factor $b=b_*$ such that
$\UQ(b)$ is in the strongly pinned regime, where $\UQ(b_*)=\mu a^2$,
Eq.\ref{matching1} becomes
\begin{equation}
\langle\rho_{\bf G}(\UQ)\rangle=
\left({\UQ\over\mu a^2}\right)^{\etaG/(4-\etaK)}
\langle\rho_{\bf G}(\mu a^2)\rangle\;.\label{matching2}
\end{equation}
Since the right hand side is in the strong coupling regime, it can be
easily computed using the coarse-grained Hamiltonian,
Eq.\ref{H_strong-pinning}. Doing this we find
\begin{equation}
\langle\rho_{\bf G}(\UQ)\rangle=
\left( \frac{\UQ}{\mu a^2}\right)^{\etaG/(4-\etaK)}
e^{ \frac12 \etaG \ln(\Qlaserscalar a)}\;,\label{matching3}
\end{equation}
which in the weakly pinned regime is equivalent to the result given in
Eq.\ref{rhoG_final1}.

Note that the nontrivial nonlinear power-law response of the
translational order parameter to the periodic laser potential,
predicted by Eq.\ref{matching3} is only a nonanalytic piece of the
full response, which includes an analytical background. Hence,
although at low temperatures, such that $\etaG/(4-\etaK)<1$,
the full response in the $\UQ\rightarrow0$ limit is dominated by the
nonanalytical part, Eq.\ref{matching3}, at higher temperatures, the
ever-present linear piece of the analytical part will dominate, and
experimentally one should instead observe
\begin{equation}
\langle\rho_{\bf G}(\UQ)\rangle\sim \UQ\;.\label{rhoUlinear}
\end{equation}

For our highly anisotropic system, the connected part of the
correlation function $C_{\bf G}({\bf r})$, is given by
\begin{equation}
C_{\bf G}^{(c)}({\bf r})=
e^{-{1\over2}[G_x^2 G_{xx}({\bf r})+2 G_x G_y G_{xy}({\bf r})
+G_y^2 G_{yy}({\bf r})]}\;,\label{Cani}
\end{equation}
where $G_{ij}({\bf r})\equiv \langle(u_i({\bf r})-u_i(0))(u_j({\bf
  r})-u_j(0))\rangle $, is the connected phonon correlation function
computed with the full Hamiltonian. In the weakly pinned regime, for
small length scales, all phonon correlation functions display the
usual 2d logarithmic growth, which, in the isotropic approximation,
i.e., using Hamiltonian $H_0$, Eq.\ref{H_0} leads to the power-law
correlation for $C_{\bf G}^{(c)}({\bf r})$ that we found in
Eq.\ref{Cc} for the FS phase.  However, for length scales longer than
$\xi$, Eq.\ref{xi}, while $G_{xx}({\bf r})$ will continue to grow
logarithmically, such growth in $G_{yx}({\bf r})$ and $G_{yy}({\bf
  r})$ will be cutoff by the pinning length $\xi$. Consequently, in
the LFS we find
\begin{equation}
C_{\bf G}^{(c)}({\bf r})\approx \left({a\over r}\right)^{\eta_{G_x}}
\left({a\over \xi}\right)^{{\overline \eta}_{G_y}}\;,
\label{CcLFS}
\end{equation}
where
\begin{equation}
\eta_{G_x} = \frac{G_x^2}{2 \pi} \, \frac{\kT}{\sqrt{B_{xx} B_{yx}}}
\end{equation}
which reduces to $\eta_{G_x} = \frac{G_x^2}{2 \pi} \,
\frac{\kT}{\sqrt{\mu(2 \mu+\lambda)}}$ when the effect of the periodic
  potential on elasticity and renormalizations due to dislocation
  pairs on the effective elastic coefficients are neglected.  A
  discrete Fourier transform of this correlation function gives the
  corresponding structure function
\begin{equation}
S({\bf q})\approx\sum_{\bf G}\left[
{B_G\over|{\bf q}-{\bf G}|^{2-\eta_{G_x}}}
+ A_G\delta^{(2)}({\bf q}-{\bf G})\right]\;,
\label{SqLFS}
\end{equation}
where the quasi-Bragg peak amplitude $B_G$ is given by
\begin{mathletters}
\begin{eqnarray}
B_G&\propto& 
\left({a\over \xi}\right)^{{\overline \eta}_{G_y}}\;,\label{B_Ga}\\
&\propto&\cases{
\UQ^{{\overline \eta}_{G_y}/2}, & \text{for ${\UQ\over\mu a^2}\gg1$,} \cr
\UQ^{2{\overline \eta}_{G_y}/(4-\etaK)}, & \text{for ${\UQ\over\mu a^2}\ll1$.}}
\label{B_Gb}
\end{eqnarray}
\end{mathletters}
and the Bragg peak amplitude $A_G$
\begin{mathletters}
\begin{eqnarray}
A_G&\propto&\delta_{G_x,0} \, \left({a\over \xi}\right)^{\etaG}\;,\label{A_Ga}\\
&\propto&\delta_{G_x,0}
\cases{
\UQ^{\etaG/2}, & \text{for ${\UQ\over\mu a^2}\gg1$,} \cr
\UQ^{2\etaG/(4-\etaK)}, & \text{for ${\UQ\over\mu a^2}\ll1$,}}
\label{A_Gb}
\end{eqnarray}
\end{mathletters}
which is finite {\em if and only if} ${\bf G}$ is parallel to
$\Qlaser$. As a consequence of the discussion after
Eq.\ref{matching3}, the amplitude $A_G$ will also have a background
analytic in $\UQ$, which in a weak pinning limit scales as $\UQ^2$
(see Eq.\ref{rhoUlinear}), and therefore at higher temperatures will
dominate over the nonanalytical part predicted in Eq.\ref{A_Gb}.

\begin{figure}[bth]
\narrowtext
  \narrowtext \centerline{\epsfxsize=0.33\columnwidth
    \epsfbox{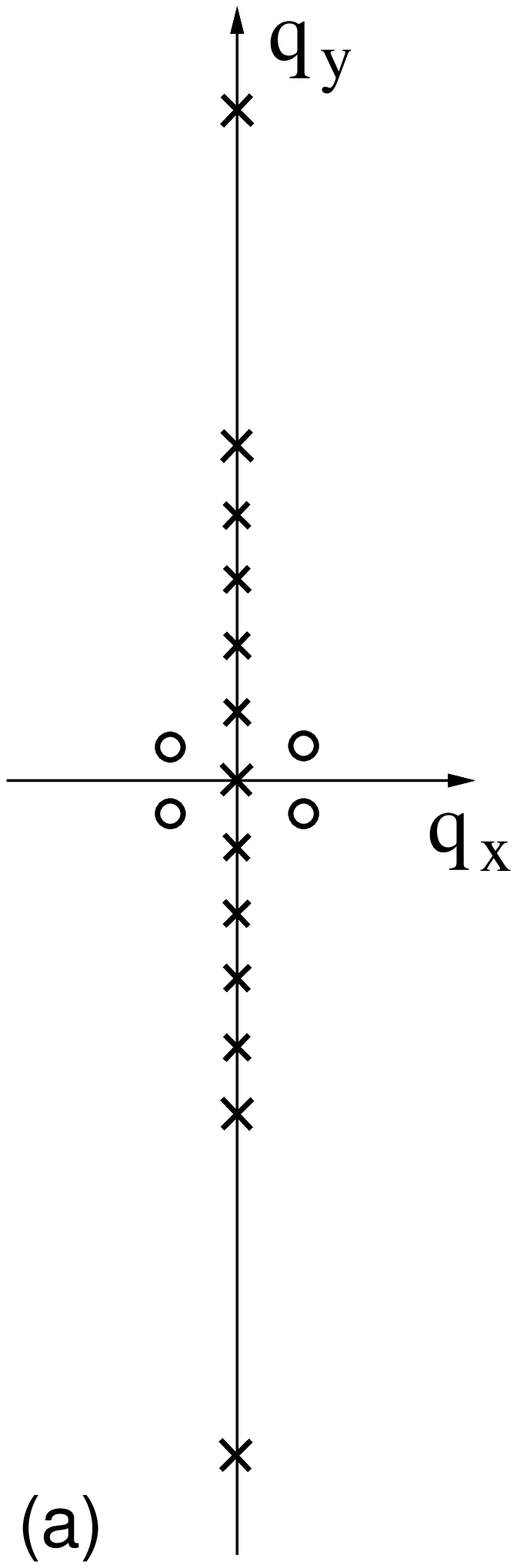} \hspace{0.04\columnwidth}
    \epsfxsize=0.33\columnwidth \epsfbox{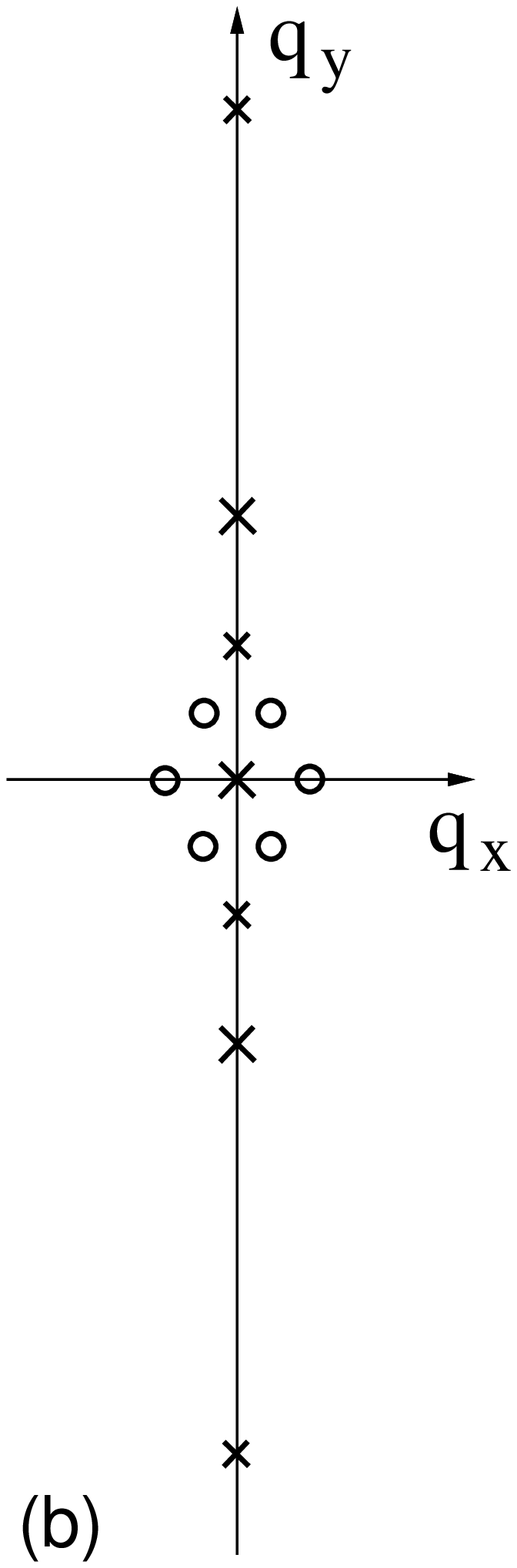}} \vspace{0.5cm}
\caption{Schematic structure function for the LFS phase with the 
  commensurability vector (a) $\vec{p}=(5,0)$ and (b) $\vec{p}=(2,-2)$,
  illustrating a combination of the quasi- and true Bragg peaks, given
  by Eq.\ref{Sq}.}
\label{fig:SqLFSfig}
\end{figure}

We illustrate schematically $S({\bf q})$ in Fig.\ref{fig:SqLFSfig} for
the commensurability vectors $\vec{p}=(5,0)$ and $\vec{p}=(2,-2)$,
respectively, with the $y$-axis chosen to point along $\Qlaser$.
These predictions for the structure function of the LFS, displaying
amplitudes that vanish as nontrivial powers (determined by a
continuously varying exponent $\overline{\eta}_{G_y}$) of the periodic
potential strength (Eqs.\ref{B_Gb},\ \ref{A_Gb}) provide the first
theoretical explanation for observations of Clark, et
al.\cite{chowdhury-ackerson-clark:85}.

\subsection{Smectic Phases}
\label{sec:SmecticPhases}

As was first pointed out by Ostlund and
Halperin\cite{ostlund-halperin:81}, in uniaxial two-dimensional
lattices, dislocations with Burgers vector along and perpendicular to
the uniaxial axis will generically have different core energies and
will therefore proliferate at different temperatures. This will
consequently allow the possibility of a phase that is intermediate
between a fully ordered crystal and a completely disordered liquid.

\begin{figure}[bht]
  \narrowtext \centerline{ \epsfxsize=0.9\columnwidth
    \epsfbox{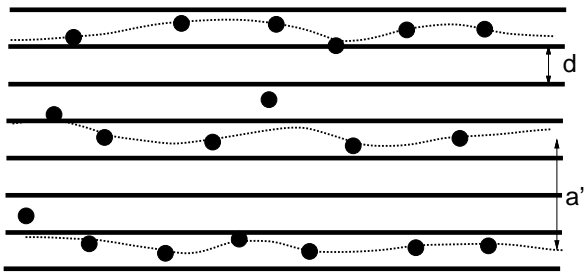}} \vspace{0.5cm}
\caption{2d colloidal smectic phase in the presence of a commensurate
  1d periodic potential with period $d$, commensurability parameter
  $p=3$, and potential maxima indicated by full horizontal lines.
  Dashed lines denote the maxima in the smectic density, which are
  pinned inside the minima of the periodic laser potential.}
\label{fig:SmecticCartoon}
\end{figure}

In a commensurate orientation, such that Bragg rows coincide with the
periodic potential troughs, we would expect dislocation pairs with
Burgers vectors parallel to the potential minima, to unbind first.
We call the resulting class of thermodynamically distinct phases,
smectics.  Their main common characteristic is that they display a
finite elastic modulus for shear deformations perpendicular to the
Burgers vector of unbound dislocations, but do not resist shear
parallel to them, possessing only liquid-like correlations between the
corresponding ``atomic'' rows. Consequently, such 2d smectics display
1d periodicity perpendicular to the Burgers vector of unbound
dislocations, and, as illustrated in Fig.\ref{fig:SmecticCartoon}, can
be equivalently described as a periodic stack of 1d liquids.

It is important to note, that despite of their name, the smectics
discussed here are fundamentally distinct from the smectic phases
found in liquid crystal materials and substrate-free smectics
discussed in Ref.\onlinecite{ostlund-halperin:81}. The most important
distinction is that in liquid crystal smectics and those without an
underlying pinning substrate, the orientational symmetry is broken
{\em spontaneously} (uniaxial anisotropy notwithstanding; see
Sec.\ref{sec:rotational-symmetry}), leading to a soft
Laplacian-curvature (rather than gradient-tension) elasticity, which
preserves this underlying symmetry even in the smectic phase, where it
is nonlinearly realized.\cite{deGennes,ChaikinLubensky} In fact such
substrate-free 2d smectics, because of the softness of their
elasticity, are well known to be unstable to thermally-driven
unbinding of dislocations, and at scales longer than the distance
between these free dislocations are therefore indistinguishable from a
nematically-ordered 2d liquid.\cite{TonerNelson} As was recognized by
the authors of Ref.\onlinecite{ostlund-halperin:81}, such thermal
instability of substrate-free 2d lattices precludes the existence of
thermodynamically distinct intermediate 2d smectic phase in which only
one set of Burgers vectors (e.g., along the uniaxial direction)
unbind. However, in strong contrast to those rotationally invariant
systems, in 2d lattices studied here the periodic (laser) potential
{\em explicitly} breaks rotational symmetry, binding by a linear
potential dislocation pairs with Burgers vector having components {\em
  along} $\Qlaser$. Consequently, such dislocations remain bound even
when those with Burgers vectors {\em perpendicular} to $\Qlaser$
unbind and therefore allow the existence of 2d smectic phases that are
thermodynamically distinct from a liquid.

Deep in such a smectic phase, the $u_y({\bf r})$ phonon field, which
(see Fig.\ref{fig:SmecticCartoon}) describes local fluctuations in
the maxima positions of the 1d density wave, is the only remaining
important degree of freedom.  The ever-present bound dislocation pairs
and the density of vacancies and interstitials are ``massive'' degrees
of freedom. They can be easily integrated out, leading only to a
finite renormalization of elastic constants for $u_y$ deformations,
and therefore are unimportant in a static theory.

In close analogy to the translational order parameter of the 2d
crystal, the smectic is distinguished from a liquid by a finite
translational order parameter $\rho_{\bf G}=e^{i{\bf G}\cdot {\bf
    u}}$, but with a {\em single} (rather than a set,
Eq.\ref{reciprocal_lattice_vectors}) reciprocal vector ${\bf
  G}=G\haty=(2\pi/a) \haty$. It is related to the total molecular
density via a standard relation\cite{deGennes}
\begin{equation}
\rho({\bf r})= \mbox{Re}[\rho_0+e^{i G y}\rho_{\bf G}({\bf r})]\;,
\label{rho}
\end{equation}
where $\rho_0$ is the mean density of the smectic.

Of course, in the presence of a 1d periodic potential, a smectic is a
thermodynamically distinct phase only if ${\bf G}=(2\pi/a) \haty$
differs from the wavevector $\Qlaser$ characterizing the external
potential and the modulated liquid.  Commensurate smectics, which we
focus on here, are equivalently characterized by the ratio of their
period $a$ to that of the periodic potential $d$, with
commensurability ratio $a/d\equiv p\in{\cal Z}$. A $p$-Smectic then
spontaneously breaks the discrete translational symmetry $T^y_d\otimes
T^x$ of the modulated liquid, with its equal occupancy of each
potential minima down to $T^y_a\otimes T^x$, with only every $p$th
minima equivalently populated. Clearly then $p=1$ is indistinguishable
from the fully disordered modulated liquid.

Above symmetry considerations uniquely specify the Hamiltonian that
characterizes the $p$-Smectic phase
\begin{eqnarray}
\HSm&=&\int d^2 r 
\biggl[ \frac12 
\Bigl( B_{xy}(\partial_x u_y)^2+ B_{yy}(\partial_y u_y)^2 \Bigr)
\nonumber \\
&&\qquad-\UQ a^{-2} \cos\bigl(\Qlaserscalar u_y({\bf r})\bigr) \biggr]\;,
\label{H_Sm}
\end{eqnarray}
which, not surprisingly, is an anisotropic scalar Sine-Gordon model in
the phonon field $u_y({\bf r})$.

Given the form of the Hamiltonian in Eq.\ref{H_Sm}, there is a close
similarity between the properties of the smectic and the 2d crystal,
studied in the previous section. The quantitative differences between
these phases are due to the distinction between the vector (${\bf
u}=(u_x,u_y)$) and scalar ($u_y$) nature of elastic degrees of freedom
in the 2d solid and smectic, respectively. More specifically, in close
analogy to the 2d solid, we find that for a fixed integer
commensurability ratio $p$, there exist a low temperature ``locked''
and higher temperature ``floating smectic'' phase.  These are
distinguished by the importance of the periodic pinning potential,
which is relevant (in the RG sense) in the LSm phase, acting as a
``mass'' for $u_y$, and irrelevant in the FSm phase, where for most
static properties it can be ignored.

\subsubsection{Floating Smectic (FSm)}
\label{sec:FSm}

In the ``floating smectic phase'' (FSm), thermal fluctuations in the
position of the layers are sufficiently large that at long length
scales they average away many effects of the periodic pinning
potential. Hence, many of the static properties of the FSm phase can
be well described by the Hamiltonian, Eq.\ref{H_Sm}, with $\UQ=0$.
However, as we discussed in detail in our analysis of the FS phase,
despite of the RG irrelevance of the periodic potential, continuous
translational symmetry is still explicitly broken by it, which leads
to true long-ranged translational order in the smectic order parameter
$\rho_{\bf G}$ for $G\haty$ at multiples of the reciprocal lattice
vector $\Qlaserscalar \haty$, characterizing the laser potential.

Calculations that closely parallel those of Sec.\ref{sec:FS} for the FS,
lead to power-law correlations in the connected part of the
translational two-point correlation function
\begin{equation}
C_{\bf G}^{(c)}({\bf r})\sim {1\over|\bf r|^{\eta_{\rm FSm}}}\;,\label{CcSm}
\end{equation}
where
\begin{equation}
\eta_{\rm FSm} = {\kT G^2\over 2 \pi \sqrt{B_{xy} B_{yy}}}\;.
\label{etaFSm}
\end{equation}
is the exponent characterizing the FSm phase, in analogy to $\etaG$,
Eq.\ref{eta_G}, of the FS. 

The disconnected part of the smectic translational correlation
function is finite only at $G=n \Qlaserscalar$ ($n\in{\cal Z}$). The
corresponding floating smectic structure function is given by an
expression similar to the FS, Eq.\ref{Sq2}. The only difference is
that $\etaG$ of the FS is replaced by $\eta_{\rm FSm}$ of the FSm and
the summation over $G$ is a sum over integer multiples of $2\pi/a$.
Consequently, one expects to see sharp peaks only on the $q_y$ axis,
with power-law peaks at $G\neq n \Qlaserscalar$, and true Bragg peaks
at $G=n \Qlaserscalar$.  This FSm structure function is schematically
displayed in Fig.\ref{fig:FSmLSm} below.

\subsubsection{Locked Smectic (LSm)}
\label{sec:LSm}

\begin{figure}[bth]
  \narrowtext \centerline{\epsfxsize=0.33\columnwidth
    \epsfbox{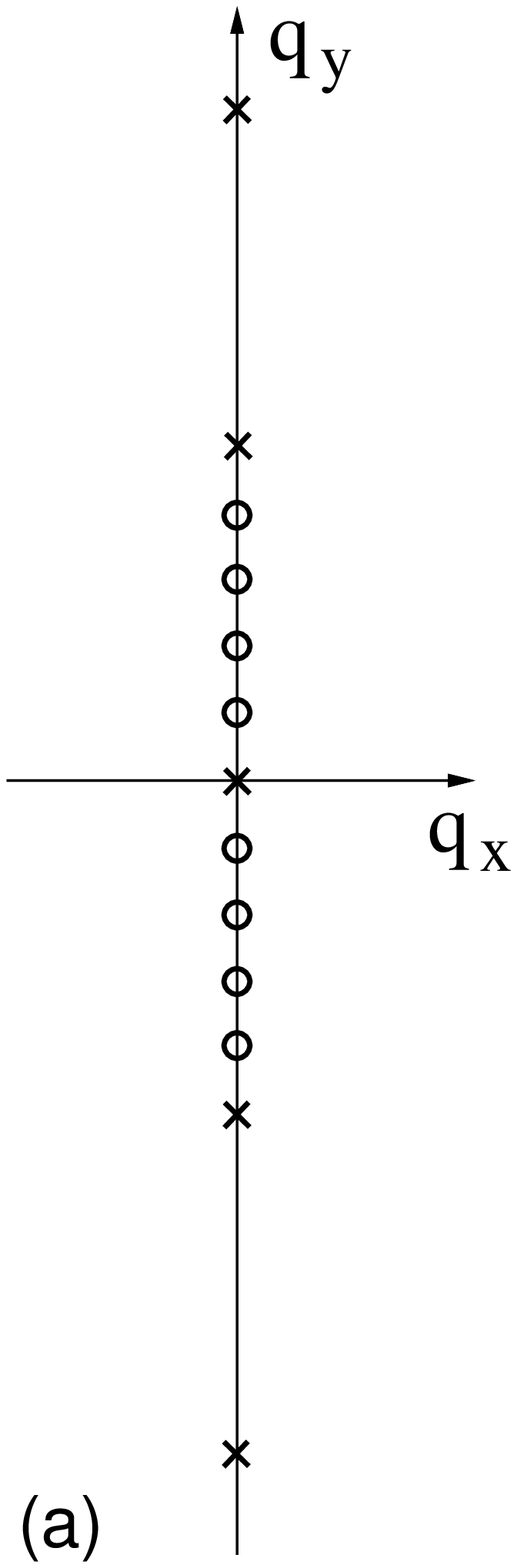} \hspace{0.04\columnwidth}
    \epsfxsize=0.33\columnwidth \epsfbox{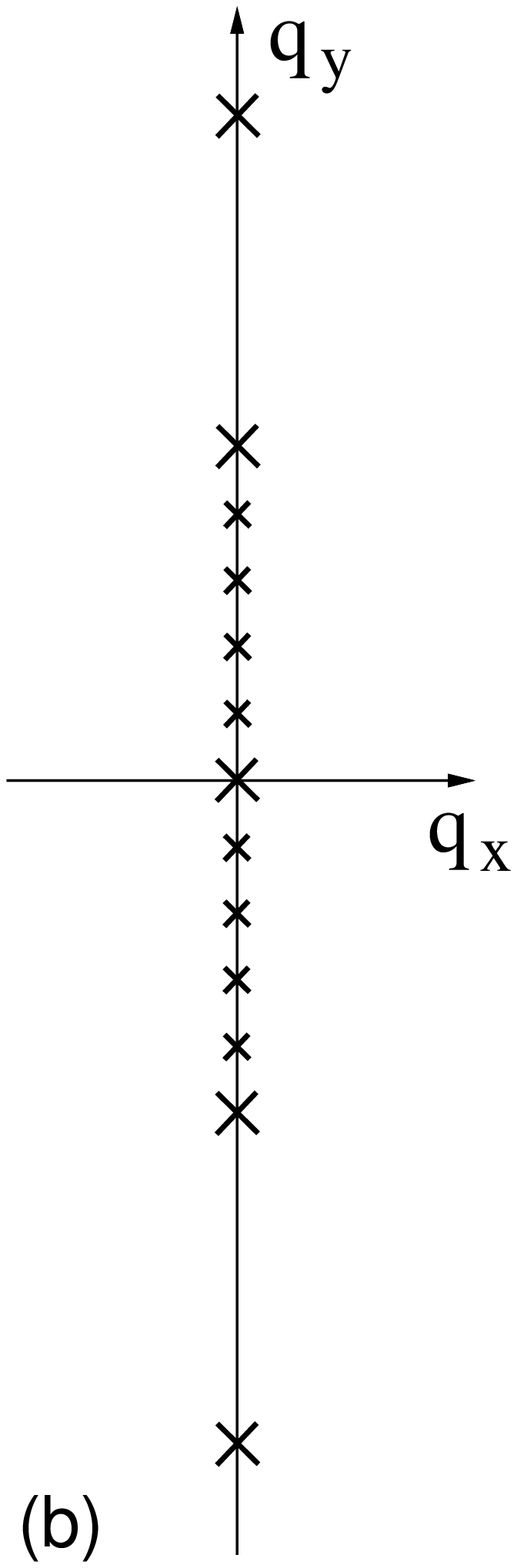}} \vspace{0.5cm}
\caption{(a) Schematic of the structure function for the floating 
  smectic phase, characterized by on-$q_y$-axis quasi-Bragg peaks
  (open circles) and true Bragg peaks (crosses). (b) Schematic of the
  structure function for the locked smectic phase, characterized by
  on-$q_y$-axis true Bragg peaks, with small and large crosses
  indicating spontaneously and directly induced translational order.}
\label{fig:FSmLSm}
\end{figure}

As the temperature is lowered, the periodic potential becomes
relevant, pinning the smectic layers. The resulting ``locked smectic''
(LSm) phase is characterized by long-range translational order, and,
as illustrated in Fig.\ref{fig:FSmLSm}, displays true Bragg peaks at
all values of the on-$q_y$-axis reciprocal lattice vectors $G=n
2\pi/a$.  At long scales, the effective elastic Hamiltonian that
characterizes this phase is simply
\begin{equation}
\HLSm={1\over2} \ {\mu\over\xi^2}\int d^2r\ u_y^2\;,
\label{H_LSm}
\end{equation}
with $\xi$ given by Eq.\ref{xi}.

\subsubsection{Modulated Liquid (ML)}
\label{sec:L}

\begin{figure}[bth]
  \narrowtext \centerline{\epsfysize=0.9\columnwidth
    \epsfbox{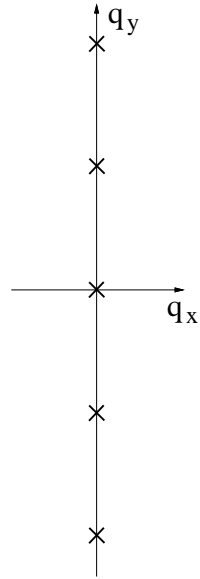}} \vspace{0.5cm}
\caption{Schematic of the structure function for the modulated 
  liquid phase characterized by on-$q_y$-axis true Bragg peaks located
  at $n (2\pi/d)$.}
\label{fig:MLiquid}
\end{figure}

The modulated liquid is the most disordered phase, which occurs at
highest temperatures and does not {\em spontaneously} break any
symmetries.  It is characterized by a vanishing shear modulus, unbound
dislocations, absence of massless Goldstone modes, and a discrete
symmetry of translations along the y-axis by periodic potential
constant $d$. The corresponding structure function of this explicitly
orientationally ordered phase, illustrated in Fig.\ref{fig:MLiquid},
is a set of true Bragg peaks at multiples of the reciprocal lattice
vector $\Qlaserscalar=2\pi/d$ of the periodic potential.
{\em Finite linear} translational order parameter susceptibility
guarantees that the average order parameter is linear in the strength
of the periodic potential.  Therefore, as is clear from
Eq.\ref{I_onaxis} the strength of the Bragg peaks scales as a {\em
  cube} of the input laser intensity, proportional to $\UQ$, as
observed in experiments by Clark, et
al.\cite{chowdhury-ackerson-clark:85,IinIout_liquid}.

\section{Phase Transitions}
\label{sec:phase_transitions}

Phase transitions that take place in our system fall into two broad
classes: roughening and melting. However, for high values of the
commensurability ratio $p$ ($p>p_c$) these classes are mathematically
related to each other by the duality
transformations\cite{jose-etal:77,Radzihovsky-Toner}, and are both
examples of the Kosterlitz-Thouless type of transitions, with kinks
and dislocations unbinding, respectively. For $p<p_c$, the roughening
transitions are in a different (Ising model and other models with a
discrete symmetry) universality class.

\subsection{Roughening Transitions}
\label{sec:roughening}

Phase transitions that fall into the roughening transition
universality class separate a low temperature ordered phase, in which
a potential Goldstone mode is strongly pinned by an external periodic
potential, from a quasi-long-range ordered phase, in which the
periodic potential is irrelevant in a renormalization group sense.
The locked floating solid to floating solid and the locked smectic to
floating smectic transitions, discussed in Sec.\ref{sec:phases} fall
into this broad universality class, although differ in details that we
discuss below.

Despite these small differences the analysis of these transitions are
quite similar and can be done via standard perturbative momentum-shell
renormalization group (RG) transformation.\cite{Wilson,jose-etal:77}
Since the smooth (locked) and rough (floating) phases are
distinguished by the relevance and irrelevance of the periodic
potential, respectively, we can find the transition temperature by
analyzing the behavior of $H_\Qlaser$, Eq.\ref{H_Q}, as a function of
length scale. We separate the phonon field, which for a solid phases
is a two component vector and a scalar for a smectic, into the high
and small wavevector modes
\begin{equation}
{\bf u}({\bf r})={\bf u}^<({\bf r})+{\bf u}^>({\bf r})\;,
\label{u_shell}
\end{equation} 
and integrate perturbatively in $\UQ$ the high wavevector part ${\bf
  u}^>({\bf r})$, with nonvanishing Fourier components inside a thin
momentum shell
\begin{equation}
\Lambda e^{-\ell}<|{\bf q}|<\Lambda\;.
\label{shell}
\end{equation}
We then rescale the lengths and long wavelength part of the fields
with
\top{-2.5cm}
\columnwidth3.4in
\narrowtext
\noindent
\begin{mathletters}
\begin{eqnarray}
{\bf r}&=&e^{\ell}{\bf r}'\;,\label{r}\\
{\bf u}^<({\bf r})&=&e^{\phi\ell}{\bf u}'({\bf r'})\;,\label{u}
\end{eqnarray}
\label{rescale}
\end{mathletters}
so as to restore the ultraviolet cutoff back to $\Lambda=2\pi/a$.
Because the pinning potential nonlinearity is a periodic function, it
is convenient (but not necessary) to take the arbitrary field
dimension to be 
\begin{equation}
\phi=0\;,
\end{equation}
thereby preserving the period $a=2\pi/\Lambda$ under the
renormalization group transformation.\cite{Qflow} Under this
transformation the resulting effective Hamiltonian,
$H=\Hel+H_\Qlaser$, can be restored into its original form with
effective $\ell$-dependent elastic and $\UQ$ couplings.

For the periodic pinning potential coupling $\UQ$, we find in a
standard way\cite{Wilson,roughenning}
\begin{equation}
\UQ(\ell)=\UQ e^{2\ell-{1\over2}\Qlaserscalar_y^2\langle u_y^2\rangle_>}\;,
\label{U_Q_relevant}
\end{equation}
where $\langle u_y^2\rangle_>$ is to be computed with the elastic
Hamiltonian appropriate to the phase being analyzed, keeping only
modes within an infinitesimal momentum shell near the zone boundary
$\Lambda$.  Hence the nature of the pinning by the substrate potential
and the transition temperature obviously depend on the degree of the
translational order in the system, i.e. whether the phase is a solid,
or a smectic.

\subsubsection{Locked Floating Solid to Floating Solid Transition}
\label{sec:LFS-FS-transition}

To determine the critical temperature for the LFS-to-FS transition, we
compute the $\langle u_y^2\rangle_>$ average using the anisotropic
elastic Hamiltonian $\Hel$, Eq.\ref{H_el_anisotropic}, describing the
2d solid phase in the presence of a 1d periodic potential. Rewriting
$\Hel$ in terms of Fourier transformed phonon fields ${\bf u}({\bf
  q})$, we find
\end{multicols}
\widetext
\begin{equation}
\Hel=\int{d^2q\over(2\pi)^2}\Bigg[
{1\over2}\left(B_{xx} q_x^2+B_{yx} q_y^2\right)|u_x({\bf q})|^2+
{1\over2}\left(K_{xy} q_x^2+K_{yy} q_y^2\right)|u_y({\bf q})|^2+
\delta q_x q_y u_x({\bf q}) u_y(-{\bf q})\Bigg]\;,
\label{H_el_FT}
\end{equation}
\bottom{-2.7cm}
\begin{multicols}{2}
\columnwidth3.4in
\narrowtext
\noindent
where,
\begin{mathletters}
\begin{eqnarray}
B_{xx}&\equiv&\lambda_{xx}\;,\\
B_{yx}&\equiv&\mu-\alpha+\gamma\;,\\
K_{yy}&\equiv&\lambda_{yy}\;,\\
K_{xy}&\equiv&\mu+\alpha+\gamma\;,\\
\delta&\equiv&\mu+\lambda_{xy}-\gamma\;,
\end{eqnarray}
\end{mathletters}
which, after a simple Gaussian integration leads to
\begin{eqnarray}
\langle u_y^2\rangle_> 
=  \int_q^>
\frac{\kT}{K_{yy}q_y^2+K_{xy}q_x^2-
\frac{\delta^2q_x^2q_y^2}{B_{xx}q_x^2+B_{yx}q_y^2}}\;,
\end{eqnarray}
where we have introduced the shorthand notation $\int_q^> \equiv
\int^> \frac{d^2q}{(2 \pi)^2}$ for the integral over the momentum
shell.  In the dilute limit and neglecting effects of the periodic
potential on the elastic coefficients this reduces to
\begin{eqnarray}
  \langle u_y^2 \rangle_> = \frac{\kT}{2 \pi \overline\mu} \, \ell
\end{eqnarray}
with $\overline\mu = 2 \mu ( 2 \mu + \lambda )/ (3 \mu + \lambda)$.
In order to compute $\langle u_y^2\rangle_>$ in general we use an
elliptical (volume conserving) momentum shell defined by major and
minor axes $\Lambda_x = \Lambda \sqrt{K_{yy}/K_{xy}}$ and $\Lambda_y =
\Lambda \sqrt{K_{xy}/K_{yy}}$.  We find
\begin{eqnarray}
 \langle u_y^2 \rangle_> ={\kT c_1\over 2\pi\sqrt{K_{yy}K_{xy}}} \, \ell\;,
\label{uu_solid}
\end{eqnarray}
where we defined a dimensionless number $c_1$ given by
\begin{equation}
c_1 \equiv \int_0^{2\pi} \frac{d\theta}{2\pi}
\frac{a_x + (a_y-a_x)\sin^2{\theta}}
{a_x + (a_y-a_x-a_{xy})\sin^2{\theta}+a_{xy}\sin^4{\theta}}\;,
\end{equation}
and
\begin{mathletters}
\begin{eqnarray}
a_x&\equiv& {B_{xx}\over K_{xy}}\;,\\
a_y&\equiv& {B_{yx}\over K_{yy}}\;,\\
a_{xy}&\equiv& {\delta^2\over K_{yy}K_{xy}} \;.\\
\end{eqnarray}
\end{mathletters}

Upon combining Eq.\ref{uu_solid} with Eq.\ref{U_Q_relevant}, we find
the eigenvalue of the substrate potential to be
\begin{equation}
\lambda_p \equiv 2-\Qlaserscalar_y^2{\kT c_1\over 4\pi\sqrt{K_{yy}K_{xy}}}\;,
\end{equation}
which after setting
\begin{equation}
\lambda_p (\TpS)=0
\end{equation}
gives us the depinning transition temperature $\TpS$
\begin{equation}
k_{\rm B} \TpS={8\pi\over c_1}\sqrt{K_{yy}K_{xy}}
\left({d\over2\pi}\right)^2\;,
\label{TpS}
\end{equation}
which separates the LFS and FS phases. In the dilute limit the 
transition temperature reduces to
\begin{eqnarray}
 k_{\rm B} \TpS = 8 \pi \overline\mu \left( \frac{d}{2 \pi} \right)^2 \; .
\end{eqnarray}

\subsubsection{Locked Smectic to Floating Smectic Transition}
\label{sec:LSm-FSm-transition}

As discussed in Sec.\ref{sec:SmecticPhases}, at low colloidal
densities our system can exhibit LSm and FSm phases, and therefore
undergo a phase transition between them in the roughening universality
class.  Analogously to the LFS-FS transition analyzed above, we can
calculate the pinning temperature for the LSm-FSm transition by
computing the $\langle u_y^2\rangle_>$ that goes into
Eq.\ref{U_Q_relevant} and finding the temperature at which this RG
eigenvalue vanishes. Using the Hamiltonian $\HSm$, Eq.\ref{H_Sm},
appropriate for the smectic phases and computing to zeroth order in
the pinning potential $\UQ$, we find
\begin{eqnarray}
\langle u_y^2\rangle_>&=&\int_q^> \, 
\frac{\kT}{B_{xy} q_x^2+B_{yy} q_y^2}\\
&=&\frac{\kT}{2\pi\sqrt{B_{yy} B_{xy}}} \, \ell\;,
\end{eqnarray}
where for convenience we again used an elliptical momentum shell with
axes $\Lambda \sqrt{B_{yy} / B_{xy}}$ and $\Lambda \sqrt{B_{xy} /
  B_{yy}}$.

After combining this result with Eq.\ref{U_Q_relevant} we find that
translational pinning by the periodic potential is relevant in a
floating smectic phase for $T<\TpSm$, with $\TpSm$ given by
\begin{equation}
k_{\rm B} \TpSm=8\pi\sqrt{B_{yy} B_{xy}}\left({d\over2\pi}\right)^2\;.
\label{TpSm}
\end{equation}

As discussed in more detail in Sec.\ref{sec:reentrance}, the elastic
moduli in Eqs.\ref{TpS} and \ref{TpSm} are functions of the strength
of the pinning potential $\UQ$, which in turn is proportional to the
input laser intensity $\Iin$. Hence the resulting functions
$\TpS(\Iin)$ and $\TpSm(\Iin)$ in principle determine the LFS-FS and
LSm-FSm phase boundaries displayed in Fig.\ref{fig:phase_diagram_pc_p}
for colloidal densities commensurate with the 1d periodic potential.

\subsection{Dislocation Unbinding Transitions}
\label{sec:dislocation_unbinding_transition}

In the analysis of the preceding Sec.\ref{sec:roughening}, where we
studied a thermal depinning transition within the solid phase, we
implicitly assumed that the dislocations that distinguish the 2d solid
and the smectic phases from the higher temperature disordered phases
remain bound.  Hence, these calculations for the pinning transition
and Eqs.\ref{TpS},\ref{TpSm} remain valid only if they fall below the
corresponding dislocation unbinding melting transition temperatures,
which we now compute.

\subsubsection{Locked Floating Solid to Locked Smectic Transition}
\label{sec:LS-LSm-transition}

It is easy to see from the effective Hamiltonian $\HLFS$,
Eq.\ref{H_LFS}, that the most striking consequence of the 1d periodic
potential is that it leads to the LFS phase, in which, the phonon
degree of freedom, $u_y$, corresponding to displacements transverse to
the potential troughs acquires a ``mass'', Eq.\ref{xi}, and as a
consequence are effectively suppressed.  Therefore, this phonon mode
can be safely integrated out, leaving an effective anisotropic 2d XY
Hamiltonian, with temperature and potential strength dependent {\em
  effective} elastic constants,
\begin{equation}
\HLFS = 
\frac{1}{2} \int d^2 r 
\Bigl( B_{yx}(\partial_y u_x)^2+B_{xx}(\partial_x u_x)^2 \Bigr) \;,
\label{H_LFS_effective}
\end{equation}
that describes a locked floating solid at scales longer than the
correlation length $\xi$ introduced in Sec.\ref{sec:LFS}.

Melting of the LFS can be understood in terms of dislocation
unbinding.  However, in contrast to melting in the absence of an
external (e.g., substrate or laser)
potential,\cite{nelson-halperin:79} here only the so-called type I
dislocation pairs (in the notation of
Ref.\onlinecite{ostlund-halperin:81}) with Burgers vectors $\pm{\bf
  b}_1=\pm b_{\vec n} \hatx$ (see Sec.\ref{sec:commensurability})
aligned parallel to the trough direction (which we continue to take
along the $x$-axis) can be thermally unbound.  In the presence of a
periodic potential, oppositely charged dislocations, with Burgers
vectors not satisfying the above condition (type II dislocations) are
bound by a potential which grows linearly with the separation and
therefore cannot thermally unbind. This discussion is consistent with
the mapping onto {\em scalar} Coulomb gas Hamiltonian, expected to
describe logarithmically bound type I dislocations, embodied in the 2d
anisotropic XY model Hamiltonian, Eq.\ref{H_LFS_effective}.

Away from the dislocation core, for a commensurate orientation defined
by the shortest direct lattice vector pointing parallel to the
troughs, ${\bf R}_{\vec n} = n_1 {\bf e}_1 + n_2 {\bf e}_2$, labelled
by direct lattice Miller indices $n_1$ and $n_2$ defined by
Eq.\ref{commensurateQ}-\ref{eq:n1n2}, the displacement vector ${\bf
u}$ for the active type I dislocation is given by
\begin{equation}
{\bf u}=\hatx{b_{\vec n}\over2\pi}\tan^{-1}\left({y B_{xx}^{1/2}\over x
B_{yx}^{1/2}}\right)\;,
\end{equation}
with 
\begin{equation}
b_{\vec n}=\mid {\bf R}_{\vec n} \mid = a\sqrt{n_1^2+n_2^2+n_1 n_2}\;.
\label{b_n}
\end{equation}
Melting of the LFS via unbinding of these defects is identical to the
vortex unbinding transition of an anisotropic 2d XY model. A standard
calculation \cite{nelson-halperin:79} leads to the prediction for the
LFS melting temperature
\begin{equation}
k_ {\rm B} T_{\rm LFS-LSm}={b_{\vec n}^2\over 8\pi}\sqrt{B_{xx} B_{yx}}\;,
\label{T_LFS-LSm}
\end{equation}
and all other concomitant Kosterlitz-Thouless phenomenology.  This
implies an exponential growths of the translational correlation
length\cite{kosterlitz-thouless:73}
\begin{equation}
\xi_t\approx a e^{c/|T-T_{\rm LFS-LSm}|^{1/2}}\;,
\label{xi_t}
\end{equation}
with $c$ a nonuniversal parameter and a {\em universal} ratio of the
jump in the geometric mean of the shear and bulk moduli,
$B_{yx}(T_{\rm LFS-LSm}^-)$ and $B_{xx}(T_{\rm LFS-LSm}^-)$ to $T_{\rm
LFS-LSm}$.\cite{Nelson-Kosterlitz}

The resulting high temperature phase is the LSm\cite{commentLSm}, for
low colloidal densities (i.e., high commensurability ratio $p$), and a
modulated liquid for high densities ($p\leq1$, see below), for which
the smectic is indistinguishable from a liquid. Because of the
unusually strong growth of the translational correlation length
$\xi_t$, Eq.\ref{xi_t}, the phenomenology of the LSm-FSm transition
that we studied in Sec.\ref{sec:LSm-FSm-transition} will be modified
for $T\rightarrow T_{\rm LFS-LSm}^+$ by a long crossover from the
crystal to smectic (or liquid) elasticity.

It is important to note the distinction between this anisotropic 2d XY
melting of a LFS into a LSm and an analogous type I melting mechanism
of Ostlund and Halperin for melting of uniaxially anisotropic, but
substrate-free 2d solids.\cite{ostlund-halperin:81} In the later case,
thermal fluctuations destabilize the resulting 2d smectic by further
unbinding type II dislocations, asymptotically converting it into a
liquid.  Here, because of the pinning potential, type II dislocations
(e.g.  $\pm{\bf b}_{2,3}$ for $\vec{p}=(p,0)$) remain bound by a
linear potential.  The resulting LSm phase is therefore distinct from
the (orientationally ordered) modulated liquid (in which type II
dislocations are also unbound), separated from it by a
thermodynamically sharp phase transition.

\subsubsection{Floating Solid to Floating Smectic Transition}
\label{sec:FS-FSm-transition}

A floating solid can melt {\em continuously} via unbinding of the type
I dislocations. However, in contrast to the similar melting of a
locked floating solid, here the dislocation unbinding in the
displacement $u_x$ proceeds in the presence of another spectator
massless phonon mode $u_y$, which is coupled to it. Consequently, as
we will show below, this transition is a {\em nontrivial} extension of
the Kosterlitz-Thouless theory and to our knowledge is, heretofore
unexplored.  Once these type I dislocations unbind the most likely
resulting phase is the floating smectic.\cite{commentLSm}

The phenomenology of the FS-FSm melting transition can be most easily
analyzed by the following steps. We (i) introduce dislocation degrees
of freedom into the elastic Hamiltonian $\Hel$,
Eq.\ref{H_el_anisotropic}, (ii) perform a duality transformation to
convert the resulting Coulomb gas Hamiltonian into a modified
Sine-Gordon model, and (iii) compute the dislocation unbinding
temperature by analyzing the resulting dual model.

To execute these standard steps, it is convenient to first perform the
following rescalings of spatial coordinates:
\begin{mathletters}
\begin{eqnarray}
x &\rightarrow&  x (B_{xx}/B_{yx})^{1/4}\;,\\
y &\rightarrow&  y (B_{yx}/B_{xx})^{1/4}\;,
\end{eqnarray}
\label{eq:rescaling}
\end{mathletters}
which leads to the Hamiltonian
\begin{eqnarray}
\HFS&=&\int d^2r
\frac12
\Bigl\{
K_x({\bbox\nabla} u_x)^2+
c_x(\partial_x u_y)^2+c_y(\partial_y u_y)^2\nonumber\\
&+&2\lambda_{xy}(\partial_x u_x)(\partial_y u_y)
+2(\mu-\gamma)(\partial_x u_y)(\partial_y u_x)
\Bigr\}
\label{H_FSresc}
\end{eqnarray}
where we dropped the prime on the rescaled coordinates and defined
elastic constants
\begin{mathletters}
\begin{eqnarray}
K_x&\equiv&\sqrt{B_{xx} B_{yx}}\;,\\
c_x&\equiv& K_x K_{xy}/B_{xx}\;,\\
c_y&\equiv& K_x K_{yy}/B_{yx}\;.
\end{eqnarray}
\end{mathletters}
Because in the presence of dislocations the displacement field $u_x$
is a multivalued function, it is essential to distinguish the last two
terms in Eq.\ref{H_FSresc}. In contrast to conventional elastic
theory, where dislocations are bound and $u_x$ is a well-defined
function, here these terms {\em cannot} be transformed into each other
by an integration by parts. Keeping track of this distinction ensures
the proper form for the elastic constants of the resulting smectic
phase.

In this new rescaled coordinate system, a type I dislocation located
at the origin, with a Burgers vector ${\bf b}=b_{\vec n}\hatx$, can
be represented by a displacement field
\begin{equation}
{\bf u}_s=\hatx{b_{\vec n}\over2\pi}\tan^{-1}\left({y\over x}\right)\;,
\label{u_s}
\end{equation}
However, in contrast to the analysis of the melting of the LS phase
above, in the presence of a finite $\partial_y u_y$ deformation, the
form of type I dislocation given in Eq.\ref{u_s} does not correspond
to a relaxed $u_x$ displacement which minimizes the energy.
Consequently, we expect (see Eq.\ref{H_FS_d}) a bilinear coupling
between the dislocation density and the $u_y$ distortion.  For a
finite density of dislocations, we define a singular strain ${\bf
  v}_s\equiv{\bbox\nabla}u_x^s$ due to a dislocation density $b({\bf
  r})$, with the standard relation
\begin{eqnarray}
{\bbox\nabla}\times{\bf v}_s&=&\hat{\bf e}_z b({\bf r})\;,\\
&=&\hat{\bf e}_z\sum_{{\bf r}_i} b_{\vec n} n_{{\bf r}_i}
\delta^{(2)}({\bf r}-{\bf r}_i)\;,\\
&\equiv&\hat{\bf e}_z b_{\vec n} n({\bf r})\;,
\end{eqnarray}
where the $\{n_{{\bf r}_i}\}$ are integer dislocation charges. A
general solution to the above equation is given (in Fourier space) by
\begin{equation}
{\bf v}_s({\bf q})={i{\bf q}\times\hat{\bf e}_z\over q^2}b({\bf
q})+i{\bf q}\chi({\bf q})\;,
\end{equation}
where $\chi({\bf q})$ is an arbitrary, single-valued function, which
for convenience and without loss of generality we can set to zero.
After expressing the gradient of the total displacement field ${\bf
  u}_t$ in terms of the dislocation part ${\bf v}_s$ and a single
valued phonon field $\bf u$
\begin{mathletters}
\begin{eqnarray}
{\bbox\nabla} u_x^t&=&{\bf v}_s+{\bbox\nabla} u_x\;,\label{grad_ut}\\
{\bbox\nabla} u_y^t&=&{\bbox\nabla} u_y\;,
\end{eqnarray}
\end{mathletters}
and inserting it into $\HFS$, we obtain a Hamiltonian that includes
both the elastic and dislocation degrees of freedom
\begin{eqnarray}
\HFSd&=&
\frac12 
\int \frac{d^2q}{(2\pi)^2}
\Biggl\{
{2b_{\vec n}\over q^2}
\biggl[ \lambda_{xy}q_y^2+(\gamma-\mu)q_x^2 \biggr]
n({\bf q})u_y(-{\bf q})\nonumber\\
&+&b_{\vec n}^2K_x{|n({\bf q})|^2\over q^2}
\Biggr\}
+ \HFS[{\bf u}]\;.
\label{H_FS_d}
\end{eqnarray}
After putting the system on the lattice, going to the grand canonical
ensemble for dislocations, and adding the dislocation core energy
$E_c$ to account for the energy coming from short length scales, (not
included in above analysis), the total partition function is given by
\begin{equation}
{\cal Z}=\int[d{\bf u}]
\sum_{\{n_r\}}e^{-\HFSd-\sum_r E_c n_r^2}\;,
\label{Zdislocations}
\end{equation}
In above, for convenience we chose to measure all the energies in
units of $\kT$.

To analyze the dislocation unbinding transition, it is convenient to
perform a duality transformation on the above Hamiltonian
$\HFSd$.\cite{jose-etal:77,Radzihovsky-Toner} To do this we introduce
an auxiliary Gaussian field $\phi$ to decouple the Coulomb interaction
between dislocations and use the Poisson summation formula to perform
the summation over the set of lattice integers $\{n_r\}$, obtaining
\begin{equation}
{\cal Z}=\int[d\phi][d{\bf u}]e^{-H_{d}}\;,
\end{equation}
where,
\begin{equation}
\Hd=\int d^2r
\Bigl\{ 
{K_x^{-1}\over2}|{\bbox\nabla}\phi|^2-V_V[b_{\vec n}(\phi+i\theta)]
\bigr\} 
+ \HFS[{\bf u}]\;.
\label{H_d}
\end{equation}
To obtain $\Hd$, Eq.\ref{H_d}, above, we defined a field $\theta({\bf
  r})$, whose Fourier transform is given by
\begin{equation}
\theta({\bf q})={1\over q^2}\left(\lambda_{xy}q_y^2+(\gamma-\mu)q_x^2\right) 
u_y({\bf q})\;,
\end{equation}
and used $V_V(\phi)$ to denote the well-known $2\pi$-periodic Villain
potential defined by
\begin{equation}
e^{-V_V(\phi)}=\sum_{n=-\infty}^\infty e^{-E_c n^2+i n\phi}\;.
\end{equation}
At low fugacity (large core energy), this potential reduces to a
cosine function, leading to
\begin{equation}
\Hd=\int d^2r
\Biggl\{ 
{K_x^{-1}\over2}|{\bbox\nabla}\phi|^2-g\cos{[b_{\vec n}(\phi+i\theta)]}
\Biggr\}
+ \HFS[{\bf u}]\;,
\label{Hd}
\end{equation}
with $g\equiv 2e^{-E_c}$.

Now the dislocation unbinding transition in the original model of the
floating solid is determined by the vanishing of the RG eigenvalue of
$g(\ell)$ cosine nonlinearity in this dual model, defined by
\begin{equation}
g(\ell)=g e^{(2-\eta_g/2)\ell}\;,
\end{equation}
where $\eta_g$ is determined by
\begin{equation}
\eta_g \ell=b_{\vec n}^2\left[\langle\phi^2\rangle_>-
\langle\theta^2\rangle_>\right]\;,
\end{equation}
with the right hand side easily computed from the quadratic part of
the dual Hamiltonian $\Hd$, Eq.\ref{H_d}. Specifically,
\begin{eqnarray}
\langle\phi^2\rangle_>&=&K_x\int_> {d^2q\over(2\pi)^2}{1\over q^2}\;,\\ 
&=&{\sqrt{B_{xx} B_{yx}}\over2\pi}\ell\;,
\end{eqnarray}
and
\end{multicols}
\widetext
\begin{mathletters}
\begin{eqnarray}
\langle\theta^2\rangle_>&=&\int_> {d^2q\over(2\pi)^2}
{(\lambda_{xy} q_y^2+(\gamma-\mu)q_x^2)^2\over q^4}
\, \langle|u_y({\bf q})|^2\rangle
\;,\\ 
&=&\int_> {d^2q\over(2\pi)^2}
{\lambda_{xy}^2q_y^4+2\lambda_{xy}(\gamma-\mu)q_y^2 q_x^2
+(\gamma-\mu)^2 q_x^4\over q^4
\left[c_x q_x^2+c_y q_y^2-\delta^2 q_x^2 q_y^2/(K_x q^2)\right]}\;,\\
&=&{\lambda_{xy}^2 c_2+\lambda_{xy}(\gamma-\mu)c_3+(\gamma-\mu)^2 c_4
\over2\pi\sqrt{K_{yy}K_{xy}}}\ell\;,
\end{eqnarray}
\end{mathletters}
where,
\begin{mathletters}
\begin{eqnarray}
c_2&\equiv&\int_0^{2\pi} {d\theta\over2\pi}{a_y^2\sin^4 \theta\over
\left[a_x + (a_y-a_x)\sin^2 \theta\right]
\left[a_x + (a_y-a_x-a_{xy})\sin^2 \theta+a_{xy}\sin^4 \theta\right]}\;,\\
c_3&\equiv&\int_0^{2\pi} 
{d\theta\over2\pi}{2a_{xy}^2\cos^2 \theta \sin^2 \theta\over
\left[a_x + (a_y-a_x)\sin^2 \theta\right]
\left[a_x + (a_y-a_x-a_{xy})\sin^2 \theta+a_{xy}\sin^4 \theta\right]}\;,\\
c_4&\equiv&\int_0^{2\pi} 
{d\theta\over2\pi}{a_x^2\cos^4 \theta\over
\left[a_x + (a_y-a_x)\sin^2 \theta\right]
\left[a_x + (a_y-a_x-a_{xy})\sin^2 \theta+a_{xy}\sin^4 \theta\right]}\;,
\end{eqnarray}
\end{mathletters}

Upon combining these results, we find that a floating solid melts into
a floating smectic at
\begin{equation}
T_{\rm FS-FSm}={b_{\vec n}^2\over8\pi}\left(\sqrt{B_{xx} B_{yx}}-
{\lambda_{xy}^2 c_2+\lambda_{xy}(\gamma-\mu)c_3+
(\gamma-\mu)^2 c_4\over\sqrt{K_{yy}K_{xy}}}\right)\;,
\label{T_FS-FSm}
\end{equation}
\begin{multicols}{2}
\columnwidth3.4in
\narrowtext
\noindent
which reduces to the melting temperature $T_{\rm LFS-LSm}$,
Eq.\ref{T_LFS-LSm} of the LFS in the limit
$K_{xy},K_{yy}\rightarrow\infty$, in which the spectator phonon $u_y$
mode is frozen out. Not surprisingly, we find that the extra $u_y$
fluctuations of the FS always {\em suppress} the melting temperature
of the FS relative to that of the LFS, i.e., for all range of
parameters, $T_{\rm FS-FSm} < T_{\rm LFS-LSm}$.

We now demonstrate that once type I dislocations unbind, the resulting
Hamiltonian is that of a floating smectic, described by the
Hamiltonian $H_{FSm}$, given in Eq.\ref{H_Sm}. To see this return to
the Hamiltonian $\HFSd$, Eq.\ref{H_FS_d}, and note that once
dislocations unbind and therefore appear in large densities, the
discrete dislocation field $n_{\bf r}$ can, to a good approximation,
be treated as a continuous density $n({\bf r})$.  Within this
Debye-H{\"u}ckel approximation the dislocation degrees of freedom can
be easily integrated out of the partition function
Eq.\ref{Zdislocations} by replacing the summation over $n_{\bf r}$ in
$\cal Z$ by an integration. Simple Gaussian integrations over
dislocation density $n({\bf r})$ and the single valued field $u_x$
then lead, in the long wavelength limit to an effective floating
smectic Hamiltonian
\begin{equation}
\HSm={1\over2}\int d^2 r\Bigl\{ \kappa(\partial_x^2 u_y)^2
+B_{xy}(\partial_x u_y)^2+B_{yy}(\partial_y u_y)^2 \Bigr\}\;,
\label{H_Sm2}
\end{equation}
where we have restored the original scaling of the spatial
coordinates, Eqs.\ref{eq:rescaling} and derived the effective elastic
constants for the resulting FSm phase
\begin{mathletters}
\begin{eqnarray}
\kappa&=&{(\mu-\gamma)^2
\over(\mu+\gamma-\alpha)^2} \left({2E_c\over b_{\vec n}^2}\right)\;,\\
B_{xy}&=&{4\mu\gamma-\alpha^2\over\mu+\gamma-\alpha}\;,\\
B_{yy}&=&\lambda_{yy}-{\lambda_{xy}^2\over\lambda_{xx}}\;.
\end{eqnarray}
\label{smectic_elastic_constants}
\end{mathletters}
We note that $B_{xy}$ vanishes as $\gamma,\alpha\rightarrow0$, as it
must in this rotationally invariant limit, in which one must recover
the rotationally invariant 2d liquid crystal smectic
elasticity.\cite{deGennes}

Another equivalent but considerably more straightforward way to obtain
the smectic Hamiltonian is to note that in the presence of unbound
type I dislocations the ${\bbox\nabla} u_x^t$, Eq.\ref{grad_ut}
contains both the longitudinal and transverse components, and
therefore, despite of its appearance is no longer a conservative
vector constrained to be a gradient of a single-valued function. This
observation allows us to incorporate unbound type I dislocations into
the Hamiltonian $\Hel$, Eq.\ref{H_el_anisotropic}, by the replacement
\begin{equation}
{\bbox\nabla}u_x\rightarrow {\bf v}\;,
\end{equation}
with ${\bf v}$ an arbitrary 2d vector field. Under this substitution
$\Hel$, Eq.\ref{H_el_anisotropic} transforms into
\begin{eqnarray}
\HFSd&=&\int d^2r
\Bigl\{{\mu\over2}(\partial_x u_y+v_y)^2 + 
{\lambda_{xx}\over2}v_x^2+{\lambda_{yy}\over2}(\partial_y u_y)^2
\nonumber\\
&+&\lambda_{xy}v_x\partial_y u_y+{\alpha\over2}[(\partial_x u_y)^2-v_y^2]
+{\gamma\over2}(\partial_x u_y-v_y)^2\Bigr\}\;.\nonumber\\
\label{H_el_dislocations}
\end{eqnarray}
After performing a simple Gaussian integration over the two
independent components of $\bf v$, we immediately obtain a Hamiltonian
for the floating smectic, which in the long wavelength limit agrees in
form and with the expressions for the elastic constants $B_{xy}$ and
$B_{yy}$, obtained in Eqs.\ref{H_Sm2} and
\ref{smectic_elastic_constants}.

\section{Shape of the melting curve}
\label{sec:reentrance}

\subsection{Strong pinning limit and reentrant melting}

One of the most interesting observations in the colloidal experiments
by Wei \etal\cite{wei-bechinger-rudhardt-leiderer:98}, which in fact
stimulated our interest in this problem, is the light-induced {\em
  reentrant} melting. As we shall explicitly demonstrate, this melting
reentrance is a {\em generic} consequence of short-ranged screened
colloidal interactions and thermal fluctuations, and hence should be
prevalent in such 2d systems.

To demonstrate the reentrance as a function of laser intensity, we
study the shape of the melting curves for the LFS-ML, LFS-LSm and
FS-FSm transitions, which we generally denote by $\Tm(\UQ)$. The
common feature of these transitions is that they are all driven by the
unbinding of type I dislocations, with $\Tm(\UQ)$ (see
Eqs.\ref{T_LFS-LSm},\ref{T_FS-FSm}) at least in part determined by the
renormalized values of the bulk modulus $B_{xx}$ for compression along
the troughs and the corresponding shear modulus $B_{yx}$.  Our goal
then is to detemine how these moduli depend on the potential amplitude
$\UQ$.

We first note that these melting boundaries $\Tm(\UQ)$ are constrained by
their limiting values
\begin{mathletters}
\begin{eqnarray}
\Tm(0)&=&{b_{\vec n}^2\over 4\pi}{\mu(\mu+\lambda)\over2\mu+\lambda}\;,
\label{Tm_0}\\
\Tm(\infty)&=&{b_{\vec n}^2\over 8\pi}\sqrt{\mu(2\mu+\lambda)}\;,
\label{Tm_infty}
\end{eqnarray}
\label{T_m}
\end{mathletters}
where $\Tm(0)$ is the well-known result in the absence of an external
potential\cite{kosterlitz-thouless:73,nelson-halperin:79,Young:79}.
In the opposite limit of {\em infinite} potential strength,
$\Tm(\infty)$ is given by Eq.\ref{T_LFS-LSm}, with
$B_{xx}(\UQ\rightarrow\infty)\approx2\mu+\lambda$ and
$B_{yx}(\UQ\rightarrow\infty)\approx\mu$. These results follow from
comparing $\HLFS$, Eq.\ref{H_LFS}, with $H_0$, Eq.\ref{H_0}, after
freezing out the $u_y$ degree of freedom ($u_y=0$) in $H_0$, as is
appropriate in this $\UQ\rightarrow\infty$ limit. Although in general
there is no universal relation between $\Tm(0)$ and $\Tm(\infty)$, in
a dilute colloidal limit, relevant to the experiments of Wei
\etal\cite{wei-bechinger-rudhardt-leiderer:98}, the two Lam{\'e}
coefficients are equal, $\mu\approx\lambda$, and Eqs.\ref{T_m} reduce
to
\begin{mathletters}
\begin{eqnarray}
\Tmdil (0)&=&\mu{b_{\vec n}^2\over 6\pi}\;,\label{Tm_0_dilute1}\\
\Tmdil (\infty)&=&\sqrt{3}\ \mu{b_{\vec n}^2\over 8\pi}
\label{Tm_infty_dilute2}\\
&\approx& 1.3\  \Tmdil (0)\;.
\label{Tm_0_dilute2'}
\end{eqnarray}
\end{mathletters}

One might have thought that the melting temperature simply increases
monotonically with $\UQ$ from $\Tm (0)$ to $\Tm (\infty)$.  However,
as we will now show explicitly, the $u_y$-mode thermal fluctuations,
enhanced as the periodic potential is {\em lowered} from infinity,
generically {\em increase} the melting temperature for $\kappa a>>1$.
Consequently, the melting curve, $\Tm(\UQ)$, must have a maximum in
this limit, implying reentrant melting for a band of temperatures as a
function of the potential amplitude.

The origin of the reentrance effect can be understood on a heursitic
level as follows. Clearly, at {\em small} $\UQ$, we expect that the
increase in the strength of the periodic potential suppresses thermal
fluctuations in $u_y$, thereby lowering the entropy of the liquid (or
the smectic) state, and therefore making freezing into a lattice
free-energetically less costly.  This naturally leads to an {\em
  increase} of $\Tm(\UQ)$ with $\UQ$ at low laser intensities.
However, for potential strengths $\UQ \gg k_B T$, this entropic
contribution to the free energy becomes unimportant. In this large
$\UQ$ limit, the behavior of $\Tm(\UQ)$ is dominated by a different
mechanism having to do with the reduction of the elastic constants
with increasing $\UQ$ and decreasing temperature. To see this, note
that the effective shear modulus $B_{yx}(\UQ)$ which enters $\Tm(\UQ)$
(see Eqs.\ref{T_LFS-LSm} and \ref{T_FS-FSm}), is determined by the
screened Coulomb interaction, $V(r) = V_0\exp(-\kappa r) / r$ ,
between colloidal particles in neighboring troughs. In order to find
an effective shear modulus for the $u_x$-modes, one needs to integrate
out the massive modes corresponding to displacements perpendicular to
the troughs of the laser potential.  This will be the route taken
further below. Heuristically, one should get roughly the same result
by assuming that the dominant effect comes from the shear modulus
$B_{xy}$ and simply averaging the potential over the massive $u_y$
degrees of freedom, which yields
\begin{eqnarray}
  B_{yx}(\UQ) \sim 
  \langle e^{-\kappa \mid {\bf r}_{n+1}-{\bf r}_{n} \mid} 
  \rangle_{u_y} \; ,
\end{eqnarray}
where ${\bf r}_n$ and ${\bf r}_{n+1}$ are positions of nearest
neighbor colloidal particles belonging to the $n$-th and $n+1$-st
Bragg planes, running parallel to the laser potential troughs. This
gives to lowest harmonic order in the fluctuations $u_y$,
\begin{eqnarray}
B_{yx}(\UQ) 
&\sim& \langle e^{- \kappa a - \kappa [u_y(n+1)-u_y(n)]} \rangle
\nonumber \\ 
&\sim&  e^{-\kappa a} e^{\kappa^2\langle u_y^2 \rangle}
\nonumber \\
&\approx& B_{yx}(\infty)e^{\text{c}\, k_B T/\UQ} \; .
\label{B_xy_UQ}
\end{eqnarray}
with $c$ a dimensionless number of order $1$. Such a thermal {\em
enhancement} of the effective shear modulus $B_{yx}(\UQ)$, which
decreases as thermal fluctuations in $u_y$ are suppressed by
increasing $\UQ$, is easy to understand: Even though, in the presence
of $u_y$ fluctuations colloidal particles in neighboring troughs spend
as much time closer together as further apart, because of the concave
form of the interaction potential the enhancement of the effective
shear modulus is larger from particles being closer together than the
corresponding suppression when they are further apart.

The above simple physical argument for reentrance is supported by
detailed microscopic lattice calculations, in which we compute {\em
  both} the effective shear $B_{yx}(\UQ)$ and bulk $B_{xx}(\UQ)$
moduli.  To do this we start with a microscopic model with a screened
repulsive Coulomb interaction $V(r) = V_0 \exp (- \kappa r) / r$,
where the screening length $\kappa^{-1}$ is typically much smaller
than $a$ and $V_0$ depends on the dielectric constant, $\kappa$ and
the sphere radius~\cite{wei-bechinger-rudhardt-leiderer:98}.  Upon
integrating out the $u_y$-modes using the screened Coulomb potential
to leading order in $\kT / \UQ$ the calculation in
Appendix\ref{effective_elastic_constants_appendix} gives (for
orientation $\vec p=(1,0)$),
\end{multicols}
\widetext
\begin{eqnarray}
  B_{yx}(\UQ) &\approx& B_{yx}(\infty) 
  \left\{ 1 + \frac{9 (\kappa a)^2}{64 \pi^2} 
              \, \left( 1+ \frac{17}{3 \kappa a} \right) 
              \, \frac{k_B T}{p^2 \UQ}
  \right\} \, ,\\
  B_{xx}(\UQ) &\approx& B_{xx}(\infty) 
   \left\{ 1 + \frac{(\kappa a)^2}{64 \pi^2} 
              \left( 1-8v-\frac{23+104v}{3\kappa a}  
              \right) \frac{k_B T}{p^2 \UQ}
  \right\} \, .
\end{eqnarray}
where $v=V_0 e^{-\kappa a} / k_B T$, $B_{yx}(\infty) = \frac38 v k_BT
\kappa^2$ and $B_{xx} (\infty) =3 \mu$. Lowering the potential
strength $\UQ$ always increases the shear modulus, whereas the
behavior of the compressional modulus depends on the magnitude of $v$
and $\kappa a$. When combined with Eq.~(\ref{T_LFS-LSm}), these
expressions imply that the melting temperature increases with
decreasing $\UQ$ for $\kappa a \gtrsim 5.6$ (in
Ref.~\cite{wei-bechinger-rudhardt-leiderer:98} $\kappa a \approx 10$),
\begin{eqnarray}
  T_{\rm LFS-LSm}(\UQ) = T_{\rm LFS-LSm}^\infty 
                 \left\{ 
                        1 + \frac{5[(\kappa a)^2\!-\!31]}{64 \pi^2}  
                           \left(1\!+\! \frac{13}{3 \kappa a} \right) 
                          \, \frac{\kT_{\rm LFS-LSm}^\infty}{p^2 \UQ} 
                 \right\} \, ,
\label{T_reentrance}
\end{eqnarray}
\begin{multicols}{2}
\columnwidth3.4in
\narrowtext
\noindent
thus implying reentrant melting for a band of temperatures as a
function of potential strength observed in experiments and illustrated
in Figs.\ref{fig:phase_diagram_p=1},\ref{fig:phase_diagram_1_p_pc} and
\ref{fig:phase_diagram_pc_p}.  Clearly given the dependence of the
$T_{\rm FS-FSm}$ on the elastic moduli, Eq.\ref{T_FS-FSm}, we expect
FS-FSm transition to display reentrance, although quantitative
predictions of the size of the reentrant band are much more difficult.

In obtaining Eq.\ref{T_reentrance} we have clearly ignored additional
renormalization of the effective elastic constants by phonon
nonlinearities and by bound dislocation pairs, which need to be taken
into account for more precise estimate of the phase boundary.  Based
on general structure of Kosterlitz-Thouless-like RG flows, the latter
renormalizations generically reduce the elastic moduli and therefore
drive the melting temperature down. Since $u_y$ modes fluctuations and
therefore the renormalizations that they induce are suppressed by the
increasing periodic potential, we expect that $\Tm(\UQ)$ experiences
larger reduction at small $\UQ$ than at large $\UQ$. The known values
for the the potential-free 2d melting and the 2d XY model downward
renormalization constrain the extreme $\UQ=0$ and
$\UQ\rightarrow\infty$ ends of the melting curve. Furthermore, since
thermal downward renormalization of elastic constants is obviously
enhanced with increasing temperature, we expect the suppression of the
melting temperature due to these effects to be most pronounced near
the maximum in $\Tm(\UQ)$.  Clearly, such a $\UQ$-dependent downward
renormalization of the elastic constants will generically tend to
reduce the range of temperatures over which there is laser-induced
reentrant melting.  However, these effects are small\cite{fisher:82}
and we therefore expect reentrant melting to persist even in their presence.

\subsection{Weak pinning: universal shape of the melting curve at 
  small potential strength}

In addition to a maximum displayed by the melting curve as a function
of laser intensity, we also find that the shape of the melting
temperature is universal in the limit of a vanishing periodic
potential strength $\UQ$. This can be seen most easily from the RG
scaling theory applied to the potential-free critical point.  More
specifically, consider the behavior of the translational correlation
length $\xi(t,\UQ)$ above the melting transition as a function of
$\UQ$ and the reduced temperature $t\equiv
[T-T_m(\UQ=0)]/T_m(\UQ=0)$. The power of the renormalization group
transformation is that it allows us to relate a difficult calculation
very close to the transition, where fluctuations are large and
perturbation theory is divergent, to a calculation outside of the
critical region, where perturbation theory is convergent. Applying
this idea to the computation of $\xi(t,\UQ)$ we find
\begin{mathletters}
\begin{eqnarray}
\xi(t,\UQ)&=& b_*\xi(t(b_*),\UQ b_*^{\lambda_K})\;,
\label{xi_match_a}\\
&=& e^{c/t^{\overline\nu}} \xi(1,\UQ
e^{c\lambda_K/t^{\overline\nu}})\;,
\label{xi_match_b}
\end{eqnarray}
\end{mathletters}
where we have chosen the RG rescaling parameter $b_*$ such that the
rescaled reduced temperature $t(b_*)$, given by the RG flow equations
of Halperin and Nelson\cite{nelson-halperin:79} is of order unity
\begin{equation}
t(b_*)=1\;.
\end{equation}
$\lambda_K=2-\eta_K/2$ is the renormalization group eigenvalue of the
1d periodic potential $U_K$ At the primary potential-free fixed point
with $\UQ=0$, we recover the well-known\cite{nelson-halperin:79}
exponential growth of the correlation length $\xi(t,\UQ=0)$ with the
exponent ${\overline\nu}$ given by
\begin{equation}
{\overline\nu}\approx0.36963\;,
\end{equation}
where an overbar denotes critical exponents at this fixed point.

The primary critical behavior is unstable for arbitrarily small $\UQ$.
Hence sufficiently close to the melting temperature $\Tm(0)$, the
periodic potential always becomes important. This is the case even for
the melting of the FS, where it leads to a marginal crossover from a
fixed line of isotropic rotationally invariant elasticity to the fixed
line characterizing elasticity given by Eq.\ref{H_el_anisotropic},
where the rotational symmetry is {\em explicitly} broken by $U_K$.
There, despite the fact that the periodic potential is irrelevant for
the translational order parameter, it is always important for the
orientational degrees of freedom, since (see
Sec.\ref{sec:rotational-symmetry}) it explicitly breaks orientational
symmetry.

In {\em locked} phases, it is clear from Eq.\ref{xi_match_b}, that for
a given small $\UQ$, the effects of this weak periodic potential will
be felt at $\Tm(\UQ)>\Tm(0)$, such that the $\UQ$-dependent argument
on the right hand side of Eq.\ref{xi_match_b} is large, i.e.  grows
beyond order $T_m$,
\begin{equation}
\UQ\approx \kT_{\rm m} e^{-c\lambda_K/t^{\overline\nu}_m}\;.
\end{equation}
This then predicts a {\em universal cusp} for the melting curve
$\Tm(\UQ)$ in the limit $\UQ\rightarrow0$ in any phase in which
$\lambda_K > 0$, i.e., the periodic potential is relevant and the
phase is locked and $\Tm(\UQ)$ given by
\begin{equation}
\Tm(\UQ)\sim \Tm(0)\Big[1
+\lambda_K[\ln (k_B \Tm / \UQ)]^{-1/{\overline\nu}}\Big]\;,
\end{equation}
as depicted in
Figs.\ref{fig:phase_diagram_p=1},\ref{fig:phase_diagram_1_p_pc} and
\ref{fig:phase_diagram_pc_p}. For {\em floating} phases, such as the
FS and FSm, where the periodic potential is irrelevant (in the RG
sense), we expect the convergent perturbation theory in $\UQ$ to lead
to a melting temperature $T_m(U_K)$ that instead grows {\em linearly}
with $U_K$.

\section{Response of the translation and hexatic 
  order parameter to an external potential} 
\label{sec:response}

In this section we use a renormalization group scaling analysis to
determine the response of the translational order parameter $M_{\bf K}
= \langle \rho_{\bf K} \rangle$ and the bond orientational order parameter
$\psi_6 = \langle e^{6 i \theta ({\bf r})}\rangle$ to the amplitude
$\UQ$ of the external laser potential. In the absence of an external
potential, $\UQ = 0$, there are only algebraic peaks in the static
structure function of the crystalline phase and the translational
order parameter $M_{\bf K} \equiv \langle \rho_{\bf K} \rangle$
vanishes like
\begin{eqnarray}
  M_{\bf K} \sim 
  L^{- {\overline\eta}_{\bf K}/2} \rightarrow 0
\end{eqnarray}
as the system size $L \rightarrow \infty$, where
\begin{eqnarray}
{\overline \eta}_{{\bf K}}= \frac{k_B T}{4
  \pi} \frac{3 \mu + \lambda}{\mu (2 \mu + \lambda)} {\bf K}^2
\end{eqnarray}
is the critical exponent of the potential-free
case~\cite{nelson-halperin:79}. For small values of the external
potential $\UQ$ we can use standard crossover scaling analysis to
determine how the translational order parameter depends on the
amplitude of the laser potential. We start from the scaling behavior
of the free energy density under a renormalization group
transformation
\begin{eqnarray}
  f(\UQ,T) = e^{-2 \ell} f (e^{\lambda_{\bf K} \ell} \UQ, T(\ell) )\; ,
\end{eqnarray}
where $\lambda_{\bf K}$ is the renormalization group eigenvalue for
the periodic potential, and $T(\ell)$ is the renormalized temperature
which characterizes the crystalline phase. Since in the free energy
density the laser potential $\UQ$ couples linearly to $\rho_{\bf G}$
we have
\begin{eqnarray}
 M_{\bf K} &=& - \frac{\partial}{\partial \UQ} f(\UQ,T) \; ,
\end{eqnarray}
and dimensional analysis tells us that the exponent of the correlation
function $\langle \rho_{\bf K} ({\bf r}) \rho_{\bf K}^\star ({\bf 0})
\rangle \sim r^{- {\overline \eta}_{\bf K}}$ is related to
$\lambda_{\bf K}$ by
\begin{eqnarray}
\lambda_{\bf K} = 2 - \frac12 {\overline \eta}_{\bf K} \; ,
\end{eqnarray}
a result consistent with standard perturbative calculation of
$\lambda_{\bf K}$. Hence we get the following scaling relation for the
translational order parameter
\begin{eqnarray}
  M_{\Qlaser} (\UQ,T) =
  e^{- \frac12 {\overline \eta}_{\Qlaser} \ell} 
  M_{\Qlaser}  (e^{(2 - {\overline \eta}_{\bf K}/2)\ell} \UQ, T(\ell) ) \; ,
\end{eqnarray}
where we expect $T(\ell)$ to approach a finite value as
$\ell\rightarrow\infty$.  Upon choosing $\ell=\ell_*$ such that $e^{(2
  - {\overline \eta}_{\bf K}/2)\ell_*} \UQ = \mu a^2$, i.e., is
comparable to the elastic energy for deformation at the lattice
cutoff $a$, we obtain
\begin{eqnarray}
  M_{\Qlaser} (\UQ,T) \sim \mid \UQ \mid^{{\overline \eta}_{\Qlaser} / (4
  - {\overline \eta}_{\Qlaser} )}\;.
\end{eqnarray}
For ${\overline \eta}_{{\bf K}}>2$, $M_{{\bf K}}$ vanishes linearly
with $\UQ$, with a singular correction.  In contrast, $M_{{\bf K}}$
should always vanish {\em linearly} with $\UQ$ in the liquid and
hexatic\cite{murray-springer-wenk:90,zahn-etal:99} phases of the
unperturbed colloid.

The laser potential will also induce long-range bond orientational
order in $\psi_6 = \langle e^{6i\theta({\bf r})}
\rangle$~\cite{balents-nelson:94}. Along similar lines as above, one
can show that the bond order parameter $\psi_6$ vanishes linearly with
$\UQ$ in the liquid, vanishes like a power of $\UQ$ in the hexatic
phase
\begin{eqnarray}
    \psi_6 \sim 
    \mid\UQ \mid^{6{\overline \eta}_{6}/(4-{\overline \eta}_{6})} \;,
\end{eqnarray}
where ${\overline \eta}_6$ is the exponent describing the algebraic
decay of bond order, and approaches a nonzero constant as $\UQ
\rightarrow 0$ in the solid phase.\cite{delta_psi6comment}

\section{Discussion and experimental implications}
\label{sec:discussion}

\subsection{Melting temperatures and critical commensurability 
  ratios in the dilute limit}

\begin{figure}[bht]
  \narrowtext \centerline{ \epsfxsize=0.9\columnwidth
  \epsfbox{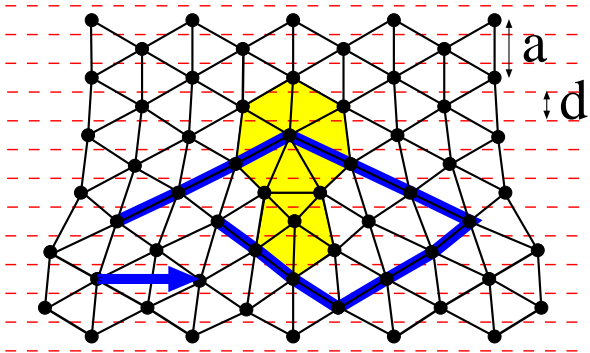}} \vspace{0.5truecm}
\caption{Triangular lattice with lattice constant $a$ subject to a
  periodic potential (maxima indicated by dashed lines) for a
  dual-primary orientation with $p d=a^\prime$, where $a^\prime = a/2$
  is the Bragg plane spacing and the commensurability ratio is $p =
  1$.  Also shown is the low energy dislocation with Burgers vector
  ${\bf b}$ parallel to the corrugation of the potential.}
\label{fig:dislocation_B}
\end{figure}

\begin{figure}[bht]
  \centerline{ \epsfxsize=0.7\columnwidth
  \epsfbox{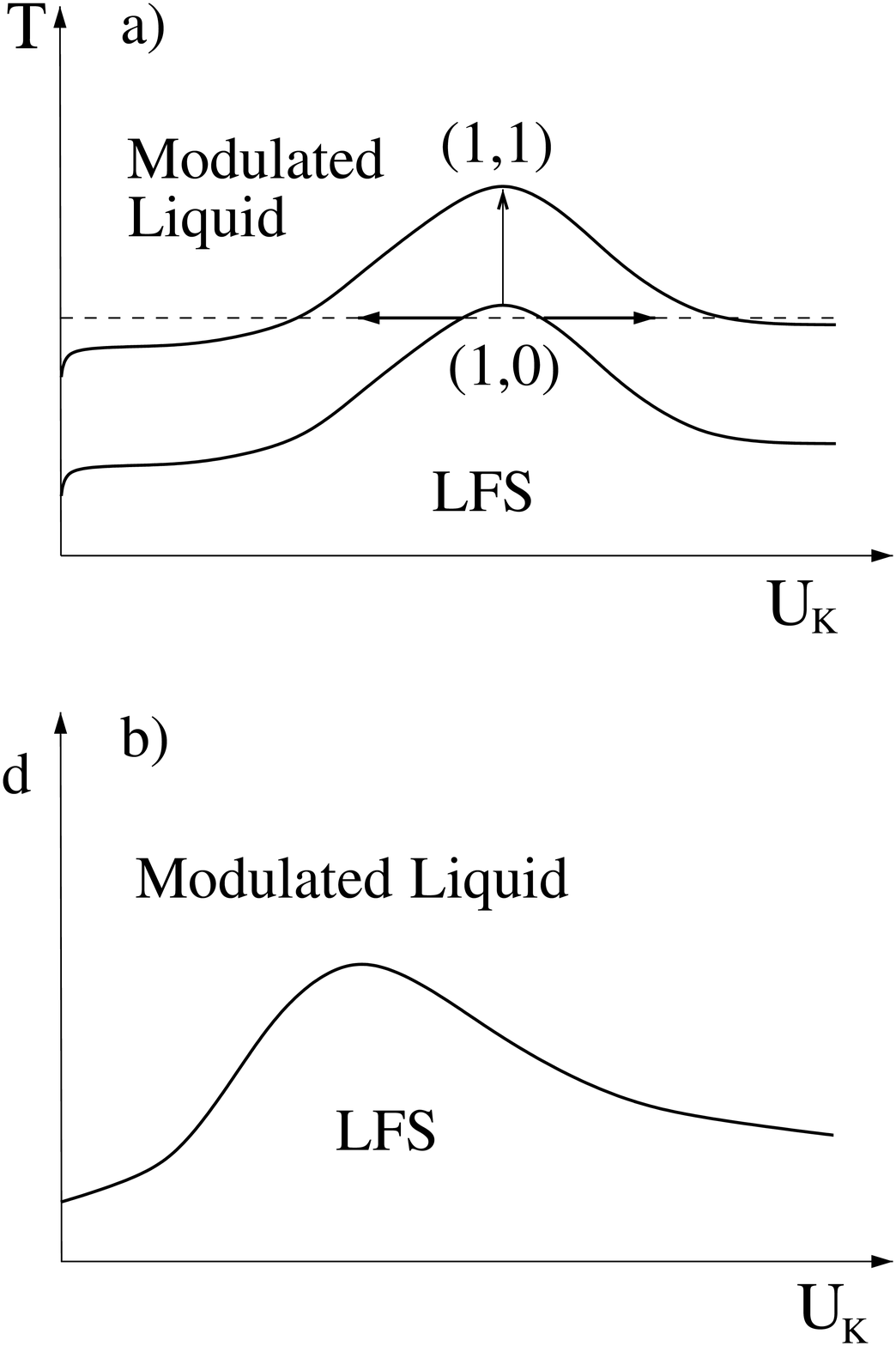}} \vspace{0.5truecm}
\caption{a) Schematic $p=1$ phase diagram as a function of potential
  strength $\UQ$ and relative orientation between the laser potential
  and the 2d solid; a change in orientation from $(1,0)$ to $(1,1)$ is
  generated by keeping the particle density (and hence the mean
  particle spacing $a$) fixed and varying the distance $d$ between the
  minima of the external potential. b) Schematic $p=1$ phase diagram
  for lattice spacing $d$ versus potential strength $\UQ$ at fixed
  temperature $T$ and fixed colloidal density; incommensurability
  effects are disregarded.}
\label{fig:sketch_melting}
\end{figure}

One of the interesting predictions of our work is that the LFS-ML,
LFS-LSm and FS-FSm transition are all mediated by the unbinding of
type-I dislocations with Burgers vectors parallel to the troughs of
the external potential, ${\bf b} = b_{\vec n} \hatx$. Consequently,
depending on the choice of relative orientation, the periodic
potential can be used to supress the unbinding of a set of
dislocations that would otherwise unbind in a ``substrate''-free
experiments.  For example, in the dual-primary orientation shown in
Fig.\ref{fig:dislocation_B}, all {\em six} fundamental Burgers
vectors are confined by a linear potential and therefore cannot unbind
entropically. It is therefore the unbinding of non-fundamental
dislocations with Burgers vector of charge $\sqrt{3} a$, illustrated
in Fig.\ref{fig:dislocation_B} that will control the melting
transition.
In general, the magnitude $b_{\vec n}^2 = a^2 (n_1^2 + n_2^2 + n_1
n_2)$ of the lowest energy Burgers vector and hence the melting
temperature $\Tm \propto b_{\vec n}^2$ depends strongly on the
relative orientation between the 2d solid and the laser potential,
e.g., for $n_2=1$ and $n_1 = 0, \, 1, \, 2, \, 3$ one finds $b_{\vec
  n}^2 / a^2 = 1, \, 3, \, 7, \, 13$.  In particular, if one keeps the
mean particle spacing $a$ (i.e., the density) and the potential
strength fixed and reduces the spacing $d$ between the laser troughs
(by e.g., varying the angle between the two interfering laser beams)
$d \rightarrow d/\sqrt{3}$ such that one goes from preferred lattice
orientation A (${\vec n}_A = (1,0)$) to preferred lattice orientation
B (${\vec n}_B = (1,1)$) the melting temperature should increase by a
factor of $3$ (see vertical arrow in Fig.~\ref{fig:sketch_melting}a).
This appears to be consistent with preliminary data of Bechinger
\etal\cite{bechinger-leiderer:unpubl}.  They find that for exactly
such a change in trough spacing the onset of light induced freezing at
{\em fixed temperature} is shifted to smaller laser intensities also
by roughly a factor of $3$ (see horizontal arrow in
Fig.~\ref{fig:sketch_melting}a).

More detailed experimental studies of $\Tm(\UQ)$ for various
commensurate orientations and trough spacings would clearly be
desirable in order to systemmatically test our predictions for the
orientation dependence of the melting transition temperature for the
LIF.  In performing such studies one must keep in mind considerable
irreversibility effects that are expected to plague
``zero-laser-field'' cooled experiments. In order to avoid dealing
with long equilibration times, one would need to warm up into the
liquid state, change the laser potential period $d$ and only then
``field-cool'' back into the solid.

Since trough spacing $d$ (controlled by the angle between the
interfering laser beams) and laser intensity appear to be convenient
experimentally tunable parameters, it is valuable to derive the shape
of the melting curve in the $d$-$\UQ$ plane (for a given temperature
and a fixed density of colloidal particles). However, since an
arbitrary value of $d$ will in general not be commensurate with the
spacing between a particular fixed set of Bragg planes, a detailed
study of incommensurate potentials would need to be done in order to
fully understand the behavior as a function of trough spacing $d$.  We
hope to discuss some of the ensuing physics in a forthcoming
publication~\cite{fnr_future}. But, for the following we would like to
restrict ourselves to values of $d$ which are commensurate. Hence,
strictly speaking, our results will be not be valid for a continuous
set of layer spacings but only for a discrete commensurate subset of
values. With this precaution in mind we expect the melting curve (for
a given temperature and particle density) in the $d$-$\UQ$ plane to
have the shape illustrated in Fig.~\ref{fig:sketch_melting}b.  We note
that in the LIF regime the critical potential strength for melting
decreases with decreasing distance between the laser fringes, whereas
in the LIM regime the critical potential strength increases as the
interference fringes become narrower.

Let us now specialize to the dilute limit, $\kappa a \gg 1$ , relevant
to the experiments of Wei
\etal\cite{wei-bechinger-rudhardt-leiderer:98}. Then the two Lam\'e
coefficients (characterizing the continuum elastic theory of the
hexagonal crystal in the absence of a laser potential) become equal,
$\mu \approx \lambda$, and the melting temperature for the LFS reduces
to
\begin{eqnarray}
  \Tm^0 &=& \mu \, \frac{b_{\vec n}^2}{6 \pi} \;, 
\label{eq:TmS_dilute_zero} \\
  \Tm^\infty &=& \sqrt{3} \, \mu \, \frac{b_{\vec n}^2}{8 \pi} 
  \approx 1.3 \Tm^0 \;,
\label{eq:TmS_dilute_inf}
\end{eqnarray}
in the limit of zero and infinite potential, respectively.  For small
values of the commensurability ratio, $p<p_c$, the LFS melts into a
modulated liquid or a locked smectic. If $p>p_c$, a floating solid
with two soft phonon modes can intervene between the LFS and a
modulated liquid or floating smectic phase. As discussed in
Sec.\ref{sec:roughening}, the transition from the LFS into the
intermediate FS phase is in a roughening universality class where the
laser potential becomes irrelevant. In the dilute limit (and
neglecting effects of the periodic potential on the elastic
coefficients) the corresponding critical temperature is aproximately
\begin{eqnarray}
  \TpS^{\rm dil} = 
  \frac{3}{\pi} \, \mu d^2 \, ,
\label{eq:TpS_dilute}
\end{eqnarray}
where $d=a'/p$ with $a' = \sqrt{3} a / (2 \sqrt{n_1^2+n_2^2+n_1n_2})$
the distance between the Bragg planes parallel to the troughs of the
laser potential. Upon combining Eq.~\ref{eq:TpS_dilute} with
Eq.~\ref{eq:TmS_dilute_zero}, the critical commensurability ratio reads
\begin{eqnarray}
  p_c^{\rm dil} = 3 \, \sqrt{\frac{3}{2}} \, \frac{1}{n_1^2+n_2^2+n_1n_2} \, . 
\end{eqnarray}
Note that only for the primary ($p_c^{\rm dil} = 3 \sqrt{3/2} \approx 3.7$) and
for the dual-primary orientation ($p_c^{\rm dil} = \sqrt{3/2} \approx
1.2$) is this critical value larger than $1$. For any
other orientation $p_c$ is less than $1$ and hence we expect that
there will always be an intervening floating solid phase. A
configuration with ${\vec n} = (2,1)$ (see also
Fig.~\ref{fig:lattice_pinned2_-1}) and hence $p_c = (3/7) \sqrt{3/2}
\approx 0.5$ is likely to be within the range of parameters accessible
to experiments with colloidal particles.

For $p>p_c$ there is a roughening transition from a locked floating
solid into a uniaxially anisotropic floating solid described by
$\Hel$, Eq.\ref{H_el_anisotropic}, which subsequently melts (by
unbinding of type I dislocations) into either a modulated liquid or a
floating smectic.  Since the melting and the roughening transition for
a locked smectic phase are respectively given by $\TmSm =
\frac{1}{8\pi} B a^2$ and $\TpSm = \frac{2}{\pi} B d^2$, where $B =
\sqrt{B_{xy} B_{yy}}$ and $a$ is the smectic layer spacing, there
exists a {\em universal} commensurability ratio $p_c^\prime = 4$
\cite{jose-etal:77} above which a floating smectic phase intervenes
between a locked smectic or a floating solid and the modulated liquid.
This universal value $p_c' = 4$ should be contrasted with the
nonuniversal critical commensurability ratio $p_c$ for the existence
of the floating solid phase, which depends on the relative magnitude
of the elastic constants and strongly on the relative orientation
between the colloidal lattice and the 1d periodic potential. Current
experiments find it difficult to access large commensurability ratios
$p$.  We hope that our theoretical results will inspire
experimentalists to overcome present obstacles and map out the rich
phase diagram shown in Fig.\ref{fig:phase_diagram_pc_p}.

\subsection{Phase diagrams as a function of the Debye screening length}

Recent Monte-Carlo simulation studies of melting in the presence of a
1d periodic external potential have explored the phase diagram in the
parameter space of $\UQ/\kT$ and $\kappa a$ with particle density and
temperature fixed.\cite{chakrabarti-etal:95,das-sood-krishnamurthy:99}
Although one might question whether such simulations are in
equilibrium with respect to dislocation climb (or even glide), it is
important to tabulate the predictions of our defect-mediated melting
theory in this parameter space in order to be able to compare with the
results of these simulations. In addition, it also seems to be more
feasable experimentally to map out the phase diagram as a function of
potential strength and particle density.

\begin{figure}[bht]
  \vspace{0.5cm}
  \centerline{ 
    \psfrag{ko}{$\kappa_{\text m}^0 a$}
    \psfrag{kdiff}{$(\kappa_{\text m}^\infty- \kappa_{\text m}^0) a$} 
    \epsfxsize=0.9\columnwidth \epsfbox{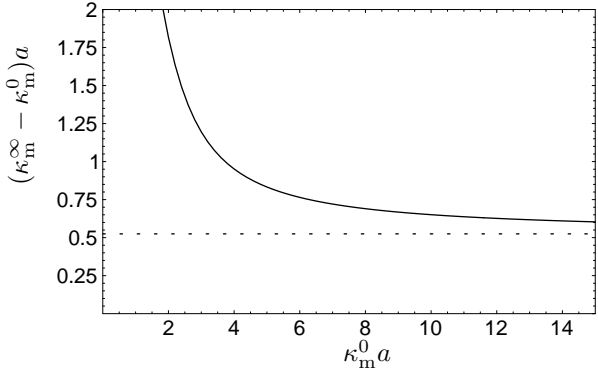}}
  \vspace{1.5truecm}
\caption{Difference between $\kappa_{\text m} a$ at infinite and zero potential strength
  as a function of $\kappa_{\text m}^0 a$. The vertical dashed line
  gives the asymptotic value $2\ln 1.3 \approx 0.52$ for very large
  $\kappa_{\text m}^0 a$.}
\label{fig:kappa}
\end{figure}

Adapting our results from Sec.\ref{sec:reentrance} we find the
following behavior.  Since the melting temperature is proportional to
the elastic moduli, which in turn are proportional to the potential
strength, for $\kappa a \gg 1$ we expect $\Tm$ to display the
following dependence on the screening length $\Tm \propto (\kappa a)^2
e^{-\kappa a}$. As an immediate consequence one gets (in the dilute
limit) the following implicit equation (see also
Eq.~\ref{eq:TmS_dilute_zero} and \ref{eq:TmS_dilute_inf})
\begin{equation}
   ( \kappa_{\rm m}^\infty - \kappa_{\rm m}^0 ) a    
    \approx 2 \ln \left( 1.3  
    \frac{\kappa_{\rm m}^\infty}{\kappa_{\rm m}^0} \right)  > 0 \; .
\label{eq:implicit}
\end{equation}
In particular this implies that the difference in the critical values
of the inverse screening length at infinite and zero potential
strength, $\kappa_{\rm m}^\infty$ and $\kappa_{\rm m}^0$, is positive.
In the limit $\kappa_{\text m}^0 a \gg 1$, Eq.~\ref{eq:implicit}
reduces to $(\kappa_{\text m}^\infty - \kappa_{\text m}^0 )a \approx 2
\ln 1.3 \approx 0.52$. The full solution of Eq.~\ref{eq:implicit}
together with the asymptotic result is shown in Fig.\ref{fig:kappa}.
We find Eq.~\ref{eq:implicit} to be consistent with experimental
results \cite{bechinger-leiderer:unpubl}.  It would be interesting to
test experimentally the functional dependence of $\kappa_{\text
  m}^\infty$ on $\kappa_{\text m}^0$, Fig.\ref{fig:kappa}, predicted
here.

The results of Monte-Carlo simulations appear to disagree with
experiments and with our predictions from the dislocation-mediated
melting theory when compared for large values of the potential
strength.  Whereas we find $\kappa_{\rm m}^\infty > \kappa_{\rm m}^0$,
the simulations reported in Ref.\cite{chakrabarti-etal:95} show quite
the opposite.  More recent simulations from the same group
\cite{das-sood-krishnamurthy:99} seem to refute these earlier results
and find in agreement with our theory $\kappa_{\rm m}^\infty -
\kappa_{\rm m}^0 > 0$. Their numerical value for $(\kappa_{\rm
  m}^\infty - \kappa_{\rm m}^0 ) a \approx 1.32$ is, however, more
than two times larger than our asymptotic prediction of $0.52$.
However, because Eq.~\ref{eq:implicit} neglects finite renormalization
of elastic constants by dislocation dipoles and nonlinear elastic
effects, our prediction is an estimate, only accurate upto unknown
factors of order $1$.

\begin{figure}[bht]
  \centerline{ 
  \psfrag{ka}{$(\kappa_{\rm m} a)^{-1}$}
  \psfrag{ko}{$(\kappa_{\rm m}^0 a)^{-1}$} \psfrag{ki}{$(\kappa_{\rm
         m}^\infty a)^{-1}$} \psfrag{U}{$\UQ$}
    \epsfxsize=0.9\columnwidth \epsfbox{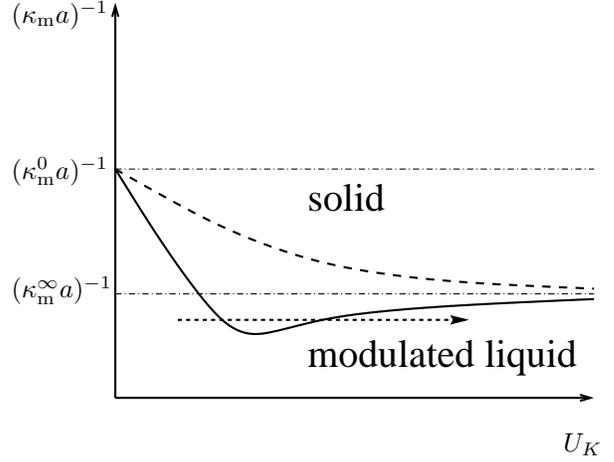}}
  \vspace{0.5truecm}
\caption{Schematic $p=1$ phase diagram as a function of potential 
  strength $\UQ$ and inverse Debye screening length $\kappa$. Solid
  and dashed curves represent the melting curves for values of
  $\kappa$ being larger or smaller respectively than the critical
  value of $\kappa_{\text crit} a \approx 5.6$.}
\label{fig:sketch_melting_mc}
\end{figure}

Next we discuss reentrance in the $\UQ/\kT$-$\kappa_{\text m} a$
phase diagram.  Upon rewriting Eq.~\ref{T_reentrance} we find
\begin{equation}
  \frac{\UQ}{\kT} = \frac{\alpha (\kappa_{\text m} a) }
                    {T/{\Tm^{\infty}(\kappa_{\text m} a)}-1} 
\label{eq:kappa_reentrance}
\end{equation}
with
\begin{equation}
  \alpha (\kappa_{\text m} a) =  
  \frac{5 \bigl( (\kappa_{\text m} a)^2-31 \bigr)}{64 \pi^2}  
  \left(1 + \frac{13}{3 \kappa_{\text m} a} \right) \, .
\end{equation}
Hence, if $\kappa_{\rm m}^0 a$ and $\kappa_{\rm m}^\infty a$ are both
smaller than the critical value $5.6$ for the existence of reentrance,
we expect $(\kappa_{\rm m} a)^{-1}$ to be a monotonically decreasing
function of the potential strength, as shown by the dashed line in
Fig.\ref{fig:sketch_melting_mc}.  If $\kappa_{\rm m}^0 a$ and
$\kappa_{\rm m}^\infty a$ are both larger than the critical value
$5.6$, we expect reentrant behavior such that with increasing
potential strength $(\kappa_{\rm m} a)^{-1}$ first decreases and
reaches a minimum $(\kappa_{\rm m}^{\rm min} a)^{-1} < (\kappa_{\rm
  m}^{\infty} a)^{-1}$ before it approaches $(\kappa_{\rm m}^{\infty}
a)^{-1}$ as an inverse power of $\UQ$ according to
Eq.\ref{eq:kappa_reentrance} (see Fig.\ref{fig:sketch_melting_mc}).
This reentrant behavior is consistent with results from experiments of
the Konstanz group
\cite{wei-bechinger-rudhardt-leiderer:98,bechinger-leiderer:unpubl}
(see the dashed arrow in Fig.\ref{fig:sketch_melting_mc}, which
describes a typical experimental path). It is also similar to what one
finds in simulations\cite{chakrabarti-etal:95} at small values of the
potential strength. However, there are significant differences. First
of all, the type of transition is very different. Whereas we discuss a
continuous dislocation mediated melting transition, simulations appear
to find a first-order transition.  Second, as discussed above, the
simulations show $\kappa_{\text m}^\infty < \kappa_{\text m}^0$ which
is opposite to what our theory predicts.  In more recent
simulations\cite{das-sood-krishnamurthy:99} $\kappa_{\rm m} a$ is
found to increase monotonically with potential strength with no sign
for reentrance. This is opposite to what was found in the earlier
simulations by the same group\cite{chakrabarti-etal:95}.

In summary, we find that our theoretical results are consistent with
recent experiments and raise strong doubts on the validity of the
Monte-Carlo results to date on melting in a 1d periodic potential.
This latter failure of simulations is not completely surprising given
difficulties of numerical methods on even larger systems to resolved
the nature of 2d melting even {\em without} an external
potential~\cite{bagchi-andersen-swope:96}.

\subsection{Static structure factor and pair correlation function}

The quantity that is most directly observed in many experiments on
colloidal systems and related simulations is the pair correlation
function, defined by
\begin{equation}
  g({\bf r}) = \frac{V}{N^2} {\sum_{ij}}^\prime 
               \left\langle 
                    \delta \bigl( {\bf r} - ({\bf r}_i-{\bf r}_j) \bigr) 
               \right\rangle \; ,
\label{pair_correlation}
\end{equation}
where the double sum is over $N$ particles but excludes the diagonal
terms where $i=j$.  It is related to the static structure factor by
\begin{equation}
  g({\bf r}) = \frac{1}{N}  \int \frac{d^2 q}{2 \pi} 
               e^{i {\bf q}\cdot{\bf r}}S ({\bf q}) \; . 
\end{equation}
Neglecting the smooth part of the structure factor and taking into
account only the center column of Bragg peaks and the two neighboring
columns of quasi-Bragg peaks with $\eta_{G_{A,x}} (T_m^-)=1/4$ and
$\eta_{G_{B,x}}(T_m^-)=1$, respectively, one finds for the pair
correlation function,
\begin{eqnarray}
  g({\bf r}) -1 =  &&\sum_{{\bf G}_1} 
                  C_{{\bf G}_1} \cos ({\bf G}_1 \cdot {\bf r})
                  \nonumber \\ 
                  &+& r^{-\eta_{G_{A,x}}} \, \sum_{{\bf G}_A} 
                  C_{{\bf G}_A} \cos ({\bf G}_A \cdot {\bf r})
                  \nonumber \\ &+&  r^{-\eta_{G_{B,x}}} \, \sum_{{\bf G}_B} 
                  C_{{\bf G}_B} \cos ({\bf G}_B \cdot {\bf r})\; ,
\end{eqnarray}
where $\eta_{G_{\alpha,x}}$ are the exponents characteristic for the LFS phase.
Note also that according to Eq.\ref{eq:LFS_exponent} these exponents
do only depend on the $x-$component of the reciprocal lattice vector
${\bf G}_\alpha$. The amplitudes $C_{{\bf G}_\alpha}$ are proportional to the
amplitudes of the corresponding Bragg peaks, ${\bf G}_1 = (2 \pi / a')
{\hat {\bf e}_y}$, and quasi-Bragg peaks, ${\bf G}_A$ and ${\bf G}_B$
with $G_{A,x} = G_x^0 = 2 \pi / a$ and $G_{B,x} = 2 G_x^0$ (see
Eqs.\ref{B_Ga}--\ref{A_Gb}).

\begin{figure}[bth]
\centerline{\epsfxsize=0.9\columnwidth 
{\epsffile{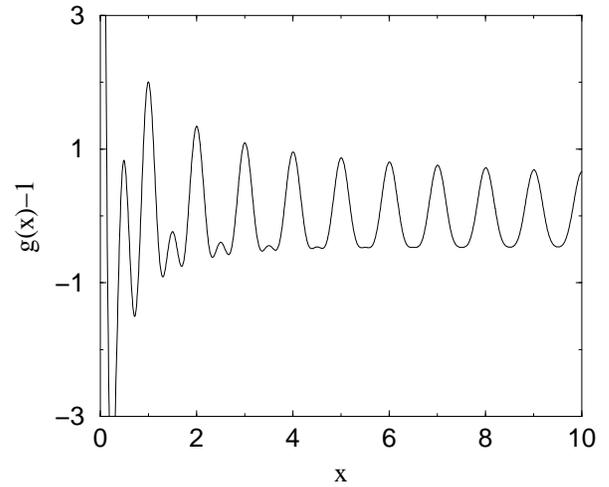}}} 
\caption{Algebraic part of the static structure factor for 
  ${\bf r} = \hatx$, i.e. looking parallel to the troughs of the laser
  potential. There are two oscillating contributions, both of which
  decay algebraically to zero.}
\label{fig:structure_factor_x}
\end{figure}

For ${\bf r} \parallel \hatx$, i.e., looking parallel to the minima of
the troughs the sum over the Bragg peaks yields a constant. This
simply reflects the effect of the laser potential to induce a periodic
modulation of the colloidal particle density with a higher density in
the minima of the troughs. Note that this trivially implies that the
pair correlation function does {\em not} approach unity as $x
\rightarrow \infty$ if $g(x)$ is normalized with respect to the mean
density.  Since the amplitudes for the quasi-Bragg peaks decay as a
power-law in the strength of the laser potential with an exponent
proportional to ${\overline \eta}_{G_y}$ a reasonable approximation for
the pair correlation function reads
\begin{eqnarray}
  g(x) -1 = const. + &g_A& \cos ( G_x^0 x) x^{-\eta_{G_{A,x}}} 
  \nonumber \\
                   + &g_B& \cos ( 2 G_x^0 x) x^{-\eta_{G_{B,x}}} \, .
\end{eqnarray}
The relative magnitude of the amplitudes $g_A$ and $g_B$ depends on
the strength of the laser potential. Whereas $g_B$ is independent of
$\UQ$ (note that the leading quasi Bragg-peak contributing to $g_B$
has $G_{B,y}=0$), $g_A$ vanishes as a nontrivial $T$-dependent power
law in $\UQ$ for $\UQ/\mu a^2 \ll 1$ (see Eq.\ref{B_Gb}) increasing
the weight of the $x^{-\eta_{G_{A,x}}}$--term with increasing
potential strength.  This prediction should be accessible to
experimental verification.  Note, that the dependence of the amplitude
$g_A$ on the potential strength may lead to $\UQ$-dependent effective
exponents when one tries to incorrectly fit the experimental data by a
single power law.  For illustration Fig.\ref{fig:structure_factor_x}
shows $g(x)-1$ for a special case, where $const. = 0$, $g_A=g_B=1$,
$\eta_{G_{A,x}}= \frac{1}{4} $, $\eta_{G_{B,x}} = 1$ and all length
are measured in units of $a$.  Due to the superposition of the two
harmonics with different power law amplitudes the minima are much
broader than the maxima of the structure factor, a feature which
appears to be present in the data of
Ref.\cite{wei-bechinger-rudhardt-leiderer:98}.
\begin{figure}[bth]
\centerline{\epsfxsize=0.9\columnwidth 
{\epsffile{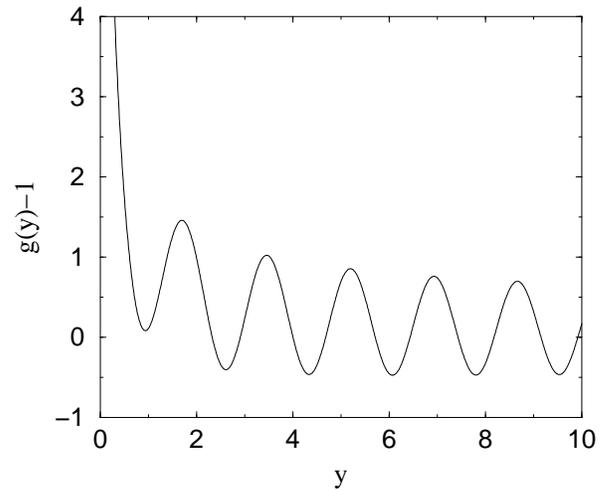}}} 
\caption{Algebraic part of the static structure factor for 
  ${\bf r} = \haty$, i.e. looking perpendicular to the troughs of the
  laser potential. Unlike Fig.\ref{fig:structure_factor_x}
  algebraically decaying oscillations are superimposed on a periodic
  contribution (not shown) which does not decay.}
\label{fig:structure_factor_y}
\end{figure} 

For ${\bf r} \parallel \haty$, i.e., looking perpendicular to the
minima of the troughs we get,
\begin{eqnarray}
  g(y) -1 = const.' \cos (2 G_y^0 y) 
            + &g_A'& \cos (G_y^0 y) y^{-\eta_{G_{A,x}}} \nonumber \\ 
            + &g_B'& y^{-\eta_{G_{B,x}}}
\end{eqnarray}
with $G_y^0 = 2 \pi / \sqrt{3} a $. Hence on top of the periodic
density modulation due to the laser potential we have again the
algebraic decay from the closest Bragg peaks. For illustration
Fig.\ref{fig:structure_factor_y} shows the algebraic part of the
static structure factor $f(x) = y^{-1/4} \cos (2 \pi y / \sqrt{3}) +
y^{-1}$ where we have again chosen the amplitudes to be equal and the
$\eta$ exponents equal to their values at the melting temperature,
$\eta_{G_{A,x}}=\frac14$, $\eta_{G_{B,x}}=1$. If one would try to fit
the envelope of this function in the regime shown in the graph using a
single power law one would find an exponent of $1/2$.  Hence caution
must be exercised in the analysis of the experimental data, and it is
essential to take into account both leading and subleading quasi Bragg
peaks.

\acknowledgements It is a pleasure to acknowledge helpful discussions
with B.I. Halperin and J. Toner. We also thank C. Bechinger, M.
Brunner and P.  Leiderer, and C.-H. Sow and C.M. Murray for
communicating their unpublished results.  L.R. was supported by the
NSF through the CAREER Award Grant No.  DMR-9625111, University of
Colorado's MRSEC Grant No.DMR-9809555 and by the A.P. Sloan and the
Packard Foundations. E.F.  acknowledges support by the Deutsche
Forschungsgemeinschaft through Grant SFB 563 and a Heisenberg
fellowship (FR 850/3-1). D.R.N.  was supported by the NSF through
Grant No. DMR97-14725, and Harvard's MRSEC Grant No.  DMR98-09363.

\appendix

\section{Variational theory of the 2d melting transition in the
presence of a 1d periodic potential}
\label{variational_appendix}

In this appendix we study the freezing transition of the modulated
liquid in the limit of a strong periodic potential. In such limit the
colloidal particles are tightly confined to the troughs of the 1d
periodic potential and our system reduces to a weakly coupled array of
1d colloidal liquids. The low energy degrees of freedom of the
resulting system are then well characterized by a scalar field
$u_n(x)$ describing particle displacements along the $n$-th trough and
an effective Hamiltonian
\begin{eqnarray}
H&=&\sum_n\int dx\bigg\{{1\over2}B
\left({d\phi_n\over d x}\right)^2\nonumber\\
&-& g\cos\left[\phi_{n+1}(x)-\phi_n(x))\right]\bigg\}\;,
\label{H-weak_coupling_appendix}
\end{eqnarray}
where for simplicity of notation we have defined rescaled phonon field
$\phi_n(x)$ and elastic couplings $B$ and $g$ related to those defined
in the Introduction through
\begin{mathletters}
\begin{eqnarray}
\phi_n(x)&=&{2\pi\over a} u_n(x)\;,\\
B&=&K d\left({a\over2\pi}\right)^2\;,\\
g&=&\mu d\left({a\over2\pi d}\right)^2\; .
\label{phi_B_g}
\end{eqnarray}
\end{mathletters}

In the Introduction we have used simple qualitative arguments to
estimate the colloidal freezing transition temperature.  Here we would
like to treat this model quantitatively and in more detail.
Unfortunately, however, as can be seen from a standard renormalization
group analysis weak coupling $g$ is always {\em irrelevant} at long
scales, with the effective coupling $g(\xi_x)$ vanishing at length scale
$\xi_x$ as
\begin{equation}
g(\xi_x)=g \left({\xi_x\over a}\right) e^{-\text{const.}(k_B T/B)\xi_x}\;.
\end{equation}
Thermal fluctuations, which are especially strong in 1d are
responsible for this effective decoupling of the colloidal system into
effectively independent one-dimensional liquids. This precludes a
description of the freezing transition in weak ($g$) coupling starting
from this model. There are two alternatives: One is to study of the
melting transition from a complementary strong coupling, fully elastic
model with topological defects (dislocations), an approach which lends
itself to a rigorous treatment that we undertake in the main part of
the paper. Alternatively, an approximate, variational treatment of the
model, Eq.\ref{H-weak_coupling_appendix} is possible and will be
presented in this appendix.

The idea behind a variational approach of a problem is that an
approximate free energy
\begin{equation}
\tilde{F}=\langle H-H_v\rangle_v + F_v
\end{equation}
is an {\em upper-bound} for the exact free energy $F$ corresponding to
the Hamiltonian $H$ of interest and where $H_v$ {\em any} other (the
so called variational) Hamiltonian, $F_v$ is the corresponding free
energy and subscript $v$ on the thermal average indicates that
Boltzmann weight with Hamiltonian $H_v$ is used. The advantage of the
variational principle can be taken if the arbitrary variational
Hamiltonian $H_v$ is judiciously chosen to be simple enough, so that
thermal averages can be calculated, but at the same time general
enough so as to be able approximately capture the physics of the full
Hamiltonian $H$.

Since, unfortunately, our abilities to compute functional integrals do
not extend beyond Gaussians, we choose a quadratic form for $H_v$
\begin{eqnarray}
H_v&=&\sum_n\int dx\left[\frac{B_x}{2} \left({d\phi_n\over d x}\right)^2
+\frac{B_y}{2} \left(\phi_{n+1}-\phi_n\right)^2\right]\;,\nonumber\\
&&
\label{H_v}
\end{eqnarray}
with $B_x$ and $B_y$ as the effective variational parameters,
respectively related to the effective long wavelength bulk and shear
moduli, latter given by
\begin{equation}
\mu=B_y d\left({2\pi\over a}\right)^2\;\label{muBy}
\end{equation}

Simple Gaussian averages then lead to the variational free energy
density $\tilde{f}(B_x,B_y)=\tilde{F}(B_x,B_y)/L_x N_y$
\begin{eqnarray}
\tilde{f}&=&\int_{\bf k}\bigg\{\left[{1\over2}(B-B_x)k_x^2-
B_y(1-\cos k_y d)\right]G_v({\bf k})\nonumber\\
&& \qquad -\frac12 k_B T \log G_v({\bf k})\bigg\} \nonumber \\
&& \quad -g\exp\left[{-\int_{\bf k}(1-\cos k_y d)G_v({\bf k})}\right]
\;,
\end{eqnarray}
where $L_x$ and $N_y(\equiv L_y/d)$ are respectively the length and
the number of laser potential troughs (i.e., the 2d dimensions of our
colloidal system) and $G_v(\bf k)$ is the Fourier transform of the
intra-trough displacement correlation function given by
\begin{equation}
G_v({\bf k})=k_B T\left[B_x k_x^2 + 2 B_y(1-\cos k_y d)\right]^{-1}\;.
\end{equation}

To find the upper-bound of the free energy density $\tilde{f}$, we now
minimize $\tilde{f}(B_x,B_y)$ over the variational parameters $B_x$
and $B_y$. Conceptually simple but tedious calculation gives
\begin{mathletters}
\begin{eqnarray}
B_x&=&B\;,\label{Bx}\\
B_y(B,g)&=& g e^{-k_B T/[\pi d(B_y B)^{1/2}]}\;.\label{By}
\end{eqnarray}
\end{mathletters}

\begin{figure}[bht]
  \narrowtext \centerline{ \epsfxsize=0.9 \columnwidth
    \epsfbox{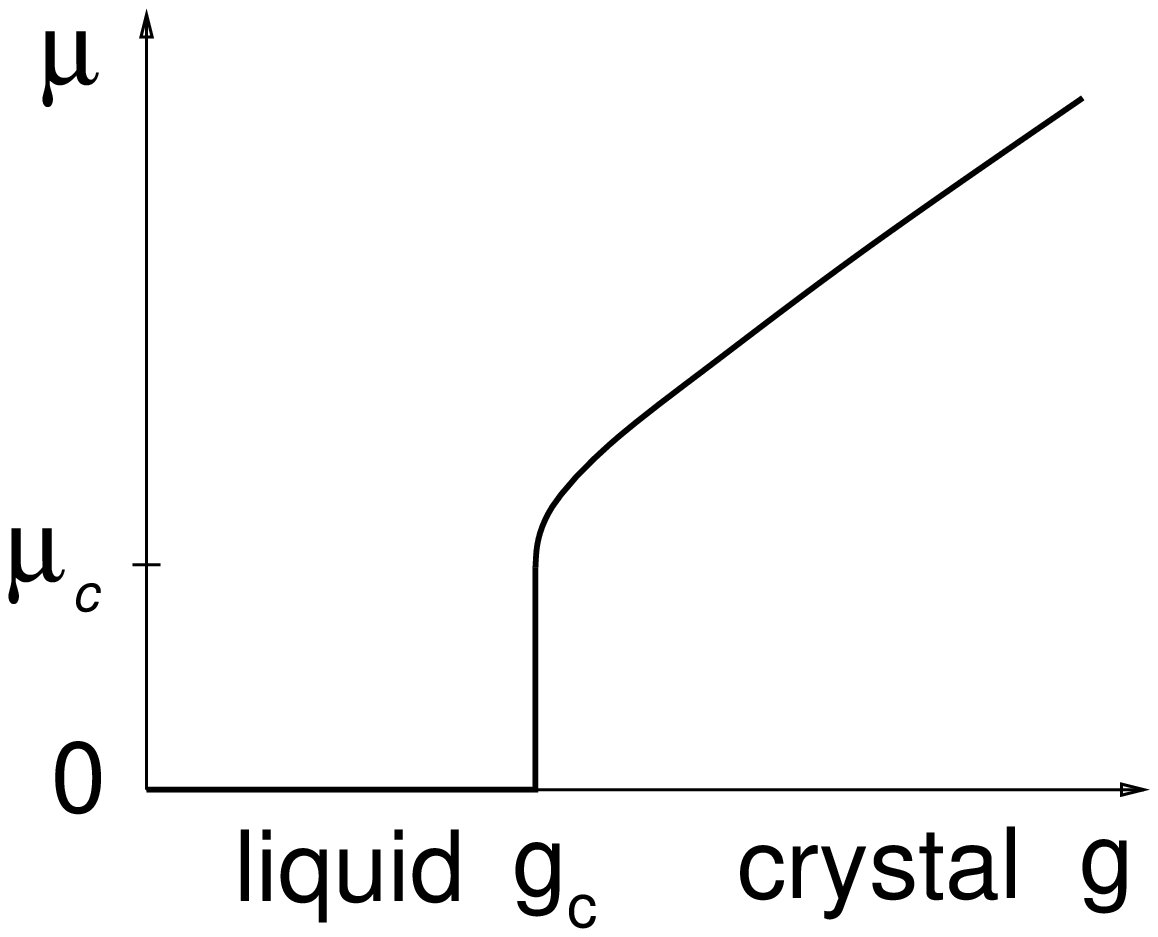}} \vspace{0.5cm}
\caption{Shear modulus $\mu$ as a function of the inter-trough coupling $g$ 
  (at fixed temperature), showing a freezing transition between
  $\mu=0$ 2d liquid and a $\mu>0$ 2d crystal, and a jump discontinuity
  at $g_c$ in the shear modulus.}
\label{fig:mu_g}
\end{figure}

Equation \ref{By}, which determines the behavior of $B_y$ and
therefore the effective shear modulus $\mu$ as a function of
temperature and inter-trough coupling $g$, illustrated in
Fig.\ref{fig:mu_g} is the main result of the variational calculation.
Simple graphical analysis of Eq.\ref{By} predicts
\begin{mathletters}
\begin{eqnarray}
B_y(g)&=&0\;,\;\;\text{for}\; g<g_c\;,\\
B_y(g)&\approx&ge^{-k_B T/\pi(g B)^{1/2}}\;,\;\;\text{for}\; g>>g_c\;,
\end{eqnarray}
\label{Bysols}
\end{mathletters}
where, the critical value of the coupling $g$ which separates the two
solutions for $B_y$ is given by
\begin{equation}
g_c=\left({\kT e\over 2\pi}\right)^2{1\over B}\;.
\end{equation}

Combining this with Eq.\ref{muBy}, we conclude that the transition
between the two solutions in Eq.\ref{Bysols} represents the freezing
of a zero shear modulus ($\mu=0$) 2d liquid into a finite shear
modulus ($\mu>0$) 2d solid. In terms of the shear modulus $\mu$ and
the bulk modulus $K$, defined by Eqs.\ref{phi_B_g} the corresponding
melting transition temperature is given by
\begin{equation}
k_B T_m={a^2\over 2\pi}\sqrt{K\mu}\;,
\end{equation}
\top{-2.5cm}
\columnwidth3.4in
\narrowtext
\noindent
a value that, upto factors of order $1$, is consistent with the
asymptotically exact prediction of our strong coupling (elastic model)
analysis given in the main text.

\section{Effective elastic constants for screened repulsive 
  Coulomb potential}
\label{effective_elastic_constants_appendix}

To calculate the effective elastic constants in the limit of large
through potential, we start from a model with a pair potential given
by a screened repulsive Coulomb potential $V(r) = V_0 a \exp (- \kappa
r) / r$, where the screening length $\kappa^{-1}$ is typically much
shorter that the mean particle spacing $a$. The total potential energy
is then given by
\begin{eqnarray}
  \Phi = \frac{1}{2} V_0 a \sum_{\langle l,l'\rangle} 
  \frac{1}{\mid {\bf R}_{ll'} \mid}
  e^{- \kappa \mid {\bf R}_{ll'} \mid } \; ,
\end{eqnarray}
where due to the short range of the potential we can safely restrict
summation to nearest neighbors, $\langle l,l'\rangle$. The distance
between the colloidal particles numbered $l$ and $l'$ can (for a
perfect lattice) be decomposed into a distance between the equilibrium
positions ${\bf r}_l$ and the displacement vectors ${\bf u}_l$:
\begin{eqnarray}
   {\bf R}_{ll'} = {\bf r}_l-{\bf r}_{l'} 
                 + {\bf u}_l - {\bf u}_{l'} \nonumber \\
                 \equiv  {\bf r}_{ll'} + {\bf u}_{ll'}
\end{eqnarray}
In the following we restrict ourselves to the primary configurations
and write the potential energy as sums over Bragg ``planes'' (i.e.,
rows of particles in $d=2$) indexed by an integer $r$ and particles
within these rows indexed by $l$,
\end{multicols}
\renewcommand{\thesection}{\Alph{section}}
\renewcommand{\theequation}{\thesection\arabic{equation}}

\begin{figure}[bth]
  \centerline{\epsfxsize=0.6\columnwidth {\epsffile{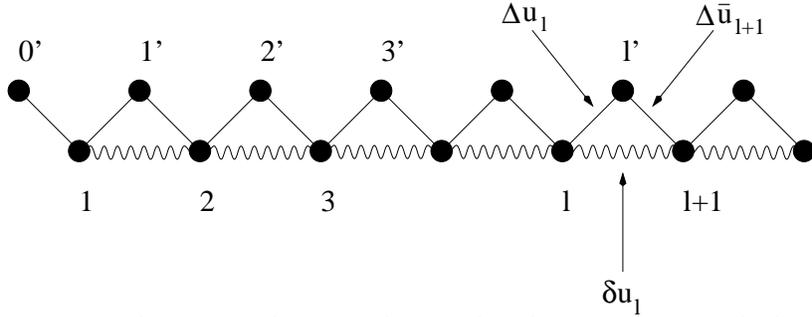}}}
\caption{Sketch of two rows of a triangular lattice of colloidal perticles 
  in a through potential illustrating two contributions to the
  effective potential energy. The sum over the lattice sites is done
  by summing along the valleys; there is one intra-valley nearest
  neighbor (wiggly line) and two inter-valley nearest neighbors one in
  the forward direction ($\Delta u$) and one in the backward direction
  ($\Delta {\bar u}$)) (solid lines).}
\label{fig:sketch_elastic_constants}
\end{figure}

\widetext
\begin{eqnarray}
\Phi = &&V_0 a \sum_{l,r}  
       \Bigl\{ \frac{1}{\left[ (a+\delta u_l)^2 + \delta h_l^2 \right]^{1/2}}
       \, \exp \left[ - \kappa \left[ (a+\delta u_l)^2 + \delta h_l^2 \right]^{1/2}
               \right] \nonumber \\
       && + \frac{1}{\left[ (a/2+\Delta u_l)^2 + (d+\Delta h_l)^2 \right]^{1/2}}
       \, \exp \left[ - \kappa \left[ (a/2+\Delta u_l)^2 + 
                                      (d+\Delta h_l)^2 \right]^{1/2}
               \right] \nonumber \\
       && + \frac{1}{\left[ (a/2+\Delta \bar u_l)^2 + (d+\Delta h_l)^2 \right]^{1/2}}
       \, \exp \left[ - \kappa \left[ (a/2+\Delta \bar u_l)^2 + 
                                      (d+\Delta h_l)^2 \right]^{1/2}
               \right] \Bigr\} \; ,
\end{eqnarray}
where the relative intra-valley and inter-valley displacement fields
are defined as follows (see Fig.\ref{fig:sketch_elastic_constants}):
\begin{eqnarray}
  \delta u_l &=& u_x(x_l+a,y_l)-u_x(x_l,y_l)\;, \\
  \delta h_l &=& u_y(x_l+a,y_l)-u_y(x_l,y_l)\;,
\end{eqnarray}
and
\begin{eqnarray}
  \Delta u_l &=& u_x(x_l+a/2,y_l+d)-u_x(x_l,y_l)\;, \\
  \Delta \bar u_l &=& -u_x(x_l-a/2,y_l)+u_x(x_l,y_l)\;, \\
  \Delta h_l &=& u_y(x_l+a/2,y_l+d)-u_y(x_l,y_l)\;.
\end{eqnarray}
\begin{multicols}{2}
\columnwidth3.4in
\narrowtext

In the strong pinning limit the laser potential $H_\Qlaser$ can be expanded
in powers of the phonon fields in the y-direction,
\begin{eqnarray}
  \beta H_\Qlaser = &&\frac{\UQ}{\kT} 
      \sum_l \cos \left( \frac{2 \pi}{d} h_l\right) \nonumber \\
      &&\approx p^2 \, \frac{8 \pi^2}{3} \, \frac{\UQ}{\kT} 
      \sum_l \left(\frac{h_l}{a}\right)^2 \nonumber \\
      &&\equiv \kT w \sum_l \left(\frac{h_l}{a}\right)^2 
\end{eqnarray}
where we have used $p d = \sqrt{3} a /2$. In the following we shall
(in order to simplify notation) measure all lengths in units of the
mean lattice spacing.

We proceed as follows: (i) first we expand all terms in the total
potential energy $\Phi$ to quadratic order in the out-of valley
displacement fields, (ii) integrate out the massive out-of-valley
modes, and (iii) take the continuum limit. Note, that it is only step
(i) which explicitly depends on the particular form of the pair
potential. For simplicity, we will limit our derivation to the leading
order in $V_0/\UQ$ and $\kT/\UQ$.
\end{multicols}
\renewcommand{\thesection}{\Alph{section}}
\renewcommand{\theequation}{\thesection\arabic{equation}}

\widetext
Step (i) gives
\begin{eqnarray}
  \beta \Phi [u,h] = \beta (\Phi_1 + \Phi_2 + \Phi_3)
\end{eqnarray}
with
\begin{eqnarray}
  \Phi_1 [u,h] &=& v \sum_l 
  \Bigl\{ 
          - \frac12 (\kappa + 1) \delta h_l^2 
          + \frac12 (\kappa^2+2\kappa+2) \delta u_l^2 
          + \beta_1 (\kappa) \delta h_l^2 \delta u_l 
          + \delta_1 (\kappa) \delta h_l^2 \delta u_l^2 
  \Bigr\} \\
  \Phi_2 [u,h] &=& v \sum_l 
  \Bigl\{ 
          - \frac18 (3\kappa^2+5\kappa+5) \Delta h_l^2 
          + \frac18 (\kappa^2-\kappa-1) \Delta u_l^2 
          + \alpha_2 (\kappa) \, \delta h_l \Delta u_l 
          \nonumber \\
          &&\qquad + \beta_2 (\kappa) \Delta h_l^2 \Delta u_l 
          + \gamma_2 (\kappa) \Delta h_l \Delta u_l^2 
          + \delta_2 (\kappa) \Delta h_l^2 \Delta u_l^2 
  \Bigr\}
\end{eqnarray}
\begin{multicols}{2}
\columnwidth3.4in
\narrowtext
\noindent
and $\Phi_3 [\bar u,h]$ obtained from $\Phi_2 [u,h]$ by the
replacement $\Delta u_l \rightarrow \Delta \bar u_l$. Here we have also introduced
\begin{eqnarray}
 \beta_1 (\kappa)  &=& \frac12 (\kappa^2 + 3\kappa + 3)\;, \\
 \delta_1 (\kappa) &=& - \frac14 (\kappa^3 + 5\kappa + 12\kappa + 12)\;,
\end{eqnarray}
and
\begin{eqnarray}
 \alpha_2 (\kappa) &=& \frac{\sqrt{3}}{4} (\kappa^2 + 3\kappa +3)\;, \\
 \beta_2 (\kappa)  &=& - \frac{1}{16} (3\kappa^3 + 14\kappa^2 +  
                                       33\kappa + 33)\;, \\
 \gamma_2 (\kappa) &=& - \frac{\sqrt{3}}{16} (\kappa^3 + 2\kappa^2 + 
                                              3\kappa + 3)\;, \\
 \delta_1 (\kappa) &=& \frac{1}{64} (3\kappa^4 + 14\kappa^3 + 55\kappa^2 + 
                                     123\kappa + 123) \; .
\end{eqnarray}
The dimensionless ratio
\begin{equation}
v \equiv e^{-\kappa} \, \frac{V_0}{\kT}
\end{equation}
measures the strength of the pair potential relative to a typical
thermal energy.  Next we integrate out the massive phonon fields $h_l$
with a Boltzmann weight given by the external potential $H_\Qlaser$,
\begin{eqnarray}
  \exp \left[ - \beta H_{\rm eff} \right]
  = \int [dh]
    \exp \left[ - w \sum_l h_l^2 - \beta \Phi [u,h] \right] \; ,
\end{eqnarray}
where $\int [dh]$ denotes an integration over the $\{ h_l \}$.
We find
\begin{eqnarray}
  \beta H_{\rm eff} =
  &&\sum_l 
  \Bigl\{ \delta u_l^2 \left[ \frac{v}{2} (\kappa^2+2\kappa+2)+\frac{v}{w}
  \delta_1(\kappa) - \frac{v^2}{w} \alpha_2^2 (\kappa) \right] \nonumber \\
  && + (\Delta u_l^2 + \Delta \bar u_l^2) 
       \left[ \frac{v}{8} (\kappa^2-\kappa-1) + 
              \frac{v}{w}\delta_2 (\kappa)
       \right] 
  \Bigr\} 
\end{eqnarray}
In the continuum limit (and reindroducing the scale $a$), we have
\begin{eqnarray}
  \delta u_l^2 &&\rightarrow a^2 (\partial_x u_x)^2 \\  
  (\Delta u_l^2 + \Delta \bar u_l^2) &&\rightarrow
  a^2 \left( \frac12 (\partial_x u_x)^2 + \frac32 (\partial_y u_x)^2 \right)
  \\
  \sum_l &&\rightarrow \frac{1}{a^2} \int d^2 x
\end{eqnarray}
we find finally our desired result, namely
\begin{equation}
H_{\rm eff}={1\over2}\int d^2 r\left[\mu_{\rm eff}(\partial_y u_x)^2+
K_{\rm eff}(\partial_x u_x)^2\right]\;,
\end{equation}
with
\begin{equation}
  \mu_{\rm eff} \approx \mu_{\rm eff}^\infty 
  \left\{ 1 + \frac{9 (\kappa a)^2}{64 \pi^2} 
              \, \left( 1+ \frac{17}{3 \kappa a} \right) 
              \, \frac{k_B T}{p^2 \UQ}
  \right\} \, ,
\end{equation}
\begin{equation}
  K_{\em eff} \approx K_{\rm eff}^\infty 
   \left\{ 1 + \frac{(\kappa a)^2}{64 \pi^2} 
              \left( 1-8v-\frac{23+104v}{3\kappa a}  
              \right) \frac{k_B T}{p^2 \UQ}
  \right\} \, ,
\end{equation}
where
\begin{eqnarray}
  \mu_{\rm eff}^\infty =
  \frac38 \, ((\kappa a)^2 - \kappa a - 1)  \, V_0 e^{-\kappa a}\; ,\\
  K_{\rm eff}^\infty =
  \frac18 \, (9 (\kappa a)^2 + 15 \kappa a + 15) \, V_0 e^{-\kappa a} \;.
\end{eqnarray}

\end{multicols} 

\begin{thebibliography}{10}
\bibitem{IinIout_liquid} The scattered intensity $\Iout$ is clearly
  proportional to the product of the scattering cross-section and the
  input laser intensity $\Iin$. Since the former is related to the
  density-density correlation function, the cross-section itself must
  be proportional to the square of the input laser intensity $\Iin$,
  which explicitly induces finite density modulation. Consequently
  $\Iout\propto\Iin^3$.
\bibitem{comment_p=1} Because an applied periodic potential will
  generically include the higher harmonics, all of our results
  for $p=1$ also directly extend to all rational fractions $1/q$, with
  $q\in Z$.
\bibitem{d_comment} The factor of $1/d^2$ is inserted into the
  definition of the discrete model purely for convenience, so as to
  have the interchannel coupling $\mu$ have the same dimensions as the
  shear modulus $\mu_{\rm eff}$ of the continuum elastic model in
  Eq.\ref{eq:KT_free_energy}, which then allows its identification as
  the microscopic shear modulus.
\bibitem{comment_otherSm} For $1<p\leq p_c$, our system in principle
  admits more exotic locked smectic phases, characterized by
  invariance under discrete translations by $n d$, where $n\in Z$, but
  $n\neq p$. Although these phases are thermodynamically distinct,
  their properties are similar to those that we consider in the main
  text.  We therefore do not study them any further here.
\bibitem{comment_longscales} Strictly speaking, the elastic
  Hamiltonian, Eq.\ref{H_0} is only valid at scales longer than the
  typical size of a thermally excited dislocation pair, such that the
  effects of topological defects can be safely incorporated into
  effective elastic constants $\mu$ and $\lambda$.
\bibitem{comment_irrelevant} It is important to appreciate that even
  in the FS phase, where the periodic potential is irrelevant in the
  renormalization group sense, its effects are nontrivial and
  experimentally observable. For example, in the presense of a
  periodic potential, even if it is irrelevant, the continuous
  translational and rotational symmetry is explicitly broken and FS
  displays true Bragg peaks in its structure function at multiples of
  $\Qlaser$.
\bibitem{Qflow} The choice of the phonon field-rescaling exponent
  $\phi$ is of course arbitrary. If instead of the convenient $\phi=0$
  choice we left it arbitrary, then to return the Hamiltonian into the
  form it had before the RG transformation, we would have to also
  allow the periodic potential wavevector $\Qlaserscalar$ to flow as
  $\Qlaserscalar(\ell)=\Qlaserscalar e^{\phi\ell}$. In the end the
  flow equations for the {\em dimensionless} coupling constants would
  involve $\Qlaserscalar(\ell)$ and would be independent of the choice
  of $\phi$.
\end{thebibliography}
\end{document}